\documentclass{lmcs}
\pdfoutput=1

\usepackage{lastpage}
\lmcsdoi{17}{4}{7}
\lmcsheading{}{\pageref{LastPage}}{}{}%
{Oct.~08,~2020}{Nov.~11,~2021}{}

\usepackage[utf8]{inputenc}

\keywords{asymptotically-tight, multivariate, worst-case, polynomial bounds}

\usepackage{amsthm}
\usepackage[inline]{enumitem}
\usepackage{blkarray}
\usepackage{amsmath}
\usepackage{amssymb}
\usepackage{fancyvrb}
\usepackage{stmaryrd}
\usepackage[draft]{fixme}
\fxsetup{targetlayout=color}
\usepackage[all]{xy}
\usepackage[algoruled,vlined,english]{algorithm2e}
\usepackage{color}
\usepackage{tikz}
\usetikzlibrary{patterns}

\newcommand{\tool}[1]{\textsc{#1}\xspace}  %for names of tools

\newcommand{\firstpaper}{\cite{BH:LMCS:2019}\xspace}
\newcommand{\ourthm}[1]{\cite[#1]{BH:LMCS:2019}\xspace}

\newcommand{\pgt}[1]{{\texttt{\upshape #1}}}
\newcommand{\lsem}{\llbracket}  %from stmaryrd math font
\newcommand{\rsem}{\rrbracket}
\newcommand{\llpar}{\llparenthesis}
\newcommand{\rrpar}{\rrparenthesis}
\newcommand{\sempar}[1]{\lsem\mbox{\pgt{#1}}\rsem}
\newcommand{\pass}{\texttt{:=}\,}
\newcommand{\semi}{\texttt{;}}

\newcommand{\X}{\texttt{X}}
\newcommand{\C}{\texttt{C}}
\newcommand{\D}{\texttt{D}}
\newcommand{\Y}{\texttt{Y}}

\newcommand{\gbox}[1]{\ensuremath{\,{#1}\,}}
\newcommand{\askip}[0]{\gbox{\texttt{skip}}}
\newcommand{\areset}[1]{\gbox{\X_{#1}\pass \pgt{0}}}
\newcommand{\acopy}[2]{\gbox{\X_{#1}\pass\X_{#2}}}
\newcommand{\ampcopy}[0]{\textbf{set}}
\newcommand{\asum}[3]{\gbox{\X_{#1}\pass\X_{#2}+\X_{#3}}}
\newcommand{\ampsum}[0]{\textbf{add}}
\newcommand{\amul}[3]{\gbox{\X_{#1}\pass\X_{#2}*\X_{#3}}}
\newcommand{\ampmul}[0]{\textbf{mul}}

\newcommand{\setS}{{\mathcal S}}
\newcommand{\setT}{{\mathcal T}}
\newcommand{\setB}{{\mathcal B}}

\newcommand{\PB}{{\texttt{PB}}}

\renewcommand{\vec}[1]{{\mathbf{#1}}}
\newcommand{\ppol}{{\texttt{\upshape MPol}}}

\newcommand{\tppol}{{\textbf{\upshape $\tau$MPol}}}
\newcommand{\abspol}{{\texttt{\upshape APol}}}
\newcommand{\tabspol}{{\texttt{\upshape $\tau$APol}}}
\newcommand{\absppol}{{\texttt{\upshape AMPol}}}
\newcommand{\tabsppol}{{\texttt{\upshape $\tau$AMPol}}}
\newcommand{\idppol}{\mathit{Id}}
\newcommand{\acc}[1]{\mathit{Acc}({#1})}
\newcommand{\acirc}{\mathop{\bullet}}   %composition of abstract multi-poly's

\newcommand\lepoly[0]{\sqsubseteq}  %polynomial syntactic order relation
\newcommand\gepoly[0]{\sqsupseteq}  %polynomial syntactic order relation
\newcommand\subsumes[0]{\rhd}  %subsumption relation
\newcommand\subsumed[0]{\lhd}  %univariate-subsumption relation

\newcommand{\evs}{{\textsc{Ev}_{\!S}}}  %simple univariate valuation
\newcommand{\evm}{{\textsc{Ev}_{\!M}}}  %matrix-based univariate valuation
\newcommand{\evmm}{{\textsc{Ev}_{\!M\!M}}}  %multivariate evaluation
\newcommand{\mtimes}{{\cdot}}  %matrix product
\newcommand{\dfm}{{\texttt{\upshape DFM}}}
\newcommand{\ldfm}{{\texttt{\upshape LDFM}}}

\newcommand{\decProgDeg}[0]{\textsc{Deg}}  %name of the decision problem

\newcommand\closure[1]{{Cl(#1)}}

\newcommand\sd[1]{{{\textsc{SD}}(#1)}}

\newcommand\itker[1]{{\llpar {#1} \rrpar}}

\newcommand\procSDL{\textsc{Solve}\xspace}

\newcommand\myforall[1]{\forall #1\ .\ } % chktex 26
\newcommand{\eqdef}{\stackrel{def}{=\;}}
\newcommand\tuple[1]{{\langle #1\rangle}}
\newcommand\fourtuple[4]{\langle #1,\ #2,\ #3,\ #4\rangle}
\newcommand\fivetuple[5]{\langle #1,\ #2,\ #3,\ #4,\ #5\rangle}

\newcommand{\is}{{\vec{i}}}  %initial state % chktex 7

\newcommand{\bools}{\mathbb{B}}
\newcommand{\ints}{\mathbb{Z}}
\newcommand{\nats}{\mathbb{N}}
\newcommand{\rats}{\mathbb{Q}}
\newcommand{\reals}{\mathbb{R}}

\newcommand{\convhull}[0]{\mathrm{convhull}}
\newcommand{\cone}[0]{\mathrm{cone}}
\newcommand{\poly}[1]{{\mathcal #1}}
\newcommand{\intof}[1]{I({#1})}
\newcommand{\inthull}[1]{{#1}_I}
\newcommand{\bitsize}[1]{{\Vert #1 \Vert}}
\newcommand{\vtxsize}[1]{\Vert #1 \Vert_{\psi}}
\newcommand{\fctsize}[1]{\Vert #1 \Vert_{\phi}}

\theoremstyle{definition}
\newtheorem{problem}[thm]{Problem}
\begin{document}

\title{Tight Polynomial Bounds for Loop Programs in Polynomial Space}

\author[A.M. Ben-Amram]{Amir M. Ben-Amram\rsuper{a}}
\address{Department of Computer Science, Holon Institute of Technology, Israel}
\email{benamram.amir@gmail.com}

\author[G.W. Hamilton]{Geoff Hamilton\rsuper{b}}
\address{School of Computing, Dublin City University, Ireland}
\email{geoffrey.hamilton@dcu.ie}

\begin{abstract}
We consider the following problem: given a program, find tight asymptotic bounds on the values of some variables at the end of
the computation (or at any given program point) in terms of its input values.  We focus on the case of polynomially-bounded variables,
and on a weak programming language for which we have recently shown that tight bounds for polynomially-bounded variables are computable.
These bounds are sets of multivariate polynomials.
While their computability has been settled, the complexity of this program-analysis problem remained open.
In this paper, we show the problem to be PSPACE-complete.  The main contribution is a new, space-efficient analysis algorithm.
This algorithm is obtained in a few steps.  First, we develop an algorithm for
univariate bounds, a sub-problem which is already PSPACE-hard. Then, a decision procedure for multivariate bounds is achieved by
reducing this problem to the univariate case; this reduction is  orthogonal to the solution of the univariate problem and uses observations on the geometry
of a set of vectors that represent multivariate bounds. Finally, we transform the univariate-bound algorithm to produce multivariate bounds.
\end{abstract}

\maketitle

\section{Introduction}%
\label{sec:intro}

A static analysis algorithm takes, as input, program code and answers a question about its possible behaviours. A standard example is to find
the set of values that a variable can assume (at a given program point).
The algorithm seeks a set of a particular kind (for example, intervals $[a,b]$) and is \emph{complete} if it can always find the tightest such result
(e.g., the smallest interval that  contains all reachable values).
 A more complex analysis may establish a \emph{relation} among the values
of several variables. While such results cannot be fully computable for programs in general, there are algorithms that guarantee a sound and complete solution
for a class of \emph{weak} programs, defined by removing or weakening some features of full programming languages. For example,~\cite{Ouaknine-polynomial-invariants} shows an algorithm that, for a particular weak language,  finds any \emph{algebraic relation}
that exists among variable
 values. Such an algorithm can establish facts like: two variables $x$ and $y$ always satisfy, at the conclusion of the program, the equation $y=x^2$.
 By retaining a copy of the input, this allows us to answer a question like: does the program square its input, or in general, can an output
value be expressed as a polynomial in the input values and what is this polynomial?

In this paper we are interested in final values that are not necessarily expressible by polynomials, but are \emph{polynomially bounded},
and we are seeking a \emph{tight asymptotic upper bound}.
A prototypical context where such a question is asked is \emph{complexity analysis}: the quantity to be analyzed may represent a counter of steps
performed by the program and we wish to establish a result such as ``the program's worst case is $\Theta(n^2)$ steps'' (where $n$ is the size of an input parameter).

In 2008, Ben-Amram et al.~\cite{BJK08} considered a simple imperative programming language with explicitly-bounded loops,
and proved that it is decidable whether computed values are polynomially bounded. This work left open the problem of
finding an explicit \emph{tight} bound for such values.
 Recently, in \firstpaper, we solved this \emph{bound analysis problem} (the class of programs remained as in~\cite{BJK08}).
However, the solution was very inefficient (doubly-exponential space) and did not settle the complexity of this problem.
In this paper we answer this question and prove that tight polynomial bounds (for the same language) can be computed in polynomial space.
Our approach is to first develop an analysis for \emph{univariate} bounds only. This simplifies the task and makes the problem of
solving it efficiently more tractable. However the full problem is to compute multivariate
bounds (such as $O(xy+yz^2)$, equivalently $O(\max(xy,yz^2))$). In Sections~\ref{sec:multivariate} and~\ref{sec:ev-multivariate} we show how
a solution to the multivariate problem can be evolved from our solution to the univariate problem.

% The results in this paper rely on those of \firstpaper; we do not wish to replicate its contents, and we also want to make this paper readable on its own.
% To achieve this, we have quarantined those proofs, which rely most heavily on \firstpaper, in an appendix. Furthermore, the appendix defines all the
% notions required for its understanding and cites any necessary theorems from \firstpaper, so that it is technically self-contained.

We now discuss, somewhat informally, some key features of the approach. We have already noted that bounds may be multivariate, i.e., depend
on multiple variables. In \firstpaper we argued that it is also necessary to compute \emph{simultaneous bounds} on multiple variables; moreover, we
find sets of simultaneous bounds.
 The capability of computing sets of simultaneous
bounds is important even if we are ultimately interested in bounding the value of a single variable.  Consider, for example, the following piece of code,
where \pgt{choose} represents non-deterministic choice:
\begin{Verbatim}[codes={\catcode`$=3\catcode`_=8}]
   choose Y := X*X or { Y := X; X := X*X }
\end{Verbatim}
For it we have two incomparable simultaneous bounds on $\X,\Y$: $\tuple{x,x^2}$ and $\tuple{x^2,x}$ (where $x$ stands for the initial value of $\X$).
This allows us to deduce that if the above command is followed by:
\begin{Verbatim}[codes={\catcode`$=3\catcode`_=8}]
   X := X*Y
\end{Verbatim}
The final value of $\X$ is $O(x^3)$.  If, in the previous phase, we had only computed separate worst-case bounds, they would have been $x^2$ for each
of the variables, and led to a loose bound of $x^4$.

A remarkable corollary of the analysis in \firstpaper is that for our class of programs, when there is a polynomial (simultaneous) upper bound on a set of
variables, it can always be matched by lower bounds; in other words, a \emph{tight} worst-case
bound can always be found. This means that the problem of computing asymptotic
upper bounds and that of computing worst-case lower bounds are the same problem.  For technical reasons, we formulate our results
in terms of worst-case lower  bounds.  More precisely, we seek functions such that, for almost all inputs, for some execution,
the output reaches
(or surpasses) this function (up to a constant factor).  We call such a function \emph{attainable} (for a full, precise definition see Section~\ref{sec:outline}).
Thus, the program property that our analyzer computes is \emph{the set of all attainable polynomial bounds for all polynomially-bounded program variables}.
These bounds are expressed by \emph{abstract polynomials}, which are ordinary polynomials stripped of their
coefficients. This agrees, of course,
with the intention of providing asymptotic (``big-Oh'' or ``big-Omega'') bounds, where coefficients do not matter.
It is not hard to see that due to this abstraction, the set of attainable bounds (for polynomially-bounded  variables) becomes finite.

Even if finite, the size of this set, and the expression of its members,
 may be a concern; indeed, in \firstpaper we argued that the size as well as the number of maximal attainable bounds may be exponential in the program size.
Here, we overcome this problem by showing that it is always possible to do with expressions of polynomial bit-size, even if the \emph{size of the set of attainable
bounds} may be exponential.  This is achieved by redefining the task of our analyzer to either \emph{verify} a given bound, or \emph{generate}
bounds---one at a time.  We also design the algorithm in a new way for achieving the desired complexity. In fact, at first we could only see how to achieve such
a result for univariate bounds.  We later saw that it is possible to extend the solution to multivariate bounds, by exploring a reduction of the multivariate-bound
problem to the univariate-bound problem.  We present our algorithms in the same order, so that the justification for each step may be understood.

\section{Preliminaries}%
\label{sec:prelim}

In this section we make some technical definitions and present the programming language that we study.

\subsection{Notations}

The set $[n]$ is $\{1,\dots,n\}$.
For a set $S$ an $n$-tuple over $S$ is a mapping from $[n]$ to $S$;
the set of $n$-tuples is denoted by $S^n$.
Natural liftings of operators to collections of objects are used, e.g., if $t$ is a tuple of integers then
$t+1$ is formed by adding $1$ to each component. The intention should usually be clear from context.
If $S$ is ordered, we extend the ordering to $S^n$ by comparing tuples element-wise (this leads to a partial order, in general, e.g., with natural numbers,
$\tuple{1,3}$ and $\tuple{2,2}$ are incomparable).

A binary relation on a set $S$ is a subset $R\subseteq S\times S$ and we use the customary notation $xRy$ instead of $(x,y)\in R$. Functions are a subclass of
relations, and we liberally switch between functional notation $y=f(x)$ and relational notation.

The post-fix substitution operator $[a/v]$ may be applied to any sort of expression containing a variable $v$, to substitute $a$ instead; e.g.,
$(x^2+yx+y)[2z/y] = x^2+2zx+2z$.
% By extension, for a set of variables $V$, the operator $[a/V]$ substitutes $a$ for every variable in $V$.

 When we write ``polynomials,'' we are referring, unless stated otherwise, to multivariate polynomials in $x_1,\dots,x_n$
that have non-negative integer coefficients.
We presume that an implementation of the algorithm represents  a polynomial as a set of monomials (which may be empty---for the zero polynomial),
 and that numbers are stored in binary,
but otherwise we do not concern ourselves with low-level implementation details.  According to context, a polynomial may refer to the function
it expresses rather than the syntactic entity.

\subsection{The core language}%
\label{sec:semantics}

Following previous publications~\cite{BJK08,B2010:DICE,BH:LMCS:2019}, we study the bound analysis problem for programs
in a weak programming language, a ``core language.''
This is an imperative
language, including bounded loops, non-deterministic branches and restricted arithmetic
expressions; the syntax is shown in Figure~\ref{fig-syntax}.

\begin{figure}[htb]
\[ \renewcommand{\arraystretch}{1.3}
\begin{array}{rcl}
\verb+X+,\verb+Y+\in\mbox{Variable} &\;\; ::= \;\; & \X_1 \mid\X_2 \mid \X_3 \mid
 \ldots  \mid \X_n\\
\verb+E+\in\mbox{Expression} & ::= & \verb+X+ \mid \verb/X + Y/ \mid
\verb+X * Y+\\
\verb+C+\in\mbox{Command} & ::= & \verb+skip+ \mid \verb+X:=E+
                                \mid \verb+C+_1 \semi \verb+C+_2
                                \mid \texttt{loop} \; \X  \; \texttt{\{C\}}
                   \mid \texttt{choose}\;  \texttt{C}_1  \; \texttt{or} \;
                   \texttt{C}_2
 \end{array} \renewcommand{\arraystretch}{1.0}\]

\caption{Syntax of the core language (from Ben-Amram, Kristiansen and Jones, 2008).}%
\label{fig-syntax}
\end{figure}

 A program has a finite set of integer variables, $\X_1,\dots,\X_n$; by convention, $n$ will denote the number of variables
throughout this article.  In code fragments we may use other identifiers ($\X$, $\Y$\dots) for convenience.
Variables store non-negative integers\footnote{An alternative definition uses signed integers, see~\cite[remark on p.~7]{BH:LMCS:2019};
this makes little difference because of the restricted arithmetics in the language.}.

The core language is inherently non-deterministic, a property motivated by the origin of its language as a \emph{weakening} of a fuller imperative language.
This type of weakening is related to the approach of \emph{conservative abstraction} often employed in program analysis:
the {\tt choose} command represents a non-deterministic choice, and replaces deterministic conditionals found in real programs.  This reflects
the idea that the
analysis conservatively abstracts any conditional command by simply ignoring the condition.
Weakening the language is also necessary if we wish our analysis problem to be solvable; otherwise we would clearly be blocked by undecidability.
 The command $\verb+loop X {C}+$ repeats \pgt{C} a (non-deterministic) number of times
bounded by the value of the variable $\pgt{X}$ upon entry to the loop. Thus, as a conservative abstraction, it
may be used to model different forms of loops (for-loops, while-loops)
as long as a bound on the number of iterations, as a function of the program state on loop initiation,
 can be determined and expressed in the language. For convenience, we stipulate that the loop-bound variable \pgt{X} may not be assigned a new value
 in the loop body (this is easily seen to be non-restrictive, since we could add a dedicated variable for the loop bound).

 An \emph{unstructured} language in which loops are implemented by conditional branching and not by structured commands may also be amenable to
 transformation into a form that our algorithms can handle, using some front-end analysis;
we will not elaborate on this but refer the reader to~\cite{BAPineles:2016}.

\subsection{Semantics, more formally}
We define a \emph{program state} to be an $n$-tuple of non-negative integers,
$\vec x = \tuple{x_1,\dots,x_n}$, representing the values of the variables $\X_1,\dots,\X_n$.
 Note the different fonts used for program variables, vectors and scalars.

The semantics
associates with every command
\verb+C+ over variables $\X_1,\dots,\X_n$
a relation $\sempar{C} \subseteq \nats^n \times \nats^n$.
 In the expression $\vec x \sempar{C} {\vec x}'$,
vector $\vec x$ (respectively ${\vec x}'$) is the state before (after) the
execution of \verb+C+.

The semantics of {\tt skip} is the identity. The semantics of an assignment
$\X_i\verb+:=E+$  associates to each store $\vec x$ a new store ${\vec x}'$ obtained by replacing
the component $x_i$ by the value of the expression \verb+E+ when evaluated over state $\vec x$,
namely with $x_j$, $x_j+x_k$ or $x_j x_k$, corresponding to expressions $\X_j$, $\X_j+\X_k$ and $\X_j \pgt{*} \X_k$.
%  \begin{align*}
% &\sempar{{\acopy{i}{j}}} = {\ampcopy}_{ij} \\
% & \sempar{{\asum{i}{j}{k}}} = {\ampsum}_{ijk} \\
% & \sempar{{\amul{i}{j}{k}}} = {\ampmul}_{ijk}
%  \end{align*}
%
Composite commands are described by the equations:
\begin{eqnarray*}
\lsem{\tt C}_1 {\tt ;C}_2\rsem &=&
  \sempar{C$_2$}\circ\sempar{C$_1$} \\
\lsem\verb+choose C+_1\verb+ or C+_2\rsem &=&
  \sempar{C$_1$}\cup\sempar{C$_2$} \\
\lsem\verb+loop X+_i\verb+ {C}+\rsem &=&
\{ (\vec{x},\vec{y}) \mid \exists i \le x_i :
\vec{x} \sempar{C}^i \vec{y} \}
\end{eqnarray*}
where $\sempar{C}^i$ represents $\sempar{C}\circ\cdots\circ\sempar{C}$
(with $i$ occurrences of $\sempar{C}$); and $\sempar{C}^0 = \idppol$.

\subsection{Multi-polynomials}

In this section we define some concepts and notations used for expressing the analysis results (including intermediate results, internal to the algorithm).

\begin{defi}
A \emph{multi-polynomial} $\vec p = \tuple{ {\vec p}[1], \dots, {\vec p}[n] }$ is an $n$-tuple of polynomials (with non-negative coefficients);
 it represents a mapping of
$\nats^n$ to $\nats^n$.
We denote by $\ppol$ the set of multi-polynomials, where the number of variables $n$ is fixed by context.
\end{defi}
A multi-polynomial $\vec p$ expresses simultaneous bounds on the variables $\X_1 \ldots \X_n$ in terms of their respective initial values $x_1 \ldots x_n$,
where each component $\vec p[i]$ represents the bound on variable $\X_i$. For example, the command $\X_4 \pass \X_2 + \X_3$ is represented by the multi-polynomial
$\vec p = \tuple{x_1,x_2,x_3,x_2+x_3}$.

Composition of multi-polynomials, ${\vec p} \circ {\vec q}$, is naturally defined since $\vec p$ supplies $n$ values for the $n$
variables of $\vec q$, and we have:  $({\vec p}\circ{\vec q})[i]= {\vec p}[i] \circ {\vec q}$.
We define $\idppol$ to be the identity transformation, ${\vec x}' = {\vec x}$ (in MP notation: ${\vec p}[i] = x_i$ for $i=1,\dots,n$).
\begin{defi}[Univariate MP]
 A univariate MP is defined as an $n$-tuple $\vec p$ where each component ${\vec p}[i]$ is a polynomial in the single variable $x$.
\end{defi}

\begin{defi}\label{def:abspol}
$\abspol$, the set of \emph{abstract polynomials}, consists of formal sums of monomials over $x_1,\dots,x_n$,
where the set of coefficients is reduced by making all non-zero coefficients equivalent to 1. In other words,
these are polynomials where the coefficients vary over the Boolean ring $\{0,1\}$.
\end{defi}

As an example, the sum of the \emph{abstract} polynomial $x+y$ and the \emph{abstract} polynomial $y+z$ is $x+y+z$, not $x+2y+z$,
as in the ring of coefficients, $1+1 = 1$ (also $1\cdot 1=1$).
The intention is that an abstract polynomial represents all the polynomials which are formed by varying its non-zero coefficients,
thus $x+y$ represents $ax+by$ for all $a,b>0$. We can also view an abstract polynomial as a set of monomials (whose coefficients are, implicitly, 1).
We use AMP as an abbreviation for abstract multi-polynomial.
We use $\alpha(\vec p)$ for the abstraction of $\vec p$ (obtained by modifying all non-zero coefficients to 1).
Conversely, if $\vec p$ is abstract we can ``coerce'' it to a concrete polynomial by considering the 1-valued coefficients as being literally 1;
this is how one should read any statement that uses an abstract polynomial where a concrete function is needed.
Abstract polynomials are used to ensure that the analysis is finite, since there is a finite number of different abstract polynomials of any given degree.
They correspond to the use of big-O (or $\Theta$) notation in typical algorithm analysis.

The set of abstract multi-polynomials is denoted by $\absppol$; this has a composition operation
${\vec p}\acirc {\vec q}$, which relates to the standard composition ${\vec p}\circ {\vec q}$
 by $\alpha(\vec p)\acirc \alpha(\vec q) = \alpha(\vec p\circ \vec q)$;
the different operator symbol (``$\acirc$'' versus ``$\circ$'') helps in disambiguating the meaning of an expression
(referring to abstract polynomials versus concrete ones).

In the special case of \emph{abstract univariate} polynomials, it is useful to restrict them to be monomials $x^d$;
this can be done since in ``big $\Theta$'' bounds, a univariate polynomial reduces to its leading monomial.
%In particular, as is well known, $\Theta(x^a)+\Theta(x^b) = \Theta(x^{\max(a,b)})$.
%Since we do not use the $\Theta$ symbol, we define, instead, the operator symbol $\binmax$ to represent this additive operation,
%namely, $x^a\binmax x^b \eqdef x^{\max(a,b)}$.
We allow the monomial $x^0$, as well as 0, and thus we obtain a semiring
isomorphic to $\tuple{\nats\cup\{-\infty\},\max,+}$.

\begin{defi}
A \emph{univariate state}  is an $n$-tuple of univariate monomials.
%Calculations with univariate states are done with the operation $\binmax$ instead of polynomial summation.
A univariate state is called positive if it does not include a zero.
\end{defi}

Based on this definition,  a univariate state can be written as $\tuple{x^{d_1},\dots, x^{d_n}}$ and
internally represented by either a vector of integers $\tuple{d_1,\dots,d_n}$ (for a non-zero monomial) or the constant 0.
% A multivariate polynomial can be applied to a univariate state, producing a univariate monomial. Since a univariate state is intended to represent a function
% (of $x$), we use the composition operator $\acirc$ for this application (this also seems convenient graphically), for example:
% \[
% (x_1^2 + x_2 x_3)\acirc \tuple{x^3,x^1,x^4} = x^6 \binmax x^{1+4} = x^6 \,.
% \]
% An AMP applied to a univariate state produces a new univariate state, for example
% \[
% \tuple{x_1^2,\, x_2 x_3,\, x_1}\acirc \tuple{x^3,x^1,x^4} = \tuple{x^6, x^5, x^4} \,.
% \]

%\begin{defi}[ordering of polynomials]
%For $p,q\in \abspol$ we define $p \lepoly q$ to hold if every monomial of $p$ also appears in $q$.  We then say that $p$ is a \emph{fragment} of $q$.
%\end{defi}

\section{Formal Statement of the Problem and our Results}%
\label{sec:outline}

Our goal is to prove polynomial-space complexity bounds for the tight-bound problem.    In \firstpaper, our algorithm was encumbered by creating and
calculating with exponentially-large objects.   In particular, we argued that there are commands for which a set of multi-polynomials
 that provides tight bounds must include
exponentially many elements.
In order to reduce the size of objects \emph{internal} to the algorithm, we have to rewrite the algorithm, but first
we circumvent the issue of exponential \emph{output size} by
adjusting the specification of the algorithm.
  This can be done in two ways. First, as one commonly does in Complexity Theory, one can reduce the problem to a decision problem:
 \begin{problem}\label{pbm:dec}
\emph{Multivariate decision problem}: Given a core-language program $P$ and a monomial $m(x_1,\dots,x_n) =
x_1^{d_1}\dots x_n^{d_n}$, report
whether $m$ constitutes a ``big-Omega'' lower bound on the highest final value of $\X_n$ for initial state $\tuple{x_1,\dots,x_n}$.
\end{problem}

There is, of course, no loss of generality in fixing the queried variable to be the highest-numbered; neither do we lose generality by restricting the
bounds to monomials (a polynomial is an asymptotic lower bound if and only if all its monomials are).
Note that we have chosen to focus on worst-case lower bounds, unlike most publications in this area, which focus on upper bounds for worst-case results.
Our reader should keep in mind, though, that
\firstpaper shows that \emph{for our core language, whenever a variable is polynomially bounded, there is a set of
polynomials which provides tight bounds for all executions}.
This means that if we form the function ``\texttt{max} of (all the lower bounds),'' we have a tight worst-case upper bound. Thus solving for lower bounds
also solves the upper bound problem. Furthermore,
we note that if we are looking at a polynomially-bounded variable, the set of abstract monomials
(or even polynomials) that \emph{lower-bound the final value} is finite. This is convenient (for instance, the expression
``\texttt{max} of (all the lower bounds)'' can be explicitly written out, if desired).
We refer to a polynomial in this set as \emph{attainable} (since the program can compute values that reach, or surpass, this polynomial,
asymptotically).

We call a function \emph{attainable} (by a program) if its values can be reached or surpassed.
% We make a deliberately general definition, since it will be used in several contexts, as further explained below.

\begin{defi}[Attainable]\label{def:attainable}
Consider a command {\C}, and a function $f:\nats^n \to\nats^n$
we say that $f$ is \emph{attainable} by {\C} if there are constants  $d>0$ and ${\vec x}_0\in \nats^n$ such that
for all ${\vec x}\ge {\vec x}_0$ there is a $\vec y$ such that
\[
  \vec x \sempar{C} \vec y \text{ and } {\vec y} \ge d f(\vec x).
\]
Given a function $f':\nats^n\to\nats$, extend the definition to $f'$ by applying the above condition to
$f({\vec x}) = \tuple{0,\dots,0,f'({\vec x})}$, in other words we use $f'$ as a lower bound on a single variable ($\X_n$).
\end{defi}
\noindent
We remark that in the above inequation, $d$ ranges over real numbers, and may be smaller than one.

Thus, for a program that
starts with data $x_1$ and $x_2$ in its two variables, and non-deterministically chooses to exchange them, the bounds $\tuple{x_1,x_2}$
and
$\tuple{x_2,x_1}$ are both attainable, while $\tuple{x_1,x_1}$ is not.

In the current work, we are attempting to find tight bounds only for variables that are polynomially bounded in terms of the input (the initial contents
of the variables).
The problem of identifying which variables are polynomially bounded is completely solved in~\cite{BJK08}, by a polynomial-time algorithm.
We will tacitly rely on that algorithm. More precisely, we assume that the algorithm is invoked as a preprocessing step for the
analysis of each loop in the program, and allows for excluding variables that may grow super-polynomially in the given loop.
This reduces the problem of handling \emph{any} loop in our language to that of handling loops in which all variables are polynomially bounded;
the key observation here is that, due to the restricted arithmetics in our language, which only include addition and product, any value derived from
a super-polynomial value is also super-polynomial. Hence, \emph{ignoring} variables which are not polynomially bounded does not lose any information
necessary for analyzing the polynomially bounded ones.

Note that in the formulation as a decision problem, the user queries about  a given polynomial; we think that it is a natural use-case, but one can also
ask the analyzer to furnish a bound:

 \begin{problem}\label{pbm:query}
\emph{Multivariate bound generation problem}: Given a core-language program $P$, and an integer $d$, find a polynomial $p(x_1,\dots,x_n)$
with the degree in every variable bounded by $d$,
 such that $p$ constitutes a ``big-Omega'' lower bound on the highest final value of $\X_n$ as a function of
$x_1,\dots,x_n$.
\end{problem}

The function of the input parameter $d$ is to (possibly) reduce the complexity of the search---not looking further than what the user is wishing to find.
There is, of course, no loss of generality in fixing the queried variable to be the highest-numbered.

For this problem we claim a solution whose space complexity is polynomial in the input size, namely the program size plus the bit-size of $d$.
Recall that if we have a polynomial-space non-deterministic algorithm, we can also determinize it in polynomial
 space (this is the essence of the famous equality NPSPACE$=$PSPACE, which follows from Savitch's Theorem~\cite{Jo:97}). This means that it is possible to transform
 the non-deterministic solution to Problem~\ref{pbm:query} (as long as it is complete)
into a deterministic generator that writes out one monomial at a time.
Moreover, by exhaustive search it is possible to
generate all the monomials needed to bound all program executions,
 even when their number is exponential.

It should thus be clear that in a similar manner we can also search for
a bound of highest degree $d_{max}$.  The
complexity of our solution to the multivariate-bound problem (Sections~\ref{sec:multivariate} and~\ref{sec:ev-multivariate})
 is \emph{output-sensitive} in the sense that the bound is
polynomial in the program size and the bit-size of $d_{max}$.

A main idea in this paper is to reduce Problem~\ref{pbm:dec} to a problem concerning univariate bounds.
In this problem, we assume that we are given a \emph{univariate initial state} in  terms of a single input variable $x$.  The initial state
 assigns to every
variable $\X_i$ an initial symbolic value of form $x^{d_i}$ for some integer $d_i \ge 0$, e.g., $\is(x) = \tuple{x, x^3, x^2}$.
The case $d_i = 0$ intuitively represents a variable whose initial value
should be treated as an unknown constant (not dependent on $x$, but possibly large).
The univariate form seems to be rather common in textbook examples of algorithm analysis (where the single parameter in terms of which we express the bounds
is usually called $n$), and the facility of stating initial polynomial values for multiple variables
may be useful to express questions like: \emph{An algorithm processes a graph. Its running time depends on the number of
vertices $n$ and the number of edges $m$.  Bound its worst-case time, \emph{in terms of $n$}, given that $m\le n^2$}. In this case we would use
an initial state like $\tuple{n,n^2}$.
Our goal is now to find an attainable univariate state, providing a lower bound on the worst-case results of a computation from the given initial state.
Here is the formal definition; note that we are providing simultaneous bounds for all variables.
\begin{defi}%
\label{def:attainable:uv}
Given a command {\C}, a function $g:\nats \to \nats^n$
and an initial univariate state $\is$,
we say that $g$ is attainable  (by {\C}, given $\is$) if there are constants $d>0$, $x_0$ such that
for all $x\ge x_0$ there is a $\vec y$ such that
\[
  \is(x) \sempar{C} \vec y \text{ and } {\vec y} \ge d g(x).
\]
As in Definition~\ref{def:attainable}, a function with codomain $\nats$ may be used an interpreted as a lower bound for just $\X_n$.
\end{defi}

\noindent
 We thus focus on the following problem.
\begin{problem}\label{pbm:uv-decision}
\emph{Univariate decision problem}:
Given core-language program $P$, a positive%
\footnote{A zero in the initial state can be accommodated in an extension discussed in Section~\ref{sec:resets}, where an assignment \pgt{X := 0} is also allowed.} % chktex 26
 initial state $\is$, the index of a chosen variable $\X_j$ and an integer $d$,
decide whether there is an attainable univariate state whose degree in the $j$th component is at least $d$.
\end{problem}

\noindent
 We also solve the corresponding bound generation problem---even for simultaneous bounds, namely
 \begin{problem}\label{pbm:uv-query}
\emph{Univariate bound generation problem}: Given a core-language program $P$, a positive initial state $\is$,  a variable index $j$ and an integer $d$, non-deterministically find an attainable univariate state
whose degree in the $j$th component is at least $d$.
\end{problem}
We have included the parameter $d$ in this problem formulation as well, since we will be able to bound the complexity of our algorithm in terms of $d$
(instead of $d_{max}$).

Here is a summary of the results we are going to prove regarding the above problems:
\begin{itemize}
\item In Section~\ref{sec:pspace-uv}, we give a polynomial space algorithm to solve Problem~\ref{pbm:uv-query}.
\item The algorithm in Section~\ref{sec:pspace-uv}, along with the complexity analysis in Section~\ref{sec:alg-complexity},
 prove that Problem~\ref{pbm:uv-decision} and Problem~\ref{pbm:uv-query} can be solved in polynomial space.
Note that this means that the space complexity is polynomial in the input size, consisting of the size of program $P$ and the bit-size of the numbers in $\is$ and $d$.
\item In Section~\ref{sec:pspace-hard} we show that this result is essentially optimal since the problem is PSPACE-hard. This clearly means that the more involved
problems (for multivariate bounds) are also PSPACE-hard.
\item In Section~\ref{sec:multivariate} we show a reduction of the multivariate-bound decision problem (Problem~\ref{pbm:dec})
 to the univariate problem (Problem~\ref{pbm:uv-decision}), and obtain in this way a solution with space complexity polynomial in terms of the size of the program and the maximum degree.
\item In Section~\ref{sec:ev-multivariate} we solve the multivariate-bound generation problem (Problem~\ref{pbm:query}).
\end{itemize}

We'd like to mention that solving the above problems also solves some variants that can be reduced to tight-bound computation:
\begin{itemize}
\item We can find tight bounds on the number of \emph{visits} to a given set of program locations (counting all locations gives the program's
time complexity, in the ordinary unit-cost model). The reduction is to instrument the program with a counter that counts the visits to the locations of
interest \ourthm{Section 2}.
\item Sometimes we want to verify a bound for the values of a variable \emph{throughout} the program, not just at its end. This can also be
easily reduced to our problem by instrumentation.
\item Similarly, one can find a tight bound on the highest value assumed by a variable at a given program location.
\end{itemize}

\section{Properties of the Core Language}%
\label{sec:coreProperties}

We next present some important properties of the core language that follow from \firstpaper and are required for our result.
First, let us cite the main result  of \firstpaper:

\begin{thm}[\firstpaper]\label{thm:tightbounds}
There is an algorithm which, for a command \verb+C+, over variables $\X_1$ through $\X_n$,
 outputs a set $\setB$ of multi-polynomials, such that the following hold, where $\PB$ is the set of indices $i$ of variables
 $\X_i$ which are polynomially bounded under $\sempar{C}$.
\begin{enumerate}
\item (Bounding) There is a constant $c$ such that   %can be the same for all p's since there are finitely many
\begin{equation*}
 \myforall{\vec x, \vec y} \vec x \sempar{C} \vec y \Longrightarrow \exists {\vec p}\in\setB \,.\forall i\in\PB \,.\, y_i \le c{\vec p}[i](\vec x)
 \end{equation*}
\item (Tightness) There are constants $d>0$, ${\vec x}_0$ such that for every ${\vec p}\in \setB$,   %same thing here
for all ${\vec x}\ge {\vec x}_0$ there is a $\vec y$ such that
\[
  \vec x \sempar{C} \vec y \text{ and } \forall i\in\PB \,.\, y_i \ge d{\vec p}[i](\vec x).
\]
\end{enumerate}
\end{thm}

We rely on this result because our new algorithm will be proven correct by showing that it matches the bounds of the previous algorithm. But the implied
\emph{existence result} is crucial in itself: the fact that a set $\setB$ as above exists, which provides both upper bounds (clause ``Bounding'') and matching
worst-case lower bounds (clause ``Tightness'').  This property is tied to the limitations of our language: in a Turing-complete programming language this
clearly does not hold---some times no polynomial tight bound exists, as for
a program that inputs $x$ and computes $\log x$, $\sqrt x$, etc.  In other cases the bound is only ``tight'' in the sense that it is matched by a lower bound
for infinitely many inputs, but not for all inputs (this is sometimes expressed by the notation $\Omega_\infty$); e.g., a program that squares a number only
if it is odd.  These cases cannot happen with our language.

Since a polynomial is a sum of a constant number of monomials, by simple arithmetics we obtain:

\begin{cor}\label{cor:monomialbounds}
For a command \verb+C+, over variables $\X_1$ through $\X_n$, assume that $\X_n$ is polynomially bounded in terms of the variables' initial values.
Then there is a set $\setS$ of \emph{monomials} such that $\max \setS$ is a tight asymptotic bound on the highest value obtained by $\X_n$ at the conclusion
of $\verb+C+$; or in more detail,
\begin{enumerate}
\item (Bounding) There is a constant $c$ such that
\begin{equation*}
 \myforall{\vec x, \vec y} \vec x \sempar{C} \vec y \Longrightarrow  \exists m\in\setS \,.\,  y_n \le c  m(\vec x)
 \end{equation*}
\item (Tightness) There are constants $d$, ${\vec x}_0$ such that for every $m\in \setS$,
for all ${\vec x}\ge {\vec x}_0$ there is a $\vec y$ with
\[
  \vec x \sempar{C} \vec y \text{ and } y_n \ge dm(\vec x) \,.
\]
\end{enumerate}
\end{cor}

We refer to monomials satisfying the Tightness condition as attainable; they constitute valid answers to Problem~\ref{pbm:query} (subject to the degree bound);
and the fact that they also provide an upper bound is, again, an important property that depends on the limitations of the programming language.
This will be used in Sections~\ref{sec:multivariate} and~\ref{sec:ev-multivariate} where we check or compute multivariate bounds in the form of monomials.

\section{Algorithmic ideas}%
\label{sec:alg-ideas}

In this section we give a brief overview and motivation of key elements in our solution.
In particular, we relate these elements to
prior work. We discuss symbolic evaluation, the motivation for working with univariate bounds, and data-flow matrices.
This is followed by an overview of the way we prove correctness of this algorithm, and the way we finally solve the multivariate-bound problem.

\subsection{Symbolic evaluation} This is an old concept. We interpret the program, with states mapping variables not to integers, but to polynomials in
$x_1,\dots,x_n$, where $x_i$ represents the initial value of the corresponding variable $\X_i$.  Due to the restricted arithmetics in our language, we
can carry out symbolic evaluation precisely, staying within the domain of multivariate polynomials.  Thus a program state is a MP\@.
However, when there are different program paths
that reach the same point, we can get different polynomials, and therefore we construct \emph{sets} of polynomials. The abstraction to
\emph{abstract polynomials} may reduce the size of this set.  Here is an example of a piece of program code, along with a symbolic state, as a set of
AMPs, before and after each command:

\begin{align*}
&\quad \{ \fourtuple{x_1}{x_2}{x_3}{x_4} \} \\
 & \X_2 \pass \X_2 + \X_4; \\
&\quad \{ \fourtuple{x_1}{x_2+x_4}{x_3}{x_4} \} \\
 &  \X_4 \pass \X_3; \\
&\quad \{\fourtuple{x_1}{x_2+x_4}{x_3}{x_3} \} \\
 & \texttt{choose \{ X}_1 \pass \X_1 + \X_3 \texttt{\} or \{ X}_2\pass \X_2 + \X_3 \texttt{\}} \\
&\quad \{ \fourtuple{x_1+x_3}{x_2+x_4}{x_3}{x_3} , \,  \fourtuple{x_1}{x_2+x_3+x_4}{x_3}{x_3} \}
\end{align*}

This is all quite straight-forward until one comes to a loop.  Since we are analysing the program statically, we have to take into account any number of iterations.
Consequently, had we worked with concrete polynomials, they would grow indefinitely.
\begin{comment}  %I feel that the example is too complex for its end  [Amir]
suppose that the above code is put in a loop:
\begin{Verbatim}[codes={\catcode`$=3\catcode`_=8}]
loop X$_3$ {
   X$_2$ := X$_2$ + X$_4$;
   X$_4$:= X$_3$;
  choose  { X$_1$:= X$_1$ + X$_3$ } or { X$_2$:= X$_2$ + X$_3$ }
}
\end{Verbatim}
\end{comment}
\begin{exa}\label{ex:simple-motivational-ex}
Consider this loop:
\begin{Verbatim}[codes={\catcode`$=3\catcode`_=8}]
loop X$_2$ {
   X$_1$ := X$_1$ + X$_2$;
}
\end{Verbatim}
Evaluating with \emph{concrete} polynomials, the value for $\X_1$
would take the values $x_1$, $x_1+x_2$, $x_1+2x_2$, $x_1+3x_2$, etc. Working with abstract polynomials, the coefficients are ignored, and we have a finite
set, which is reached after a finite number of iterations. In our example, the value for $\X_1$ does not rise beyond $x_1+x_2$.
This is good, but unfortunately, we also lose completeness: $x_1+x_2$ is not a valid asymptotic upper bound on the
final value of $\X_1$; a correct upper bound is $x_1+{x_2}^2$, taking into account the bound $x_2$ on the number of loop iterations. The passage from
$x_1+x_2$ to $x_1+{x_2}^2$ is based on correctly \emph{generalizing} the behaviour that we see in one iteration to the long-term behaviour,
and deducing  that the \emph{increment} $+x_2$ may be applied to this variable at most $x_2$ times.
Making such generalizations precisely is the main challenge addressed in the algorithm of \firstpaper.
\end{exa}

Our algorithm in the cited work gave a precise set of bounds but suffered from efficiency problems. This is inherent in the approach of generating
symbolic polynomials. In fact, it is easy to write pieces of code that generate exponentially big sets of expressions---try the following:
\begin{Verbatim}[codes={\catcode`$=3\catcode`_=8}]
  choose  { X$_3$:= X$_1$ } or { X$_3$:= X$_2$ } ;
  choose  { X$_4$:= X$_1$ } or { X$_4$:= X$_2$ } ;
  . . .
  choose  { X$_n$:= X$_1$ } or { X$_n$:= X$_2$ }
\end{Verbatim}
% Or, alternatively, to write a single, exponentially big expression (a variant of the above---which we leave to the reader).

\subsection{Univariate bounds}
Suppose that we decide to compute only univariate  bounds. We have a single input parameter $x$ and we want to express everything in terms of $x$.
Then, if we also ignore coefficients, the only functions we need are $x$, $x^2$, $x^3$,\dots and the efficiency issue, at least the problem of big
expressions, is resolved.  An abstract state is just an $n$-tuple of degrees and takes polynomial space.
However, important information is lost.  If we start with all variables set to $x$, and the loop body sets $\X_1 \pass \X_1+\X_2$,
we get the symbolic value $2x$ in $\X_1$ after the first iteration (or just $x$ if we ignore coefficients). How do we know that this becomes
$x+tx$ after $t$ iterations?  We cannot deduce it from the final expression.  We are compelled to also keep track of how the value
was computed---so we know that there is an increment that can accumulate upon iteration.  Note that the fact that the value after one iteration is different
from the value before does not necessarily mean that the difference will accumulate---consider a loop body that simply sets $\X_1\pass \X_2$, where the
initial value of $\X_2$ is bigger than that of $\X_1$.

\subsection{Data-flow matrices.}
The discussion in the last paragraph hints that it may be useful to record data-flow: does $x_1'$ (value of $\X_1$ after an iteration) depend on $x_1$
 (its initial prior to the iteration)? On $x_2$? On both?  Such information may be represented by a \emph{bipartite graph} with arcs from $x_1,x_2$,
 $\dots,x_n$ leading to $x_1',x_2'$, $\dots,x_n'$ and showing how values propagate, or a \emph{data-flow matrix}, which is the same information
 in matrix form.   This is a very concise representation which does not record the actual expressions computed, only
 the Boolean property that some $x_j'$ depends, or does not depend, on $x_i$.
Some of the previous results in complexity analysis~\cite{KasaiAdachi:80,KN04} showed that the existence of polynomial bounds on computed values may
sometimes be deduced by examining the dependence graph.
Later works~\cite{NW06,JK08} showed that by
 enriching the information in data-flow matrices (allowing for a finite number of ``dependence types'') one has sufficient information to soundly conclude
 that a result is polynomially bounded in a larger set of programs.
As a basic example, the technique allows us to distinguish a loop body
 that sets $x_1' = 2x_1$ (doubling the value of $\X_1$, leading to exponential growth upon iteration) from one that sets $x_1'=x_1+x_2$ (which entails
 polynomial growth). Ben-Amram et al.~\cite{BJK08} derived a complete decision procedure for polynomial growth-rates in the Core Language, still by tracking
 data-flow properties (with an abstraction similar to data-flow matrices) and without explicitly computing any bounds.
 In the current work, we use data-flow matrices in addition to a symbolic state, to aid the analysis of loops, specifically the precise generalization of
iterations that seem to increase certain variables. Suprisingly, it turns out that all we need is  Boolean
 matrices which record only linear data-dependence.

 \subsection{Correctness.}
 We have motivated our work by wishing to solve the efficiency issue with the algorithm of \firstpaper. However that algorithm---the so-called
\emph{Closure Algorithm}---remains the foundation of
 our current contribution.  We both motivate the polynomial-space algorithm, and formally prove its correctness, based on the idea that it should
match the results of the Closure Algorithm, in the following sense: the Closure Algorithm computes AMPs; for instance, for the loop in Example~\ref{ex:simple-motivational-ex},
it outputs the AMP ${\vec p} = \tuple{x_1+{x_2}^2,\, x_2}$.  Our algorithm works with an initial univariate state, e.g., ${\vec x} = \tuple{x^3, x^2}$.
An abstract multi-polynomial can be applied to a univariate state, e.g., ${\vec p}\acirc {\vec x} = \tuple{x^4, x^2}$ (as already mentioned, we reduce abstract
univariate polynomials to their leading monomial).  The correctness assertion which we will prove for the polynomial-space algorithm is that it can match the
result ${\vec p}\acirc {\vec x}$ (though it never maintains $\vec p$).  Our algorithm will be non-deterministic, and in this example it will also generate
additional final abstract states, including $\tuple{x^3, x^2}$.  In the \emph{soundness}
 part of the proof, we argue that every such result is attainable, by showing that it is
bounded from above by ${\vec p}\acirc {\vec x}$ for $\vec p$ computed by the Closure Algorithm; in the \emph{completeness} part, we show that for every
$\vec p$ computed by the Closure Algorithm, we can reach (or surpass) ${\vec p}\acirc {\vec x}$ (clearly, if ${\vec p}\acirc {\vec x}$
 is a maximal element in the set of attainable bounds,
then based on soundness, we will reach but not exceed it).  Thus, our proof does not validate the new analysis by relating its results with
 the concrete semantics of the
analysed program, but with the results of the Closure Algorithm, which \firstpaper already related to concrete results.

\subsection{Multivariate bounds.}
As the restriction to univariate bounds was a key element in the development and proof of the polynomial-space algorithm, it came as a surprise that the
same algorithm (practically the same code)---with a change of domain from univariate AMPs to multivariate ones---can compute tight multivariate bounds.
But our correctness proof, as outlined above, \emph{does not} work for the multivariate algorithm (otherwise we could have skipped the univariate
stage altogether).  We arrive at the result for multivariate bounds by first establishing a \emph{reduction} among decision problems---the attainability of
a multivariate bound and that of univariate bounds.  We study the degree vectors $(d_1,d_2,\dots,d_n)$ of monomial bounds $x_1^{d_1}\dots x_n^{d_n}$,
and show that these vectors define a polyhedron in $n$-space, and that a linear-programming view of the problem allows us to decide membership in this
polyhedron using the univariate analysis.  This linear-programming approach also shows that in order to obtain a tight upper bound it is not necessary to
take a \textbf{max} of all the attainable bounds---it suffices to use those that figure as the vertices of the polyhedron.  We proceed to use this insight,
along with the linear-programming connection to univariate bounds, to prove that we can compute a sound and complete set of multivariate bounds---where
by \emph{sound} we mean that the bounds are attainable, i.e., constitute lower bounds on some possible results,
and by \emph{complete} we mean that their maximum is an upper bound on \emph{all} possible results.

\section{The Closure Algorithm}%
\label{sec:closurealg}

In this section, we present a version of the algorithm of \firstpaper, computing in exponential space and time. We call it ``the Closure Algorithm''
since an important component of the algorithm (and a cause of combinatorial explosion) is the computation of transitive closure to find the
effect of any number of loop iterations.  We need the Closure Algorithm because our polynomial-space
solution evolved from this one, and moreover our \emph{proof} for the efficient algorithm relies on the correctness of its predecessor.

In fact, the algorithm below is already a step beyond \firstpaper: even if it is still exponential, it is somewhat simplified.
Since this is not the main contribution here, we have decided to present it, in this section, concisely
and without proofs, to allow the reader to understand and proceed quickly to our new polynomial-space algorithm.
 For completeness, Appendix~\ref{sec:proof-appendix} includes proofs that show that the Closure Algorithm below is equivalent to the old one \firstpaper.

\subsection{Symbolic semantics}%
\label{sec:AMPsemantics}

We present our analysis as symbolic semantics that assigns to every command $\C$ a symbolic abstract value $\sempar{C}^S \in \wp(\absppol)$,
that is, a set of AMPs.   For atomic commands, this set is a singleton:
 \begin{align*}
& \sempar{{\askip}}^S = \{\idppol\} &&
\sempar{{\acopy{i}{j}}}^S = \{{\ampcopy}_{ij}\} \\
& \sempar{{\asum{i}{j}{k}}}^S = \{{\ampsum}_{ijk}\} &&
\sempar{{\amul{i}{j}{k}}}^S = \{{\ampmul}_{ijk}\}
 \end{align*}
where ${\ampcopy}_{ij}$, ${\ampsum}_{ijk}$ and ${\ampmul}_{ijk}$ are AMPs defined by:
\begin{alignat*}{2}  \label{eq:atomicAMPs}
& {\ampcopy}_{ij}(x_1,\dots,x_n) = \tuple{x_1,\dots,x_{i-1},x_j,x_{i+1},\dots,x_n} \\
& {\ampsum}_{ijk}(x_1,\dots,x_n) = \tuple{x_1,\dots,x_{i-1},x_j+x_k,x_{i+1},\dots,x_n} \\
& {\ampmul}_{ijk}(x_1,\dots,x_n) = \tuple{x_1,\dots,x_{i-1},x_j x_k,x_{i+1},\dots,x_n} \,.
% & {\ampcopy}_{ij}[i] = x_{j}, &\text{ and }&  {\ampcopy}_{ij}[r] = x_r \text{ for all } r\ne i \\
% & {\ampsum}_{ijk}[i] = x_{j}+x_k, &\text{ and }& {\ampsum}_{ijk}[r] = x_r \text{ for all } r\ne i \\
% & {\ampmul}_{ijk}[i] = x_j x_k, &\text{ and }& {\ampmul}_{ijk}[r] = x_r \text{ for all } r\ne i
 \end{alignat*}
For composite commands, except the loop, the definition is also straightforward:
\[ \begin{array}{lcc}
\lsem\verb+choose C+_1\verb+ or C+_2\rsem^S &=&
  \sempar{C$_1$}^S\cup\sempar{C$_2$}^S \\[1ex]
\lsem{\tt C}_1 {\tt ;C}_2\rsem^S &=&
   \sempar{C$_2$}^S\acirc\sempar{C$_1$}^S  \,. \\
\end{array}
\]
To handle a loop command, $\verb+loop X+_\ell \verb+ {C}+$, we first compute $\setS = \sempar{C}^S$, obtaining a representation of the possible effects of any
\emph{single iteration}.   Then we have to apply certain operations to calculate, from $\setS$, the effect of the entire loop:
\[   \lsem\verb+loop X+_\ell \verb+ {C}+\rsem^S  =  LC({\sempar{C}^S})[x_\ell/\tau] \,. \]
The rest of this section builds up to the definition of the function $LC$ and the explanation of the above expression.

\begin{exa}\label{ex:motivate-sdl}
Consider the following loop (also considered in Section~\ref{sec:alg-ideas}):
\begin{Verbatim}[codes={\catcode`$=3\catcode`_=8}]
loop X$_3$ {
   X$_2$ := X$_2$ + X$_4$;
   X$_4$:= X$_3$;
  choose  { X$_1$:= X$_1$ + X$_3$ } or { X$_2$:= X$_2$ + X$_3$ }
}
\end{Verbatim}
The abstraction of the \emph{loop body} is the following set of AMPs:
\[
 \setS = \{ \tuple{ x_1 + x_3,\  x_2 + x_4,\ x_3 ,\ x_3 }, \quad  \tuple{ x_1,\  x_2 + x_3 + x_4,\ x_3 ,\ x_3 } \}.
\]
We will later show how this is used to compute the effect of the whole loop.
\end{exa}

\subsection{Simple Disjunctive Loops, Closure and Generalization}

We cite some definitions from \firstpaper that we use in presenting the inference of the effect of a loop from an abstraction of its body.

\begin{defi}
A \emph{polynomial transition (PT)} represents the passage from a   state $\vec x = \tuple{x_1,\dots,x_n}$ to  a new state
${\vec x}' = \tuple{x'_1,\dots,x'_n} = {\vec p}({\vec x})$ where $\vec p = \tuple{ {\vec p}[1], \dots, {\vec p}[n] }$ is a multi-polynomial.
\end{defi}

\begin{defi}\label{def:sdl}
A \emph{simple disjunctive loop (SDL)} is  a finite set $\setS$ of AMPs, representing polynomial transitions.
\end{defi}

The loop is ``disjunctive'' because the meaning is that, in every iteration, any of the given transitions may be chosen.
The sense in which \emph{abstract} MPs represent \emph{concrete} transitions calls for a detailed definition but we omit it here, since in our
context the intent should be clear:
we expect $\setS$ to be the result of analyzing the loop body with an analysis that generates asymptotically tight bounds.
Importantly, a SDL does not specify the number of iterations; our analysis of a SDL generates results parameterized by the number of iterations as well as the initial
 state. For this purpose, we now introduce $\tau$-polynomials, where $\tau$ (for ``time'') is a parameter to represent the number of iterations.
\begin{defi}
$\tau$-polynomials are polynomials in $x_1,\dots,x_n$ and $\tau$.
\end{defi}
As $\tau$ represents the number of iterations, it is not a component of the state vector, which still has dimension $n$.
If $p$ is a $\tau$-polynomial, then $p(v_1,\dots,v_n)$ is the result of substituting each $v_i$ for the respective $x_i$;
and we also write $p(v_1,\dots,v_n,t)$ for the result of substituting $t$ for $\tau$ as well.
% The set of $\tau$-polynomials in $n$ variables ($n$ known from context) is denoted $\tpol$.

We form multi-polynomials from $\tau$-polynomials to represent the
effect of a variable number of iterations.
For example, the $\tau$-polynomial transition $\tuple{x'_1,x'_2} = \tuple{x_1,\ x_2+\tau x_1}$ represents the effect of repeating ($\tau$ times) the assignment
$\X_2 \verb/:=X/_2+\X_1$.   The effect of iterating the composite command:
$\X_2 \verb/:=X/_2+\X_1\verb/; X/_3\verb/:=X/_3+\X_2$ has an effect described by  ${\vec x}' = \tuple{x_1,\ x_2+\tau x_1, \ x_3 + \tau x_2 + \tau^2 x_1}$
(note that this is an upper bound which is not reached precisely, but is correct up to a constant factor).
We denote the set of $\tau$-multi-polynomials by $\tppol$.
We should note that composition ${\vec q}\circ {\vec p}$ over $\tppol$ is performed by substituting ${\vec p}[i]$ for each occurrence of $x_i$ in $\vec q$.
Occurrences of $\tau$ are unaffected (since $\tau$ is not part of the state).

The notion of \emph{abstract (multi-) polynomials} is  extended naturally to abstract $\tau$-(multi-) polynomials. We denote the set of abstract $\tau$-polynomials
(respectively, multi-polynomials) by $\tabspol$ (respectively, $\tabsppol$).
The \emph{SDL Problem} is to compute, from $\setS$, a set of abstract $\tau$-multi-polynomials that represent the effect of any number of
loop transitions, where $\tau$ is used to express dependence on the number of transitions.  As for general programs, we focus on
loops which are polynomially bounded, and search for attainable bounds; these notions are defined as follows.

\begin{defi}
A SDL $\setS$ is said to be \emph{polynomially bounded} when there exists a $\tau$-MP, $\vec b$,  such that for all
${\vec x}\in \nats^n$, if we start with $\vec x$ and consecutively apply $t$ arbitrary transitions from $\setS$, the final state $\vec y$ satisfies
${\vec y}\le {\vec b}({\vec x},t)$.
\end{defi}
Note that we require a polynomial bound dependent on the number of iterations $t$, considering $t$ as an independent variable.
This makes the SDL analysis independent of the program context where the SDL is extracted from. Clearly, when we plug the results of SDL analysis
back into a context where the number of iterations is also bounded by a polynomial,
we shall obtain a polynomial bound on the final state in terms of the program variables (eliminating $t$).
\begin{defi}\label{def:SDLattainable}
Given SDL $\setS$ and a function ${\vec f}:\nats^{n+1} \to \nats^n$
(in the current context, $\vec f$ will be a $\tau$-MP)
we say that $\vec f$ is \emph{attainable over $\setS$} if  there are constants $d>0$, ${\vec x}_0$ such that
for all ${\vec x}\ge {\vec x}_0$, for infinitely many values $t>0$, there exist ${\vec y}_0,\dots,{\vec y}_t$ such that
\[
 {\vec y}_0 = {\vec x} ; \quad
\forall i<t\,. \exists {\vec p}\in\setS \,.  {\vec y}_{i+1} = {\vec p}({\vec y}_i)  ;
\quad
\text{and }
{\vec y}_t \ge d{\vec f}({\vec x}, t).
\]
\end{defi}

We remark that SDLs enjoy the properties we pointed out in Section~\ref{sec:coreProperties}. In particular, a complete set of attainable lower bounds provides
a tight asymptotic upper bound as well.
In \firstpaper, we studied the SDL problem, and what we present next
is an improved version of our solution from that article; the difference is explained at the end of this section.
% (and a full proof of the equivalence of the algorithms is in Appendix~\ref{sec:proof-appendix}).  %referee says: this is unnecessary repetition
The solution consists of applying two operators to a set of ($\tau$)-AMPs, which we shall now define.
The first, \emph{abstract closure}, naturally represents the limit-set of accumulated iterations.

\begin{defi}[abstract closure]
 For finite
$P\subset \tabsppol$, we define $\closure{P}$ to be the smallest set including $\idppol$ and the elements of $P$, and closed under AMP composition.
\end{defi}
In \firstpaper we proved that when loop $\setS$ is polynomially
bounded, the abstract closure of $\setS$ is finite, and can be computed in a finite number of steps.
Briefly, the reason that the closure must be finite, when the loop is polynomially bounded,
is that the polynomials in the abstract closure are of bounded degree (since they are valid lower bounds).
Moreover, there is a finite number of different abstract polynomials of any given degree.

\begin{exa}\label{ex:closure}
In Example~\ref{ex:motivate-sdl} we have seen that
\[
 \setS = \{ \tuple{ x_1 + x_3,\  x_2 + x_4,\ x_3 ,\ x_3 }, \quad  \tuple{ x_1,\  x_2 + x_3 + x_4,\ x_3 ,\ x_3 } \}.
\]
Computing $\closure{\setS}$ adds $\idppol$ as well as the composition of the above two AMPs, which equals
\[
{\vec p} = \tuple{ x_1 + x_3,\  x_2 + x_3 + x_4,\ x_3 ,\ x_3 }.
\]
The reader is invited to verify that this gives a composition-closed set.
\end{exa}

The second operation is called \emph{generalization} and its role is to capture the behaviour of
variables that grow by accumulating increments in the loop, and make explicit the dependence on the
number of iterations.   The identification of which additive terms in a MP should be considered as increments that accumulate
is at the heart of our problem, and its solution led to the definition of \emph{iterative MPs} and \emph{iterative kernel} below.

\begin{defi}
The \emph{support} $\sup p$ of a polynomial $p$ is the set of variables on which it depends, identified by their indices,
 e.g., $\sup (x_1x_3+x_4) = \{1,3,4\}$.
%  For a multi-polynomial $\vec p$ and a set $V\subseteq [n]$,  we define $\sup {\vec p}[V]$ to be $\bigcup_{i\in V} \sup {\vec p}[i]$.
 \end{defi}

\begin{defi}
For $\vec p$ an (abstract) multi-polynomial,
we say that $x_i$ is \emph{self-dependent} in $\vec p$ if $i \in \sup{\vec p}[i]$.
We also say that the entry $\vec p[i]$ is self-dependent; the choice of term depends on context and the meaning should be clear either way.
We denote by $\sd{\vec p}$ the set $\{i \,:\, i \in \sup{\vec p}[i]\}$, i.e., the self-dependent variables of $\vec p$.
\end{defi}

It is easy to see that, given that variable $\X_i$ is polynomially bounded in a loop, if $x_i$ is self-dependent, then $\vec p[i]$ must be of the form $x_i + q$,
where $q$ is a polynomial that does not depend on $x_i$ (we say that $x_i$ is \emph{linearly self-dependent} in $\vec p$).
Otherwise, $\vec p$ represents a transition that multiplies $\X_i$ by a non-constant quantity and iterating $\vec p$
will cause exponential growth.

\begin{defi}\label{def:dublin-free}
We call an (abstract) MP $\vec p$ \emph{doubling-free} if for any $i$, if entry ${\vec p}[i]$ has a monomial ${\mathfrak m}\cdot x_i$ then ${\mathfrak m} = 1$.
\end{defi}
Assuming that we have made sure that a loop under analysis is polynomially bounded (as briefly discussed in Section~\ref{sec:outline} and more fully in \firstpaper),
  then all the loop transitions must be doubling-free.
Hence if ${\vec p}[i]$ depends on $x_i$, it must include the monomial $x_i$.

\begin{defi}\label{def:iterative}
Let $\vec p \in \absppol$.
A monomial $x_{i_1}\dots x_{i_n}$ is called \emph{iterative} with respect to $\vec p$ if all $i_k$ belong to $\sd{\vec p}$.
We call $\vec p$ \emph{iterative} if all its monomials are iterative with respect to $\vec p$.
% Moreover, its dependence graph
%  $G({\vec p})$ does not have any simple cycle longer than one arc. In other words,
% it consists of a directed acyclic graph (DAG) plus some self-loops (we call this \emph{quasi-acyclic}).
\end{defi}
Iterative AMPs are crucial in the analysis of loops as, intuitively, they represent a transformation that can happen multiple times.  To see what we mean compare
the following assignment statements:
\begin{itemize}
\item $\X_1 \verb/:=X/_2\verb/+X/_3$, represented by the AMP ${\vec p} = \tuple{x_2+x_3,x_2,x_3}$.  Both $x_2$ and $x_3$ are
self-dependent and therefore $\vec p$ is iterative.  In fact, if iterated any number of times, the resulting AMP is the same (in this case even the concrete MP
remains the same, since there is no growth in the loop).
\item $\X_1 \verb/:=X/_1\verb/+X/_3$, represented by the AMP ${\vec p} = \tuple{x_1+x_3,x_2,x_3}$.  All three variables are
self-dependent and therefore $\vec p$ is iterative.  In fact, if iterated any number of times, the resulting AMP is the same. Importantly, the concrete
MP will not be the same: increments of $x_3$ \emph{accumulate} in variable $\X_1$. The algorithm will have to correctly express this growth, generalizing
$\vec p$ to $\tuple{x_1+ \tau x_3,x_2,x_3}$.
\item $\X_1 \verb/:=X/_1\verb/+X/_3; \X_3 \verb/:=X/_2$, represented by the AMP ${\vec q} = \tuple{x_1+x_3,x_2,x_2}$.  Here $x_1$ and $x_2$
are self-dependent, but $x_3$ (used in ${\vec q}[1]$) is not,
therefore $\vec q$ is not iterative.  In fact, if iterated twice, we get $x_1' = x_1+x_3+x_2$; informally, the action of the first application is to add $x_3$ to
$\X_1$ while the action of the second is to add $x_2$.  It would have been incorrect to generalize from the first step and assume that increments of $x_3$ will
be accumulated on iteration.
\end{itemize}

\begin{defi}[ordering of multi-polynomials]
For $\vec p,\vec q\in \absppol$ we define $\vec p \lepoly \vec q$ to hold if every monomial of $\vec p[i]$ also appears in $\vec q[i]$ for $i = 1,\ldots,n$.
We then say that $\vec p$ is a \emph{fragment} of $\vec q$.
\end{defi}

\begin{lem}
Let $\vec p \in \absppol$.  % be such that $G({\vec p})$ is quasi-acyclic.
Then $\vec p$ has a unique maximal iterative fragment, which shall be denoted by $\itker{\vec p}$ (``the iterative kernel of $\vec p$'').
\end{lem}

\begin{proof}
Every monomial which is not iterative in $\vec p$ will not be iterative in any fragment
of $\vec p$, so form $\vec q$ by deleting all these monomials. Then we must have ${\vec i}\lepoly{\vec q}$ for any iterative ${\vec \imath}\lepoly {\vec p}$. % chktex 7
On the other hand, $\vec q$ is clearly iterative, so $\vec q$ is the maximal iterative fragment.
\end{proof}

Note that even if all entries of $\vec p$ are non-zero, $\itker{\vec p}$ may include zeros, e.g., for ${\vec p} = \tuple{x_3,x_2,x_1+x_2}$,
where $\itker{\vec p} = \tuple{0, x_2, x_2}$.

Consider a loop that has a transition $\vec p$ with ${\vec p}[3] = x_3 + x_1 x_2$,  where $x_1$ and $x_2$ are also self-dependent in $\vec p$.
Then the initial contents of variables $\X_1$ and $\X_2$ are preserved (or may even grow) when we iterate $\vec p$; and $\X_3$ accumulates an
increment of (at least) $x_1 x_2$ on each iteration.  This motivates the following definition.

\begin{defi}[generalization]\label{def:generalize}
Let ${\vec p}\in \absppol$ be iterative; we define the \emph{generalization} of $\vec p$ to be
\[
 {\vec p}^\tau [i] = \begin{cases}
   x_i + \tau q &  \text{if $\vec p[i] = x_i+q$} \\
   {\vec p}[i] & \text{otherwise.}
\end{cases}
\]
 \end{defi}

To continue the example presented above the definition, let
\begin{align*}
\vec p &= \tuple{x_1,\ x_2+x_1,\ x_3+x_1 x_2} \\
\intertext{then}
{\vec p}^\tau &= \tuple{x_1,\ x_2 + \tau x_1,\ x_3 + \tau x_1 x_2} \,.
\end{align*}

% \subsection{The Algorithm}
% \label{sec:closurealg-final}

% We formulate the analysis  as a set of equations defining an abstract semantics
% $\sempar{C}^S$.  For atomic commands we define:
% {
%  \begin{align*}
% & \sempar{{\askip}}^S = \{\alpha(\idppol)\} &&
% \sempar{{\acopy{i}{j}}}^S = \{\alpha({\ampcopy}_{ij})\} \\
% & \sempar{{\asum{i}{j}{k}}}^S = \{\alpha({\ampsum}_{ijk})\} &&
% \sempar{{\amul{i}{j}{k}}}^S = \{\alpha({\ampmul}_{ijk})\}
%  \end{align*}
% }
% where ${\ampcopy}_{ij}$, ${\ampsum}_{ijk}$ and ${\ampmul}_{ijk}$ are the MP representations of the basic operations (Page~\pageref{eq:atomicAMPs}}).
% Note that now, the semantics of an atomic command is a singleton set \emph{including} the corresponding AMP, while in Section~\ref{sec:semantics} we
% defined the concrete semantics $\sempar{C}$ to be \emph{equal} to the mapping represented by the MP.
% \noindent
% For composite commands,
% \[ \begin{array}{lcc}
% \lsem\verb+choose C+_1\verb+ or C+_2\rsem^S &=&
%   \sempar{C$_1$}^S\cup\sempar{C$_2$}^S \\[1ex]
% \lsem{\tt C}_1 {\tt ;C}_2\rsem^S &=&
%    \sempar{C$_2$}^S\acirc\sempar{C$_1$}^S  , \\[1ex]
%   \lsem\verb+loop X+_\ell \verb+ {C}+\rsem^S &=&
% LC({\sempar{C}^S})[x_\ell/\tau],
% \end{array}
% \]
We  now define the operators used in analysing a loop command:
\begin{align}
 Gen(\setS) &= \setS \cup \{ \itker{\vec p}^\tau \mid {\vec p}\in \setS \}, \\
 LC(\setS) &= \closure{Gen(\closure{\setS})}. \label{eq:newLC}
 \end{align}
Recall that we have defined
\[   \lsem\verb+loop X+_\ell \verb+ {C}+\rsem^S = LC({\sempar{C}^S})[x_\ell/\tau]   \,. \]
This means that one takes the set $\setS$ representing the analysis of the loop body, applies the function $LC$, which generates $\tau$-AMPs,
and concludes by substituting $x_\ell$, the maximum number of iterations, for $\tau$.

\begin{exa} In Example~\ref{ex:motivate-sdl} we have seen that
\[
 \setS = \{ \tuple{ x_1 + x_3,\  x_2 + x_4,\ x_3 ,\ x_3 }, \quad  \tuple{ x_1,\  x_2 + x_3 + x_4,\ x_3 ,\ x_3 } \}.
\]
and in Example~\ref{ex:closure} we have seen that $\closure{\setS}  = \setS \cup \{\idppol, {\vec p}\}$ where
\[
{\vec p} = \tuple{ x_1 + x_3,\  x_2 + x_3 + x_4,\ x_3 ,\ x_3 }.
\]
Next, we apply generalization.  We will take a shortcut and apply it only to $\vec p$, as it subsumes the two previous AMPs. We have:
\begin{alignat}{2}
&\itker{\vec p} &=\ &\tuple{ x_1 + x_3,\  x_2 + x_3 ,\ x_3 ,\ x_3 }, \\
&\itker{\vec p}^\tau &=\ & \tuple{ x_1 + \tau x_3,\  x_2 + \tau x_3 ,\ x_3 ,\ x_3 },
\end{alignat}
Then, in the final closure computation, we construct the composition
\[
\itker{\vec p}^\tau \acirc {\vec p} = \tuple{ x_1 + x_3 + \tau x_3,\  x_2 + x_3 + x_4 + \tau x_3 ,\ x_3 ,\ x_3 }
\]
and this is as high as we get. So we finish by substituting $x_3$ (being the loop bound) for $\tau$, obtaining the result
\[
\tuple{ x_1 + x_3 + {x_3}^2,\  x_2 + x_3 + x_4 +  {x_3}^2 ,\ x_3 ,\ x_3 } .
\]
% It is, of course, valid to simplify this  to $\tuple{ x_1 + x_3^2,\  x_2 + x_3^2 + x_4 ,\ x_3 ,\ x_3 }$.
\end{exa}

We conclude this section with some comments about the improvement of the above algorithm on
\firstpaper.  First, in \firstpaper, iterative kernels were not used.
Generalization was applied to \emph{idempotent} elements in the closure set (satisfying ${\vec p}\acirc {\vec p} = {\vec p}$);
an idempotent AMP is not necessarily iterative,
and the definition of the generalization operator was slightly more complex than the definition for
iterative AMPs.   The choice to focus on idempotent elements was guided by an algebraic theorem that we used to prove the algorithm correct
(Simon's Forest Factorization Theorem). Another difference is that in the original algorithm,
the computation of the closure and subsequent generalization had to
be performed in several rounds (until the set stabilized); in contrast, when generalization is applied to iterative kernels as done here, it suffices to compute
closure, generalize and close once more, as expressed by~\eqref{eq:newLC}.

\section{The Polynomial-Space Algorithm}%
\label{sec:pspace-uv}

In this section we provide the polynomial-space algorithm for the univariate bound problem.  We will try to present the algorithm in a way which motivates
it and clarifies why it works.
First, let us remind the reader that
we are trying to compute attainable functions: such a function describes a bound
that can be reached (for almost all inputs, on at least one possible computation for every
such input), and therefore
provides a lower bound on the worst-case result; moreover, we know by Theorem~\ref{thm:tightbounds} that a complete set
of attainable bounds exists, in the sense that it
also implies a tight upper bound (specifically, the \textbf{max} of this finite set of functions is a worst-case upper bound).
Intuitively, since an attainable function is ``something that can happen,'' to find such functions we just have to evaluate the program symbolically
and see what it computes: expressed as a function of the input, this will be an attainable function.
Our evaluator will be non-deterministic and so generate a set of attainable functions. It will be then necessary to prove that this is a complete set.

\subsection{The Simple Case:  Without Additions in Loops}%
\label{sec:evs}

As long as we do not need generalization---which is only necessary when additions are performed within a loop---our analysis (Section~\ref{sec:closurealg})
 is nothing more than
a symbolic evaluation, where instead of maintaining a concrete state as a vector of numbers and applying the operations numerically, we maintain
AMPs that represent the computation so far, and apply symbolic arithmetics to them.
While $\sempar{C}^S$ represents all the possible effects of \pgt{C}, as a set, we develop in this section a
non-deterministic evaluator, which only follows one possible path.  Thus
a \pgt{choose} command will be implemented verbatim---by non-deterministically choosing one branch
 or the other.
Similarly, in a loop, the non-deterministic evaluator literally iterates
 its body a non-deterministic number of times.  The information that it has to maintain, besides the current command, is
just the current symbolic state. Importantly, since we are provided with an initial univariate state, all our computation is with such states,
which are much more compact than AMPs. Given a degree bound $d$,
a univariate state can be represented by a vector in ${[d]}^n$.  This representation requires polynomial space.
% Because it seems more readable, we will write the elements of a univariate state in our text
%  as $x^{d_i}$
% even if the only significant piece of information is the exponent.

Next, we give a set of definitions for a simple non-deterministic interpreter $\evs$, which does not handle generalization (we give this simple version
first for the sake of presentation). Throughout this presentation,
 $\vec s$ refers to a univariate state.

The interpretation of \pgt{skip}, assignment and multiplication commands is the obvious
 \[
\evs\sempar{\askip} \vec s  \Rightarrow   \vec s \,; \quad
\evs\sempar{\acopy{i}{j}} \vec s   \Rightarrow    {\ampcopy}_{ij} \acirc \vec s \,; \quad
\evs\sempar{\amul{i}{j}{k}} \vec s   \Rightarrow    {\ampmul}_{ijk} \acirc \vec s
 \]
where $\ampcopy$, $\ampmul$ are as in the last section, and $\Rightarrow$ is the evaluation relation.
Since the addition of abstract univariate monomials $x^a $ and $x^b$ is $x^{\max(a,b)}$, we could simulate
addition $\asum{i}{j}{k}$ by replacing ${\vec s}[i]$ with $\max({\vec s}[j],{\vec s}[k])$;
however we prefer to use non-determinism and define two alternatives for evaluating this command:
\begin{alignat*}{2}
 &\evs\sempar{{\asum{i}{j}{k}}} {\vec s} &\quad \Rightarrow  \quad & \ampcopy_{ij} \acirc {\vec s}  \\
 &\evs\sempar{{\asum{i}{j}{k}}} {\vec s} &\quad \Rightarrow  \quad & \ampcopy_{ik} \acirc {\vec s}
 \end{alignat*}
This
non-determinism means that we can get the precise result as well as another one which is
lower and seems redundant. However, this redundancy turns out to be advantageous later, when we get to multivariate bounds.

The interpretation of the choice and sequencing commands is natural, using non-determinism to implement \pgt{choose}.

\begin{alignat*}{2}
%   &\evs\sempar{choose C$_1$ or C$_2$} {\vec s} &\quad = \quad &\pgt{either }  \evs\sempar{C$_1$} {\vec s} \pgt{  or } \evs\sempar{C$_2$} {\vec s}
  &\evs\sempar{choose C$_1$ or C$_2$} {\vec s} &\quad \Rightarrow \quad & \evs\sempar{C$_1$} {\vec s} \\
  &\evs\sempar{choose C$_1$ or C$_2$} {\vec s} &\quad \Rightarrow \quad & \evs\sempar{C$_2$} {\vec s} \\
  &\evs\sempar{C$_1$; C$_2$} {\vec s} &\quad \Rightarrow\quad & \evs\sempar{C$_2$} \left( \evs\sempar{C$_1$} {\vec s} \right)
 \end{alignat*}

For the loop, since we are assuming that there is no addition---hence generalization is not necessary---we just compute composition-closure,
i.e., iterate the loop body any finite number of times, using
the auxiliary function $\evs^*\sempar{C}$:
\begin{alignat*}{2}
&\evs^*\sempar{C} {\vec s} &\quad \Rightarrow \quad &  {\vec s} \\
&\evs^*\sempar{C} {\vec s} &\quad \Rightarrow \quad &  \evs^*\sempar{C} \left( \evs\sempar{C} {\vec s} \right) \\
\ \\
&\evs \lsem\pgt{loop X \{C\}}\rsem {\vec s} & = \quad & \evs^*\sempar{C} {\vec s}
\end{alignat*}
For any $\vec x$, the set $\sempar{C}^S \acirc {\vec x}$ and the set of possible results of evaluating $\evs\sempar{C} {\vec x}$
 have the same maximal elements; thus they define the same worst-case bound.

The difference between computing with full AMPs and evaluating with univariate states as above can be explained as follows: the effect
of any computation path of the program is given by an expression
\[ {\vec p}_{1} {\vec p}_{2} \dots {\vec p}_j \]
where ${\vec p}_i$ represents the $i$th atomic command (i.e., an assignment) in the path, and juxtaposition ${\vec p} {\vec q}$ means applying ${\vec q}$
after ${\vec p}$ (left-to-right composition).  In order to evaluate the result given an initial state $\is$, we compute
\[ \is  {\vec p}_{1} {\vec p}_{2} \dots {\vec p}_j \,.\]
The point is that thanks to associativity, we can parenthesize such an expression in different ways. The expression
\[ \is ( {\vec p}_{1} {\vec p}_{2} \dots {\vec p}_j )\]
represents transforming the path to an AMP (our symbolic semantics from the last section) and then applying the AMP to the given initial state.
On the other hand, $\evs$ implements the computation
\[ ((\is {\vec p}_{1} ) {\vec p}_{2} ) \dots {\vec p}_j \]
(which is our default reading of a product---left-associative), which gains efficiency, since we only maintain a univariate state.

% Note that the effect of using \emph{non-deterministic choice} to evaluate an addition instruction is that, instead of a single big polynomial, our analysis
% non-deterministically generates the set of its monomials. Here is a small example: for the code
% \begin{Verbatim}[codes={\catcode`$=3\catcode`_=8}]
%    X$_1$:= X$_2$ + X$_3$;
%    X$_3$:= X$_1$ * X$_2$
% \end{Verbatim}
% we do not get the AMP $\tuple{x_2+x_3, x_2, x_2^2+x_2x_3}$,  but, instead, we get either
%  $\tuple{x_2, x_2, x_2^2}$ or
%  $\tuple{x_3, x_2, x_2x_3}$; when applied to an initial univariate state, one of which will provide the tight bound, depending on which of $x_2$ and $x_3$
%  has the higher degree in the initial state. Technically, what we obtain is a set of \emph{fragments} of the AMPs which constitute the precise symbolic semantics
%  of the program.

\subsection{Data-flow matrices}%
\label{sec:matrices}

When we consider loops that include addition, we have to introduce generalization, in order to account for repeated increments.
From the definition of the function $LC$ (Eq.~\ref{eq:newLC}), we can represent each of the values of $LC( \sempar{C}^S)$ by an expression of this form:
\begin{equation} \label{eq:exp}
 {\vec p}_{11} {\vec p}_{12} \dots {\vec p}_{1 j_1} \,\llpar {\vec p}_{21} {\vec p}_{22} \dots {\vec p}_{2 j_2} \rrpar^\tau \,
     {\vec p}_{31} {\vec p}_{32} \dots {\vec p}_{3 j_3} \, \llpar {\vec p}_{41} {\vec p}_{42} \dots {\vec p}_{4 j_4} \rrpar^\tau \cdots
\end{equation}
where each ${\vec p}_{ij}$ is an AMP from $ \sempar{C}^S$, representing a single iteration.

Our interpreter will start with a given univariate state and repeatedly apply the body of the loop; due to non-determinism, each application is
equivalent to the application of one possible AMP from $\sempar{C}^S$, so we are effectively applying
${\vec p}_{11} {\vec p}_{12} \dots $.  However, the parenthesized
expressions need a different handling. In order to compute $\llpar {\vec p} \rrpar ^\tau$, even if we only want to evaluate it on a univariate state---i.e.,
$\is \llpar {\vec p} \rrpar ^\tau$---we need to know more about $\vec p$: namely, we have to identify self-dependent variables, and to ensure that
generalization is sound (i.e., only applied to iterative fragments).
In order to track variable dependences and verify self-dependence, we now introduce \emph{data-flow matrices} and add their maintenance to our interpreter.

 \begin{defi}[Data-Flow Matrix]
Given ${\vec p}\in \absppol$, its data-flow matrix $\dfm({\vec p})$ is a Boolean matrix $M \in M_n(\bools)$
such that $M_{ij}$ is 1 if ${\vec p}[j]$ depends on $x_i$, and 0 otherwise.
\end{defi}

 \begin{exa}\label{ex:dfm}
 A data-flow matrix can be associated with any sequence of assignments. Consider
the command $\X_1 \verb/:=X/_1\verb/*X/_2\verb/; X/_3 \verb/:=X/_2\verb/+X/_3$, in a program with $n=3$ variables,
it is represented by the AMP ${\vec p} = \tuple{x_1 x_2, x_2, x_2+x_3}$.  The data-flow matrix is
 \[\dfm({\vec p}) = \begin{bmatrix} 1 & 0 & 0\\ 1 & 1 & 1\\ 0 & 0 & 1\end{bmatrix} \,. \]
\end{exa}

In previous works~\cite{NW06,JK08,BJK08}, matrices were not Boolean, and the values of their components used to differentiate types of value dependence
(the precise meaning of ``type'' varies). For our algorithm, we find that we only need to track one type of dependence, namely linear dependence.
Non-linear dependence will not be tracked at all (it could be, but this would complicate the algorithm; we go by the rule of maintaining the minimal
information necessary for a purpose).

\begin{defi}[Linear Data-Flow Matrix]
Given ${\vec p}\in \absppol$, its \emph{linear} data-flow matrix $\ldfm({\vec p})$ is a Boolean matrix $M \in M_n(\bools)$
such that $M_{ij}$ is 1 if ${\vec p}[j]$ depends on $x_i$ \emph{linearly}, and 0 otherwise.
\end{defi}

 \begin{exa}
For $\vec p$ as in Example~\ref{ex:dfm},
 \[\ldfm({\vec p}) = \begin{bmatrix} 0 & 0 & 0\\ 0 & 1 & 1\\ 0 & 0 & 1\end{bmatrix} \,. \]
\end{exa}

 In order to compute data-flow matrices for composite commands we use ordinary Boolean matrix
product. The reader is invited to verify that, if $A$ represents a command \pgt{C} and $B$ represents command \pgt{D},
 then $A\mtimes B$ represents the data-flow in $\pgt{C;D}$. This works both for $\dfm$ and for $\ldfm$.  In terms of AMPs, we have
 \begin{lem}\label{lem:dfm-pq}
The matrix abstraction commutes with left-to-right composition: $\dfm({\vec p}{\vec q}) = \dfm({\vec p})\dfm({\vec q})$ and
$\ldfm({\vec p}{\vec q}) = \ldfm({\vec p})\ldfm({\vec q})$.
\end{lem}

Let us define some notations.  $I$ is the identity matrix (we always assume that the dimension is fixed by context; it is the number of variables in the
program under analysis).   The ``assignment'' matrix $A(j,i)$ represents the data-flow in the command $\acopy{i}{j}$.  It differs from $I$ in a single column,
column $i$, which is the $j$th unit vector:
\[
\renewcommand{\arraystretch}{1.2}
A(j,i) = \quad
\begin{blockarray}{cccccc}
 & & & i \\
\begin{block}{c[ccccc]}
    & 1   &  & 0\\
   & & \ddots & 0  \\
   &  & & 0 & \\
  j &  & & 1 & \ddots \\
    &  & & 0 &  & 1 \\
\end{block}
\end{blockarray}
 \]

We similarly define $E(i)$ (``erase'') to differ from $I$ in having the $i$th column zero.  This represents the LDFM of $\amul{i}{j}{k}$.
The following observations will be useful.
\begin{itemize}
\item Let $M$ be an LDFM representing some transformation $\vec p$.
  Then $I\land M$ (the $\land$ is applied component-wise) gives the self-dependent variables  of $\vec p$ (as the non-zero diagonal elements).
 \item $(I\land M)\cdot M$ is precisely $\ldfm(\itker{\vec p})$.
\end{itemize}

\subsection{Maintaining Data-Flow Matrices in Non-Looping Commands}%
\label{sec:evm}

Next, we give to each non-loop command a (non-deterministic) semantics in terms of both a univariate state and a data-flow matrix. The evaluator $\evm$
computes this semantics. It is invoked as $\evm \sempar{C} {\vec s}$, to evaluate command $\C$ over initial univariate state
$\vec s$.
This returns a pair: the final state and a data-flow matrix, abstracting the computation that took the given state to the final one.
See Figure~\ref{fig:evm}.  Note that in the case of addition, even if we use the fact that the sum of two abstract monomials equals one of them,
we still record the data-flow from both summands.

\begin{figure}[tb]
 \begin{alignat*}{2}
&\evm\sempar{{\askip}} {\vec s} &\quad \Rightarrow  \quad & (\vec s, I) \\
&\evm\sempar{{\acopy{i}{j}}} {\vec s} &\quad \Rightarrow \quad & ({\ampcopy}_{ij} \acirc {\vec s} , A(j,i) )    \\
&\evm\sempar{{\amul{i}{j}{k}}} {\vec s} &\quad \Rightarrow \quad & ({\ampmul}_{ijk} \acirc {\vec s},  E(i) ) \\
&\evm\sempar{{\asum{i}{j}{k}}} {\vec s} &\quad \Rightarrow  \quad & ({\ampcopy}_{ij} \acirc {\vec s}, A(j,i)+A(k,i) )  \\
&\evm\sempar{{\asum{i}{j}{k}}} {\vec s} &\quad \Rightarrow  \quad & ({\ampcopy}_{ik} \acirc {\vec s}, A(j,i)+A(k,i) )  \\
 & \evm\sempar{choose C$_1$ or C$_2$} {\vec s} &\quad \Rightarrow \quad& \evm\sempar{C$_1$} {\vec s} \\
 & \evm\sempar{choose C$_1$ or C$_2$} {\vec s} &\quad \Rightarrow \quad& \evm\sempar{C$_2$} {\vec s} \\
 &\evm\sempar{C$_1$; C$_2$} {\vec s} &\quad \Rightarrow\quad &   \pgt{let }({\vec s}_1, M_1) = \evm\sempar{C$_1$}{\vec s}  \pgt{;}
\quad   \pgt{let }({\vec s}_2, M_2) = \evm\sempar{C$_2$}{\vec s}_1\\
&&&  \pgt{in}\quad ({\vec s}_2, M_1\mtimes M_2)
\end{alignat*}
\caption{The main function of the analyzer, except for loop handling.}%
\label{fig:evm}
\end{figure}

\subsection{Handling loops}%
\label{sec:loops}

When we handle a loop, we will be (in essence) evaluating expressions of the type in Section~\eqref{eq:exp}.
We now outline the ideas used for simulating a parenthesized sub-expression $\llpar {\vec p}_{1} {\vec p}_{2} \dots {\vec p}_j\rrpar^\tau$.

We introduce the following notation
 for changing a set of components of $\vec x$:

 \begin{defi}
 For $Z\subset [n]$, and an $n$-tuple $\vec x$,
 ${\vec x}[Z\gets 0]$ is obtained by changing the components indexed by $Z$ to 0.
 \end{defi}

 We can use this notation with univariate states as well as with (abstract) MPs.  In fact, we note that ${\vec x}[Z\gets 0] = \idppol[Z\gets 0]\acirc {\vec x}$.

Let us recall how, in the Closure Algorithm, we arrive at $\itker{\vec p}^\tau$ for some ${\vec p} \in \closure{\sempar{C}^S}$.
\begin{enumerate}
\item We form ${\vec p} \in \closure{\sempar{C}^S}$ as
${\vec p} =  {\vec p}_{1} {\vec p}_{2} \dots {\vec p}_j$ where every ${\vec p}_i \in \sempar{C}^S$ represents the effect of an iteration.
\item We reduce $\vec p$ to $\itker{\vec p}$. This involves deleting monomials which are not self-dependent.
\item We apply generalization to form $\itker{\vec p}^\tau$.
\end{enumerate}
Our PSPACE algorithm never forms $\vec p$.  But it goes through ${\vec p}_{1}, {\vec p}_{2} \dots {\vec p}_j$ while simulating their action on an
initial univariate
state, say $\vec x$.  If we always do this faithfully, we end up with a result that corresponds to an application of $\vec p$ to the univariate state,
${\vec p}\acirc {\vec x}$.
Now we have two challenges:
\begin{enumerate}
 \item how to obtain $\itker{\vec p}\acirc {\vec x}$, which may differ from ${\vec p}\acirc {\vec x}$.
 \item how to simulate generalization.
\end{enumerate}

\noindent
 The solution to both challenges involves two tools: first, the data-flow matrix; second, we modify the initial state for the simulation of this sequence
 of loop iterations by non-deterministically setting a set of components to \emph{zero}.  Let us explain how this handles the challenges.

 \begin{enumerate}
 \item $\itker{\vec p}$ is obtained from $\vec p$ by deleting all non-iterative monomials.
It is easy to see that \[\itker{\vec p}\acirc {\vec x} = {\vec p}\acirc ({\vec x}[\overline{\sd{\vec p}}\gets 0]),\]
where by $\overline S$ we mean the complement of $S$.
Our interpreter will guess a set $Z$ and replace $\vec x$ with ${\vec x}[Z\gets 0]$. After simulating a number of iterations of the loop---corresponding to the
application of some ${\vec p}\in \closure{\sempar{C}^S}$---we verify that we made a valid guess, meaning that all variables not in $Z$ are
self-dependent in $\vec p$, using the associated data-flow matrix.  Note that a set larger than $\overline{\sd{\vec p}}$ is considered valid; this
is important for the next point.

\item Consider ${\vec p}[i]$ of the form $x_i + q$, where $q(x_1,\dots,x_n)$ depends only on self-dependent variables:
 such an entry is subject to generalization, multiplying $q$ by $\tau$,
which is later replaced by the loop bound $x_\ell$. In algorithm $GEN$ below, we shall multiply by $x_\ell$ directly without the intermediate use of the
$\tau$ symbol. So the expected result is
$x_i + x_\ell \cdot q$, which, when evaluated with abstract univariate states, is $\max(x_i, x_\ell \cdot q)$.
We see that we can get a correct result in the following way: we non-deterministically either set $x_i$ to 0, and at the end
multiply the $i$'th component of the result by $x_\ell$, or we modify neither $x_i$ nor the final state.
  In the first case we  compute $x_\ell \cdot q$, and in the
second $\max(x_i, q)$.  The latter result is always sound (it skips generalization), while the former is sound as long as generalization can be
rightfully applied, a condition which can be checked using the data-flow matrix (namely that we have evaluated an iterative MP and that $i\in \sd{\vec p}$).
\end{enumerate}

\noindent
The code for interpreting a loop is  presented in Figure~\ref{fig:evmoast}.
% duplicate explanation - see below figure
%  uses function $\evm^*$ for ``straight'' computation, yielding elements of $\closure{\sempar{C}^S}$,
% and function $\evm^\oast$ for applying generalization where possible (using function $GEN$, explained below).
%
\begin{figure}[t]
\[\begin{array}{lcl}
\evm \lsem\verb/loop X/_\ell \verb/{C}/\rsem{\vec s} & \Rightarrow & \evm^\oast\sempar{C}(\ell, {\vec s}) \\
\text{\ } \\
\evm^*\sempar{C} {\vec s} & \Rightarrow &  ({\vec s}, I) \\
\evm^*\sempar{C} {\vec s} & \Rightarrow &    \pgt{let }({\vec s}_1, M_1) = \evm\sempar{C}{\vec s}  \pgt{;} \quad  \pgt{let }({\vec s}_2, M_2) = \evm^*\sempar{C}{\vec s}_1  \\
&&\pgt{in}\quad ({\vec s}_2, M_1\mtimes M_2)  \\
\text{\ } \\
\evm^\oast\sempar{C}(\ell, {\vec s}) & \Rightarrow & ({\vec s}, I) \\[1ex]
\evm^\oast\sempar{C}(\ell, {\vec s}) & \Rightarrow &  \pgt{let } Z \subset [n];
  \pgt{let }({\vec s}_1, M_1) =  GEN(Z, \ell, \, \evm^*\sempar{C} ({\vec s} [Z\gets 0])) \pgt{;} \\
&&\pgt{let }({\vec s}_2, M_2) = \evm^\oast\sempar{C}(\ell,{\vec s}_1) \\
&&\pgt{in}\quad ({\vec s}_2, M_1\mtimes M_2) \\
\end{array}\]
\caption{Loop handling (up to the $GEN$ function).  The construct $\pgt{let } Z \subset [n]$ non-deterministically selects a set.}%
\label{fig:evmoast}
\end{figure}
The non-deterministic alternatives for $\evm^*$ and $\evm^\oast$ correspond to zero or more loop iterations respectively.
Note how the definition of $\evm^\oast$  corresponds to the expression
$\closure{Gen(\closure{\sempar{C}^S})}$ in Equation~\ref{eq:newLC}:  an element of the inner closure, $\closure{\sempar{C}^S}$,
 corresponds to a simulation of a finite number
of iterations, as done by function $\evm^*$; an element of the outer closure is a composition of a finite number of elements, some of which are generalized
(we actually generalize whenever possible).

Function $GEN$ checks, using the data-flow matrix, whether we have indeed simulated an iterative MP,
and if we did, it applies generalization by multiplying by $x_\ell$ the outcome of self-dependent variables that have been initialized to zero
in this sub-computation.  As explained above, the effect is that we multiply the
increments (the term $q$ where ${\vec p}[i] = x_i + q$).

\medskip
\parbox{0.9\textwidth}{  %to prevent a page break
\(
GEN(Z, \ell, ({\vec s}, M)) = \text{\ if\ } ( \forall i\notin Z : M_{ii}=1 )
 \text{\ then\ } ({\vec s}^{gen},M^{gen}) \text{\ else\ } ({\vec s},M) \)
\par\smallskip
where
\[
 {\vec s}^{gen} =  {\vec s} \prod_{i \in Z : M_{ii}=1\kern -1ex} {\ampmul}_{i i \ell} \qquad\qquad
 {M^{gen}} =   (I\land M) M (I\land \overline M) \lor (I\land M)
%  {M^{gen}} =   M \prod_{i : M_{ii}=1 \land (\exists j\ne i : M_{jj}=M_{ji}=1)\kern -1ex} C_{i i \ell} \,.
\]
}

Note that the multiplication is performed by applying ${\ampmul}_{i i \ell}$ for each such $i$.
There may be several such indices, hence the use
of $\prod$ notation in ${\vec s}^{gen}$\ (it expresses iteration of postfix composition).
The formula for $M^{gen}$ deserves some explanation.  Note that it does not depend on $Z$ at all. This
is important for the correctness proof. Informally, assuming we have reached the current state by simulating a
sequence of assignments equivalent to an AMP $\vec p$, the expression $(I\land M) M$ is $\ldfm(\itker{\vec p})$;
multiplying by $(I\land \overline M)$ results in \emph{removing} from this matrix the columns that correspond to self-dependent
entries; and then we put them back by adding $I\land M$.  The point of this exercise can be seen in the next example.

\begin{exa}
We illustrate an application of function $GEN$.
We refer back to Example~\ref{ex:motivate-sdl}. Recall that the analysis of the loop body produced the following AMPs
\[
 \setS = \{ \tuple{ x_1 + x_3,\  x_2 + x_4,\ x_3 ,\ x_3 }, \quad  \tuple{ x_1,\  x_2 + x_3 + x_4,\ x_3 ,\ x_3 } \}.
\]
We focus on the path corresponding to the first AMP, and consider its evaluation with
initial state ${\vec s} = \tuple{x^1,x^2,x^3,x^4}$.
Applying $\evm^*$, we may obtain the final result ${\vec s}' = \tuple{x^3,x^4,x^3,x^3}$ along with the matrix
\[ M = \begin{bmatrix}1 & 0 & 0 & 0 \\ 0 & 1 & 0 & 0 \\ 1 & 0 & 1 & 1 \\ 0 & 1 & 0  & 0\end{bmatrix} . \]
What $GEN$ does depends on the choice of $Z$.  In the case $Z=\emptyset$, it returns the above result, which is equivalent to skipping generalization.
Instead, $\evm^\oast$ may choose a non-empty set; for example, let $Z = \{1,4\}$.
Then $\evm^\oast$ modifies $\vec s$ by setting the first and fourth component to zero before invoking $\evm^*$; so its result changes into
${\vec s}'' = \tuple{x^3,x^2,x^3,x^3}$, while the matrix is unaffected.
Next, function $GEN$ checks the matrix: it confirms the condition for generalization, as $M_{22} = M_{33} = 1$.
Only $i=1$ satisfies $i\in Z \,\land\, M_{ii}=1$, so just this component is multiplied by the loop bound ${\vec s}''[3] = x^3$ and we get
\(
{\vec s}^{gen} = \tuple{x^6,x^2,x^3,x^3}
\).
For clarity, we now break the computation of $M^{gen}$ in parts: first,
\[
M_1 = (I\land M) M =
  \begin{bmatrix}1 & 0 & 0 & 0 \\ 0 & 1 & 0 & 0 \\ 1 & 0 & 1 & 1 \\ 0 & 0 & 0  & 0\end{bmatrix}
\]
removes the dependence on $x_4$, which is not self-dependent. In other words, $M_1$ represents the iterative part of
a corresponding AMP\@.  Then,
\[
 M_1 (I\land \overline M) \lor (I\land M) =
 \begin{bmatrix}0 & 0 & 0 & 0 \\ 0 & 0 & 0 & 0 \\ 0 & 0 & 0 & 1 \\ 0 & 0 & 0  & 0\end{bmatrix}  \lor
 \begin{bmatrix}1 & 0 & 0 & 0 \\ 0 & 1 & 0 & 0 \\ 0 & 0 & 1 & 0 \\ 0 & 0 & 0  & 0\end{bmatrix}  =
 \begin{bmatrix}1 & 0 & 0 & 0 \\ 0 & 1 & 0 & 0 \\ 0 & 0 & 1 & 1 \\ 0 & 0 & 0  & 0\end{bmatrix}
 \]
 effectively replaces the first column with a unit column.  Why?  In $M$, we had two 1's in this column, corresponding to
 the expression $x_1+x_3$.   Generalization turns this into $x_1 + x_3\cdot x_3$. But we are only interested in linear terms,
 so we replace this column with a unit column, erasing the dependence on $x_3$.
\end{exa}

 \section{Complexity}%
 \label{sec:alg-complexity}

 We now consider the space complexity of the algorithm. The size of a univariate state, implemented as a degree vector, is $O(n(1+\log d))$ bits where $d\ge 1$ is
 the highest degree reached.  In the problem formulation where a degree is specified as input, the algorithm can safely truncate any higher degree down to
 $d$, as this will still give a correct answer to the user's question.
 It remains to consider
 the space occupied by the recursion stack. This will be proportional to the depth of the program's syntax tree once we avoid the calls of
 functions $\evm^*$ and $\evm^\oast$ to themselves, or rather change them into tail calls. This can be done by routine rewriting of the functions,
using an auxiliary ``accumulator'' parameter, so for instance
 the code
\[\begin{array}{lcl}
\evm^*\sempar{C} {\vec s} & \Rightarrow &  ({\vec s}, I) \\
\evm^*\sempar{C} {\vec s} & \Rightarrow &    \pgt{let }({\vec s}_1, M_1) = \evm\sempar{C}{\vec s}  \pgt{;}
\quad   \pgt{let }({\vec s}_2, M_2) = \evm^*\sempar{C}{\vec s}_1  \\
&&\pgt{in}\quad ({\vec s}_2, M_1\mtimes M_2)  \\
\end{array}\]
changes into
\newcommand{\tevm}{{\textsc{TailEv}_{\!M}}}
\[\begin{array}{lcl}
\tevm^*\sempar{C} ({\vec s},A) & \Rightarrow &  ({\vec s}, A) \\
\tevm^*\sempar{C} ({\vec s},A) & \Rightarrow &    \pgt{let }({\vec s}_1, M_1) = \evm\sempar{C}{\vec s} \\
 &&    \pgt{in}\quad \tevm^*\sempar{C}({\vec s}_1,A\mtimes M_1) \\
\end{array}\]
and calls to $\evm^*$ from $\evm^\oast$ pass $I$ for the accumulator parameter.
Thus we have

 \begin{thm}
 The univariate polynomial-bound problem (Problem~\ref{pbm:uv-decision}) as well as the lower-bound generation problem (Problem~\ref{pbm:uv-query}) has polynomial space complexity.
 \end{thm}

 Note that the non-deterministic nature of the algorithm places the decision problem in NPSPACE, however by Savitch's theorem
 this implies PSPACE as well. The corresponding generation problem can also be determinized by essentially turning it into an exhaustive search,
 still in polynomial space.
%  In Section~\ref{sec:degree} we show, by bounding the attainable degrees, that the problem is also polynomial-space in the size of the program if the
%  program is the only input and a worst-case polynomial bound is sought.
Finally we would like to recall that a program in our language can have exponentially-growing variables as well, and emphasize that handling them,
 as described in \firstpaper, does not increase the complexity of our problem (the output in such a case will simply indicate that the variable has an
 exponential lower bound; no tight bound will be given).

\section{Proof of the PSPACE Algorithm}%
\label{sec:proof}

\newcommand\liftGEN{\overline{GEN}}

The purpose of this section is to show that our PSPACE algorithm, $\evm$, obtains correct results. How do we define correctness?
The original goal of the algorithm is to obtain symbolic expressions that tightly bound the concrete \emph{numeric} results obtained
 by a core-language program, applied to an initial integer-valued state. However, thanks to our previous work, we know that such a set of
 symbolic expressions---namely AMPs---is obtained by the abstract semantics of Section~\ref{sec:closurealg}.
 So now we are able to define $\sempar{C}^S$ as our reference and define our goal as matching the results it provides,
 specialized to a univariate initial state.  So the purpose of this section is to prove the following:
%  More formally, let us define ${\mathcal E}

 \begin{thm}\label{thm:alg-correct}
 The interpreter $\evm$ satisfies these two correctness claims:
 \begin{enumerate}
 \item Soundness:  Given any univariate state $\vec x$,  if $\evm \sempar{C} {\vec x} \Rightarrow^* ({\vec y}, M)$ (for any $M$)
 then $\exists {\vec p}\in \sempar{C}^S$ such that ${\vec y} \le {\vec p}\acirc {\vec x}$.
 \item Completeness:  Given any ${\vec p}\in \sempar{C}^S$, and
univariate state $\vec x$,  there is  ${\vec y} \ge {\vec p}\acirc {\vec x}$ such that
 $\evm \sempar{C} {\vec x} \Rightarrow^* ({\vec y},M)$ for some $M$.
 \end{enumerate}
 \end{thm}

 \noindent
 Note that we can also state this as follows: for any $\vec x$, the set $\sempar{C}^S \acirc {\vec x}$ and the set of possible
results of evaluating $\evm \sempar{C} {\vec x}$
 have the same maximal elements; thus they define the same worst-case bound.

 Obviously, the proof is inductive and to carry it out completely we also have to assert something about the matrices.
We define a notation for brevity:
\let\hra=\hookrightarrow
\let\hla=\hookleftarrow
\begin{defi}
Let $f$ be a program function that, given a univariate state, evaluates to a pair---a univariate state and a matrix.
Let $\setS$ be a set of AMPs.
We write:
\begin{itemize}
\item
$f \hra \setS$ if  $\forall {\vec x}$,  if $f  {\vec x} \Rightarrow^* ({\vec y}, M)$
 then $\exists {\vec p}\in \setS$ such that ${\vec y} \le {\vec p}\acirc {\vec x}$
 and $M = \ldfm(\vec p) $.
 \item
 $f \hla \setS$ if $\forall {\vec p}\in \setS$,
$\forall {\vec x}$,   $\exists ({\vec y}, M)$ such that
 $f {\vec x} \Rightarrow^* ({\vec y},M)$, ${\vec y} \ge {\vec p}\acirc {\vec x}$
 and $M = \ldfm(\vec p)$.
\end{itemize}
\end{defi}

\noindent
The full correctness claim will be the following:

%  \addtocounter{theorem}{-1}  %repeat the last thm. no
 \begin{thm}\label{thm:alg-correct-withM}
 The interpreter $\evm$ satisfies these two correctness claims:
 \begin{enumerate}
 \item Soundness:  Given any command \C,   $\evm \sempar{C} \hra \sempar{C}^S$.
 \item Completeness:  Given any command \C, $\evm \sempar{C} \hla \sempar{C}^S$.
\end{enumerate}
 \end{thm}

\noindent
Correctness is straight-forward for straight-line code (this basically amounts to the associativity argument proposed in Section~\ref{sec:evs},
with Lemma~\ref{lem:dfm-pq} justifying the matrices), and it
trivially extends to commands with branching, and even to loops without additions (since they are analyzed by unrolling  finitely many times).
It is where \emph{generalization} is used that correctness is subtle and so, this is the focus of this section.
Since the case of straight-line (or more generally, loop-free) code is simple, we will state the properties without a detailed proof.

\subsection{Correctness without loops}

\begin{lem}\label{lem:evm-sl}
Let {\C} be a loop-free command.
Then
$\evm \sempar{C} \hra \sempar{C}^S$ and $\evm \sempar{C} \hla \sempar{C}^S$.
\end{lem}

$\evm^*$ just iterates $\evm$ an undetermined number of times, which matches the definition of composition-closure precisely so the following
is also straight-forward.

\begin{lem}\label{lem:evmstar}
Let {\C} be a loop-free command.
Then $\evm^* \sempar{C} \hra \closure{\sempar{C}^S}$ and $\evm^* \sempar{C} \hla \closure{\sempar{C}^S}$.
\end{lem}

\subsection{Soundness of the analysis of loops}

We now wish to extend the soundness claim to loops.  The proof is by structural induction, where all loop-free commands are covered by
the above lemmas, and similarly, given correctness for commands $\C$ and $\D$, correctness for their sequential composition, or non-deterministic choice,
follow straight from definitions.  So
the main task is to prove the correctness for loops, namely the
correctness of $\evm^\oast$.  The main task is to relate the possible results of
$GEN(Z, \ell, \, \evm^*\sempar{C} ({\vec x}[Z\gets 0] ) )$, where $Z$ ranges over subsets of $[n]$,
 with members of the AMP set $Gen(\closure{\sempar{C}^S})[x_\ell/\tau]$.
Note that in the expression $Gen(\closure{\sempar{C}^S})[x_\ell/\tau]$,
unlike in Section~\ref{sec:closurealg}, we
substitute the $x_\ell$ for $\tau$ immediately after generalization;
this is done for the sake of comparing the results, since $\evm$ does not use $\tau$. This modification of the Closure Algorithm is
harmless, as for AMPs ${\vec p}$, ${\vec q}$ from this set, we can use the rule
$({\vec p}\acirc{\vec q})[x_\ell/\tau] = ({\vec p}[x_\ell/\tau])\acirc({\vec q}[x_\ell/\tau])$, due to the assumption that $\X_\ell$ is not modified inside
the loop.

\begin{lem}
For any univariate state $\vec x$, and $Z\subset [n]$,
if \[GEN(Z, \ell, \, \evm^*\sempar{C} ({\vec x}[Z\gets 0]) ) \Rightarrow^* ({\vec y},M),\]  then there is
${\vec r}\in Gen(\closure{\sempar{C}^S})$ such that ${\vec y}\le {\vec r}[x_\ell/\tau]\acirc {\vec x}$,
and $M = \ldfm({\vec r}[x_\ell/\tau])$.
\end{lem}

Note that this lemma constitutes an induction step; it is used under the inductive assumption that for the loop body $\C$, we have soundness.

\begin{proof}
Let $\evm^*\sempar{C} ({\vec x}[Z\gets 0]) \Rightarrow^* ({\vec y},M) $.
When generalization is not applied, i.e., ${\vec y} = {\vec s}'$ and $M = M'$,
we have, by the  inductive assumption,
${\vec p} \in \closure{\sempar{C}^S}$ such that ${\vec s}'  \le {\vec p}\acirc ({\vec x}[Z\gets 0]) \le {\vec p}\acirc {\vec x}$ and $M' = \ldfm({\vec p})$.
Further, ${\vec p} = {\vec p}[x_\ell/\tau]$ (since there is no $\tau$).

Next,
suppose that matrix $M'$ satisfies the condition for generalization ($\forall i\notin Z : M'_{ii}=1$).
By the inductive assumption,
we can choose ${\vec p} \in \closure{\sempar{C}^S}$ such that ${\vec s}' \le {\vec p}\acirc {\vec x}[Z\gets 0]$
and $M' = \ldfm({\vec p})$.
We now wish to show that ${\vec y} = {\vec s}^{gen} \le \itker{\vec p}^\tau[x_\ell/\tau]\acirc {\vec x}$.
We focus on the non-trivial case which is that of
an index $k$ to which generalization applies, i.e., $k\in Z$ and $M'_{kk}=1$.
Thus $x_k$ is self-dependent in $\vec p$.
Let ${\vec p}_Z = {\vec p}\acirc \idppol[Z\gets 0]$,
then ${\vec p}_Z$ is iterative. In fact,
${\vec p}_Z \lepoly \itker{\vec p}$.   Write $\itker{\vec p}[k] = x_k + q_k$, where $q_k$ is a polynomial not dependent on $x_k$.
Since $k\in Z$, ${\vec p}_Z[k] \lepoly q_k$.   Moreover, ${\vec s}'[k] \le {\vec p}[k]\acirc {\vec x}[Z\gets 0] = {\vec p}_Z[k]\acirc {\vec x}$.
Now, in the Closure Algorithm, function $Gen$ sets $\itker{\vec p}^\tau[k] = x_k + q_k\tau$
(the $\tau$ subsequently substituted with $x_\ell$),
 so multiplying ${\vec s}'[k]$ by ${\vec s}'[\ell]$, as GEN does, produces
\[{\vec s}^{gen}[k] \le  ({\vec p}_Z[k]\acirc {\vec x})\cdot {\vec s}'[\ell] \le
   \left( \itker{\vec p}^\tau[k][x_\ell/\tau] \right) \acirc {\vec x} \,.\]
We conclude that the lemma is fulfilled by  ${\vec r} = \itker{\vec p}^\tau$.

 Regarding the computation of the matrix $M^{gen}$, we refer to the text in Section~\ref{sec:loops}.  It explains why, given
 $M' = \ldfm({\vec p})$, we get $M^{gen} = \ldfm(\itker{\vec p}^\tau[x_\ell/\tau])$.
\end{proof}

$\evm^\oast$ combines an undetermined number of calls to $GEN(Z, \ell, \, \evm^*\sempar{C} ({\vec x}[Z\gets 0]) )$,
for arbitrary sets $Z$, which by straight-forward induction, and the definition of $\lsem\verb+loop X+_\ell \verb+ {C}+\rsem^S$,
 gives

\begin{lem}[Soundness for loops]
Let {\C} be a command such that
$\evm \sempar{C} \hla \sempar{C}^S$. Then
\[\lambda {\vec s}\,.\,\evm^\oast\sempar{C}(\ell, {\vec s}) \, \hra \, \closure{Gen(\closure{\sempar{C}^S})}[x_\ell/\tau] \,.\]
Hence,
\[\evm \lsem\verb+loop X+_\ell \verb+ {C}+\rsem \hra \lsem\verb+loop X+_\ell \verb+ {C}+\rsem^S \,. \]
\end{lem}

\subsection{Completeness for an unnested loop}

Again, the crux of this proof will be the application of generalization.  Here we want to show that when the Closure Algorithm generalizes,
$\evm^\oast$ can match its result.   Based on the presentation in Section~\ref{sec:loops}, we might expect that for ${\vec p}\in \closure{\sempar{C}^S}$,
given that $\evm^*$ can match ${\vec p}\acirc{\vec x}$, we will capture the results of generalization
by making a correct guess of the self-dependent variables and the accumulators.  Let us first
formalize the latter term.

\begin{defi}
Let ${\vec p}\in \absppol$ be iterative.
Recall that for $i\in \sd{\vec p}$ where $\vec p[i] = x_i + q_i$,
\[ {\vec p}^\tau[i] =  x_i + q_i \cdot \tau  \]
So that, given a univariate state $\vec x$,
\[ {\vec p}^\tau[i][x_\ell/\tau]\acirc{\vec x} =  \max((q_i \acirc {\vec x}) \cdot {\vec x}[\ell],\
 {\vec p}[i]\acirc {\vec x} ).
 \]
Let $\acc{\ell,{\vec p},\vec x}$ be the set of indices $i$ where the first term under the \textbf{max} is larger; intuitively, where generalization increases the result.
We refer to $\acc{\ell,{\vec p},\vec x}$ as the \emph{accumulators} in the computation under consideration (i.e., iterating ${\vec p}$ starting at state $\vec x$).
\end{defi}

Recall that function $GEN$ is applied to a state-matrix pair
 $({\vec s}',M)$ obtained from $\evm^*\sempar{C}{\vec s}$.   By correctness of this analysis (i.e., the inductive assumption),
we assume ${\vec s}' \ge {\vec p}\acirc {\vec s}$,
  where ${\vec p}\in \closure{\sempar{C}^S}$, and $M = \ldfm({\vec p})$.
	We shall say that this application of $\evm^*$
 simulates $\vec p$ (in the completeness statement, we quantify over $\vec p$ first, so it may be considered fixed throughout the discussion).
When $\evm^*$ is applied to a state in which some entries are set to 0, namely, ${\vec s}[Z\gets 0]$, this is equivalent to
simulating ${\vec p}\acirc \idppol[Z\gets 0]$ on $\vec s$.
 Then, $GEN$ possibly multiplies some entries of ${\vec s}'$ by ${\vec s}'[\ell] = {\vec s}[\ell]$.
 This can be seen as modifying ${\vec p}\acirc \idppol[Z\gets 0]$ into a new AMP $\vec q$ by multiplying some entries by $x_\ell$,
 before applying the AMP to $\vec s$.
 We express this by writing ${\vec q} = \liftGEN(Z, {\vec p}, \ell)$, noting that the action turning $\vec p$ into $\vec q$ depends on
${\vec p}$, $M$ (which is determined by $\vec p$) and $Z$, but not on $\vec s$.  We also note that $M^{gen}$ depends only on $M$.

Based on the presentation of the algorithm, the reader may infer that we intend that, for an iterative $\vec p$, letting $A = \acc{\ell,{\vec p},\vec x}$,
we should have $\liftGEN(A, {\vec p}, \ell)\acirc {\vec x} = {\vec p}^\tau[x_\ell/\tau]\acirc {\vec x}$.  However, this is not always the case,
as shown by the following example.

\begin{exa}
Let ${\vec p} = \fourtuple{x_1}{x_2+x_1}{x_3+x_2}{x_4}$, and $\ell=4$. Note that $\vec p$ is iterative. Let ${\vec x} = \tuple{x^2, x^2, x^1, x^1}$.
Then $\acc{\ell,{\vec p},{\vec x}} = \{2,3\}$.  Letting $A=\{2,3\}$, it is easy to check that ${\vec q} = \liftGEN(A, {\vec p}, \ell) =
\fourtuple{x_1}{x_1x_4}{0}{x_4}$.  Since we have begun our computation with ${\vec x}[A\gets 0] = \tuple{x^2, 0, 0, x^1}$,
the final state is $\tuple{x^2, x^3, 0, x^1}$.   Note that the result falls short of ${\vec p}^\tau[x_\ell/\tau]\acirc {\vec x} = \tuple{x^2,x^3,x^3,x^1}$.
\end{exa}

The solution to this mismatch is given by Lemma~\ref{lem:ub-induction} below,
which shows that by calling $GEN$ \emph{multiple times} in succession, in other words, by going through a number of recursive calls
of $\evm^\oast$, completeness is recovered.

We shall now make some preparations for this lemma.

\paragraph{\textbf{The Dependence Graph.}}
The matrix $\dfm({\vec p})$ may be seen as the adjacency matrix of a graph, which we call $G({\vec p})$.
Arcs in this graph represent data-flow in $\vec p$ (we use $\dfm({\vec p})$ and not $\ldfm({\vec p})$, so this includes non-linear dependence).
Paths in the graph correspond to data-flow
effected by iterating $\vec p$.  For instance, if we have a path $i\to j \to k$ in $G({\vec p})$, then $x_i$ will appear in the expression
$({\vec p}\circ {\vec p})[k]$.

% Like all directed graphs, $G({\vec p})$ defines a \emph{preorder} $i \prec_{\vec p} j$  on its nodes (which are the variables' indices),
%  by means of the reachability relation.

\begin{lem}\label{lem:qacyclic}
If $\vec p$ is iterative, and causes no exponential growth when iterated on concrete data, then $G({\vec p})$ has no cycles of length greater than 1.
% In other words, $\prec_{\vec p}$ is a partial order.
\end{lem}

Note that the assumption refers to concrete computation (${\vec p}\circ {\vec p} \circ \ldots$), and that we are building on the assumption that the
loop under analysis is polynomially bounded (this was presumably verified beforehand, see Section~\ref{sec:outline}).

\begin{proof}
Assume, to the contrary, that $G({\vec p})$ does have a cycle $i\to \dots \to k \to i$, where $k\ne i$. Let $r$ be the length of the cycle.
Since $\vec p$ is iterative, all these variables must be self-dependent (as some other entry depends on them). Thus we have
 ${\vec p}[i] = x_i + q(x_1,\dots,x_n)$ where $q$ involves $x_k$; but ${\vec p}[k]$ depends on $x_{k-1}$, and so on;
hence ${\vec p}^{(r-1)}[k] \ge x_i$, and
${\vec p}^{(r)}[i] \ge 2x_i$.  Thus iteration of $\vec p$
generates exponential growth, contradicting the assumption.
\end{proof}

When focusing on a particular AMP we may assume, w.l.o.g., that the variables
are indexed in an order consistent with $G({\vec p})$, so that if $x_i$ depends on $x_j$ then $j\le i$.
We shall refer to an (abstract) MP satisfying this property as \emph{neat}\footnote{Readers who like linear algebra may draw some intuition about neat MPs
from thinking about triangular matrices whose diagonal elements are in $\{0,1\}$. Interestingly, this structure has also popped up in other works
in the area~\cite{HarkFG20}.}.

We also state a rather evident property of an iterative MP:\@
\begin{lem}\label{lem:supip}
If $\vec p$ is iterative, then ${\vec p}\acirc {\vec p} \ge {\vec p}$.
\emph{A fortiori}, ${\vec p}\acirc {\vec p}^\tau \ge {\vec p}$.
\end{lem}

\begin{proof}
Consider the calculation of $({\vec p}\acirc{\vec p})[k]$; it is obtained by substituting ${\vec p}[i]$ for each occurrence of $x_i$ in ${\vec p}[k]$.
By iterativity, we know that ${\vec p}[i]$ includes $x_i$, so we
can clearly only enlarge the polynomial.
\end{proof}

\begin{lem}\label{lem:ub-induction}
Let ${\vec p}\in \absppol$ be neat and iterative, and let $\vec x$ be a univariate state.
There are sets $Z_1,\dots, Z_{n-1}$ such that
\[
{\vec q}_{n-1} \acirc {\vec q}_{n-2} \acirc \dots \acirc {\vec q}_1  \acirc {\vec x}\ge {\vec p}^\tau[x_\ell/\tau] \acirc {\vec x}
\]
where ${\vec q}_i = \liftGEN(Z_i,{\vec p},\ell)$.
\end{lem}

\begin{proof}
Let $A = \acc{\ell,{\vec p}, {\vec x}}$.
We define sets $Z_i$ for $i=1,2,\dots,n-1$ as follows:
\begin{align*}
Z_i &= A \cap \{i+1\} \cup \overline{\sd{\vec p}} \,.
\end{align*}

We now state an inductive claim and prove it. The claim is: for $1 \le i \le n$, for all $j\le i$,
\begin{equation} \label{eq:ind}
({\vec q}_{i-1} \acirc {\vec q}_{i-2} \acirc \dots \acirc {\vec q}_1 \acirc {\vec x})[j] \ge ({\vec p}^\tau[x_\ell/\tau] \acirc {\vec x})[j] \,.
\end{equation}
Note that~\eqref{eq:ind} specializes to the lemma's statement for $i=n$.

To start the induction, consider $i=1$.  Then we only have to prove that
\[
{\vec x}[1] \ge ({\vec p}^\tau[x_\ell/\tau] \acirc {\vec x})[1] \,.
\]
By neatness, ${\vec p}[1]$ can only depend on $x_1$, in fact ${\vec p}[1] = x_1$.    Also ${\vec p}^\tau[1] = x_1$. So equality holds.

Now, let $k>1$, and assume that~\eqref{eq:ind} holds when $i=k-1$. To prove it for $i=k$, we consider some sub-cases.
\begin{enumerate}
\item
$j\le k$ and $j\notin \sd{\vec p}$: in this case, ${\vec p}[j]$ only depends on $x_t$ with $t<k$  where,
 by IH,
\[
({\vec q}_{k-2} \acirc \dots \acirc {\vec q}_1 \acirc {\vec x})[t] \ge ({\vec p}^\tau[x_\ell/\tau] \acirc {\vec x})[t] \,.
\]
Since $j\notin \sd{\vec p}$, the construction of $\liftGEN(Z_{k-1},{\vec p},\ell)$ does not modify ${\vec p}[j]$, i.e.,
\begin{align*}
 {\vec q}_{k-1}[j] \acirc ({\vec q}_{k-2} \acirc \dots \acirc {\vec q}_1 \acirc {\vec x})
& =  {\vec p}[j] \acirc ({\vec q}_{k-2} \acirc \dots \acirc {\vec q}_1 \acirc {\vec x}) \\
& \ge {\vec p}[j] \acirc ({\vec p}^\tau[x_\ell/\tau] \acirc {\vec x}) \\
& \ge {\vec p}[j] \acirc {\vec x}  & \text{by Lemma~\ref{lem:supip}} \\
& = {{\vec p}[j]}^\tau[x_\ell/\tau] \acirc {\vec x} & \text{since $j\notin \sd{\vec p}$.}
\end{align*}

\item
$j < k$ where $j \in \sd{\vec p}$: here ${\vec p}[j]$ depends on $x_j$.
Yet  $j\notin Z_{k-1}$, so the construction of $\liftGEN(Z_{k-1},{\vec p},\ell)$ does not modify ${\vec p}[j]$, i.e.,
\begin{align*}
 {\vec q}_{k-1}[j] \acirc ({\vec q}_{k-2} \acirc \dots \acirc {\vec q}_1 \acirc {\vec x})
& =  {\vec p}[j] \acirc ({\vec q}_{k-2} \acirc \dots \acirc {\vec q}_1 \acirc {\vec x}) \\
& \ge  ({\vec q}_{k-2} \acirc \dots \acirc {\vec q}_1 \acirc {\vec x})[j] & \text{(since ${\vec p}[j]\ge x_j$)}\\
& \ge  ({\vec p}^\tau[x_\ell/\tau] \acirc {\vec x})[j]  & \text{by IH.}
\end{align*}

\item
$j=k\in \sd{\vec p}$ but $k\notin A_{k-1}$: here ${\vec p}[j]$ depends on $x_j$.
Yet  $j\notin Z_{k-1}$, so the construction of $\liftGEN(Z_{k-1},{\vec p},\ell)$ does not modify ${\vec p}[j]$, i.e.,
\begin{align*}
 {\vec q}_{k-1}[j] \acirc ({\vec q}_{k-2} \acirc \dots \acirc {\vec q}_1 \acirc {\vec x})
& =  {\vec p}[j] \acirc ({\vec q}_{k-2} \acirc \dots \acirc {\vec q}_1 \acirc {\vec x}) \\
& \ge  {\vec p}[j] \acirc {\vec x} & \text{(since ${\vec q_i}[j]\ge x_j$ for all $i$)}\\
& \ge  ({\vec p}^\tau[x_\ell/\tau] \acirc {\vec x})[j]  & \text{since $j\notin A_{k-1}$.}
\end{align*}

\item
$j = k \in A_{k-1}$: here, ${\vec p}[j] = {\vec p}[k]$ depends on $x_k$.
We may write ${\vec p}[k] = x_k + z$ for a polynomial $z$ in $x_1,\dots,x_{k-1}$.
We have, by IH,
\[
z \acirc ({\vec q}_{k-2} \acirc \dots \acirc {\vec q}_1 \acirc {\vec x})
\ge z \acirc ({\vec p}^\tau[x_\ell/\tau] \acirc {\vec x})  \,
\]
Since $k\in A_{k-1}$, we know that
\[
 {\vec p}^\tau[k][x_\ell/\tau]\acirc{\vec x} =  (z\acirc {\vec x})\cdot {\vec x}[\ell] = (z\cdot x_\ell)\acirc {\vec x}
\]
Also, since $k\in Z_{k-1}$,
\[
{\vec q}_{k-1}[k] = \liftGEN(Z_{k-1},{\vec p},\ell)[k] = ({\vec p}[k]\acirc \idppol[Z_{k-1}\gets 0])\cdot x_\ell = z\cdot x_\ell
\]
so
\begin{align*}
 {\vec q}_{k-1}[k] \acirc ({\vec q}_{k-2} \acirc \dots \acirc {\vec q}_1 \acirc {\vec x})
&= (z\cdot x_\ell) \acirc ({\vec q}_{k-2} \acirc \dots \acirc {\vec q}_1 \acirc {\vec x}) \\
&= {\vec p}^\tau[k][x_\ell/\tau]\acirc ({\vec q}_{k-2} \acirc \dots \acirc {\vec q}_1 \acirc {\vec x}) \\
&\ge {\vec p}^\tau[k][x_\ell/\tau]\acirc  ({\vec p}^\tau[x_\ell/\tau] \acirc {\vec x}) & \text{by IH} \\
&\ge ({\vec p}^\tau[x_\ell/\tau] \acirc {\vec x}) & \text{by Lemma~\ref{lem:supip}.} \tag*{\qedhere}
\end{align*}
\end{enumerate}
\end{proof}

\noindent
We are now ready to conclude

\begin{lem}
For any univariate state $\vec x$, for any ${\vec r} \in (Gen(\closure{\sempar{C}^S}))$,
there are ${\vec y}, M$ such that $\evm^\oast\sempar{C} (\ell,{\vec x}) \Rightarrow^* ({\vec y},M)$,
${\vec y}\ge {\vec r}[x_\ell/\tau]\acirc{\vec x}$, and $M = \ldfm({\vec r}[x_\ell/\tau])$.
\end{lem}

\begin{proof}
We focus on the interesting case, which is when generalization is applied to $\itker{\vec p}$,
for ${\vec p}\in \closure{\sempar{C}^S}$, to produce $\vec r$.
Note that $\itker{\vec p} = {\vec p}\acirc \idppol[\overline{\sd{\vec p}}\gets 0]$.
For any set $Z_i \supseteq \overline{\sd{\vec p}}$,  it is easy to see that
\[
{\vec q}_i \eqdef  \liftGEN(Z_i,{\vec p},\ell) = \liftGEN(Z_i, \itker{\vec p}, \ell).
\]
Using Lemma~\ref{lem:ub-induction} we have sets $Z_1,\dots,Z_{n-1}$, containing $\overline{\sd{\vec p}}$, such that
\[
{\vec q}_{n-1} \acirc {\vec q}_{n-2} \acirc \dots \acirc {\vec q}_1  \acirc {\vec x}\ge \itker{\vec p}^\tau[x_\ell/\tau] \acirc {\vec x}
= {\vec r}[x_\ell/\tau] \acirc {\vec x} \,.
\]
We conclude, using Lemma~\ref{lem:evmstar}, that starting with $\vec x$, and making at most $n$ recursive calls, $\evm^\oast$ can
reach ${\vec y}\ge {\vec r}[x_\ell/\tau]\acirc{\vec x}$ as desired.
The correctness of the matrix is again the equation $M^{gen} = \ldfm(\itker{\vec p}^\tau[x_\ell/\tau])$, which has been argued
in Section~\ref{sec:loops}.
\end{proof}

Straight-forward induction, and the definition of $\lsem\verb+loop X+_\ell \verb+ {C}+\rsem^S$,
 give

\begin{lem}[Completeness for loops]
Let {\C} be a command such that
$\evm \sempar{C} \hla \sempar{C}^S$. Then
\[\lambda {\vec s}\,.\,\evm^\oast\sempar{C}(\ell, {\vec s}) \, \hla \, \closure{Gen(\closure{\sempar{C}^S})}[x_\ell/\tau] \,.\]
Hence,
\[\evm \lsem\verb+loop X+_\ell \verb+ {C}+\rsem \hla \lsem\verb+loop X+_\ell \verb+ {C}+\rsem^S \,. \]
\end{lem}

\section{PSPACE-Completeness of bound analysis}%
\label{sec:pspace-hard}

In this section we complement our PSPACE upper bound with a hardness result, to show that our classification of the problem's complexity as PSPACE is tight.
The hardness proof is a reduction from \emph{termination of Boolean Programs}, a known PSPACE-complete problem.
First, we state the definition of the decision problem to which we reduce. This is a special case of the univariate-bound decision problem,
 with a fixed initial state.

\begin{defi}\label{def:decProgDeg}
The decision problem $\decProgDeg$ (for ``degree'') is defined as the set of triples
$(P,j,d)$ such that $P$ is a core-language
program, where
the maximal value of $\X_j$ at the completion of the program, in terms of the univariate
input ${\vec x} = \tuple{x,x,\dots,x}$, has a polynomial lower bound of $\Omega(x^d)$.
\end{defi}

 The complexity of the problem is classified in relation to the ``input size'' defined as: $|P|+d$,
where  $|P|$ is the size of the syntax tree representing $P$. This means that we allow $d$ to be represented in unary notation (and we will find that this
 does not affect the complexity class).

\begin{defi} A  \emph{Boolean program} is an instruction sequence
$b=$ {\tt  1:I$_1$ 2:I$_2$\ldots m:I$_m$} specifying a computation on
Boolean variables {\tt B}$_1$,\ldots,{\tt B}$_k$,
with program locations $1,2,\ldots,m$.

\emph{Instructions} {\tt
I$_\ell$} have  two forms: 	{\tt X$_i$ := not X$_i$}, and {\tt if X$_i$ % chktex 26
then goto
$\ell'$ else $\ell''$}. Here $1\leq i\leq k$ and $\ell, \ell',
\ell''\in\{0,1,2,\ldots,m\}$, where  $\ell, \ell',\ell''$ are always three different locations.
%$\ell' \ne \ell'' \ne \ell$.

Semantics: the \emph{computation} by $b$ is a finite or infinite state
sequence $b \vdash (\ell_1,\sigma_1)\to(\ell_2,\sigma_2)\to\ldots$, where
each store
$\sigma$ assigns a truth value in $\{\textit{true}, \textit{false}\}$ to each of
$b$'s variables, and
$\ell_t$ is the  program location at time $t$.

We are considering \emph{input free} programs. These programs have a fixed initial state:
$\ell_1 = 1$, and $\sigma_1$ assigns \textit{false} to every variable.
Given state $(\ell_t,\sigma_t)$, if $\ell_t=0$ then
the computation has terminated, else the following rules apply.

If  instruction
\pgt{I}$_{\ell_t}$ is \pgt{X$_i$ := not X$_i$}, then $\sigma_{t+1}$ is % chktex 26
identical to $\sigma_{t}$ except that $\sigma_{t+1}(\pgt{B}_i) = \neg
\sigma_{t}(\pgt{B}_i)$. Further, $\ell_{t+1}=(\ell_t+1)\bmod (m+1)$.

If instruction
\pgt{I}$_{\ell_t}$ is \pgt{if X$_i$ then goto $\ell'$ else $\ell''$}, then
$\sigma_{t+1}$ is
identical to $\sigma_{t}$. Further,
$\ell_{t+1}=\ell'$ if $\sigma_{t}(\pgt{B}_i)=\textit{true}$, and
$\ell_{t+1}=\ell''$ if $\sigma_{t}(\pgt{B}_i)=\textit{false}$.

Finally, program $b$ \emph{terminates}
%,  written  \sempar{$b$}$\downarrow$,
  if for some $t$: $b \vdash
 (\ell_1,\sigma_1)\to\ldots\to (\ell_t,\sigma_t) = (0,\sigma_t)$.
 \end{defi}

The following lemma is  proved in~~\cite[Chapter 28]{Jo:97} (with a trivial difference, concerning acceptance instead of termination):

\begin{lem} The following set  is complete for PSPACE:\@
\[{\mathcal B} = \{b\ |\ \mbox{$b$ is an input-free Boolean program and
$b$ terminates}\} \,. \]
\end{lem}

\begin{thm}
The problem $DEG$ is PSPACE-hard, even for programs with a single loop.
\end{thm}

\begin{proof}
Reduction from problem ${\mathcal B}$ above.

Suppose program $b = $ {\tt  1:I$_1$ 2:I$_2$\ldots m:I$_m$} has $k$ variables
{\tt B}$_1$,\ldots, {\tt B}$_k$. Without loss of generality, each variable has
value \textit{false} after execution, if
$b$ terminates (just add at the end of $b$ one test and
one assignment for each variable.)

Program {\tt p} will have   $2 + 2k(m+1)$ variables named ${\tt X}_1, {\tt X}_2$, and then ${\tt X}_{\ell,i,v}$
%and  ${\tt T}_{\ell\ i\ v}$
 for all $0 \le \ell\le m$, $1 \le i\le k$ and $v\in\{0,1\}$.
Informally, the program simulates the Boolean program, such that
the pair ${\tt X}_{\ell,i,0}$ and  ${\tt X}_{\ell,i,1}$ represent the value $b$ of variable
${\tt B}_{i}$ when program location $\ell$ was last visited. This is not a deterministic simulation, due to the
absence of deterministic branching; instead, the program may take many paths, most of which are wrong
simulations; and the contents of variables ${\tt X}_{\ell,i,v}$ will reflect the correctness of the simulation.
%Variables   ${\tt T}_{\ell\ i\ v}$ are just temporaries.
In fact, the value of each such variable will be either  $x$ or $x^2$, where the value $x$ denotes
error. That is, if the program reaches location  $\ell$ with ${\tt B}_{i} = v$,
 only the path that simulates it correctly will have ${\tt X}_{\ell,i,v} = x^2$.

We first define the initialization command \pgt{INIT}: for all $i$, it sets ${\tt X}_{1,i,0}$ to ${\tt X}_1^2$
(the initial value of $\pgt{B}_i$ is \textit{false}). All other simulation variables are set to  ${\tt X}_1$.

For every program location $\ell$ we define a command $\C_\ell$ ``simulating'' the corresponding instruction,
as follows:

\begin{itemize}
\item
For instruction $\ell : {\tt X}_i {\tt\ :=\ not\ X}_i$,\\ $\C_\ell$ modifies only variables %${\tt X}_{\ell\ i\ v}$ and
${\tt X}_{\ell',i,v}$ where $\ell' = (\ell+1)\bmod (m+1)$.  For each $v=0,1$, it sets:

$\begin{array}{ll}
 \X_{\ell',i,v}& \verb/:= X/_{\ell,i,\neg v} \\
% \intertext{and then it sets}
%  \X_{\ell\ i\ v}&\verb/:= X/_1 \\
\end{array}$

while for all $j\ne i$, and $v=0,1$, it sets

$\begin{array}{ll}
 \X_{\ell',i,v}& \verb/:= X/_{\ell,i,v} \\
% \intertext{and then it sets}
%  \X_{\ell\ i\ v}&\verb/:= X/_1 \\
\end{array}$

 \item
For instruction $\ell$ : \pgt{if X$_i$ then goto $\ell'$ else $\ell''$},\\
$\C_\ell$ modifies the variables % ${\tt X}_{\ell\ i\ v}$,
 ${\tt X}_{\ell',i,v}$ and  ${\tt X}_{\ell'',i,v}$ as follows:

$\begin{array}{lll}
 \X_{\ell',i,0}&\verb/:= X/_1 ;    & \rhd \textit{0 is definitely an error here}\\
 \X_{\ell',i,1}&\verb/:= X/_{\ell,i,1};    & \rhd \textit{1 is an error if it was so before}\\
 \X_{\ell'',i,1}&\verb/:= X/_1 ;    & \rhd \textit{1 is definitely an error here}\\
 \X_{\ell'',i,0}&\verb/:= X/_{\ell,i,0}    & \rhd \textit{0 is an error if it was so before}\\
%   \X_{\ell\ i\ v}&\verb/:= X/_1 ;    & \rhd \textit{for both values of $v$, as we left that location.}\\
\end{array}$

For all $j\ne i$, we simply have, for both  values of $v$,

$\begin{array}{ll}
 \X_{\ell',i,v}&\verb/:= X/_{\ell,i,v};\\
 \X_{\ell'',i,v}&\verb/:= X/_{\ell,i,v}\\
%  \X_{\ell\ i\ v}&\verb/:= X/_1\\
\end{array}$
\end{itemize}

\noindent
Finally, these commands are put together to make the program  {\tt p}:

\begin{Verbatim}[commandchars=\\\{\},codes={\catcode`$=3\catcode`_=8}]
INIT;
loop X$_1$ \{ choose  C$_1$ or C$_2$ or ... C$_m$ \};
X$_2$ := X$_{0,1,0}$ * X$_{0,2,0}$ * ... * X$_{0,k,0}$
\end{Verbatim}

The outcome of the reduction is
$(\pgt{p},2,2k)$. Thus, we are asking whether
the final value of \verb/X/$_2$ depends on $x$ with a degree of $2k$ or more, which will only happen if the loop
is completed with $x^2$ in each of the variables $\X_{0,i,0}$.  This means that program $b$
could be successfully simulated up to a point where the location was 0 and all variables  \textit{false}
(that is, the program really terminated).

A comment is due on the subject of the loop bound (which we chose, somewhat arbitrarily, to be $x$). Obviously we do not know in advance how long $b$ runs.
But since it is input-free, if $b$ terminates, its running time is a constant and the desired output
($x^{2k}$) will be achieved for all values of $x$ large enough to permit the complete simulation
of $b$. If $b$ does not terminate, the output will always be bounded by $x^{2k-1}$.
\end{proof}
\bigskip

What is surprising in the above proof is the simplicity of the programs; curiously, they
do not contain implicit data-flow in loops, which was the main challenge in the problem's solution.

\section{Checking Multivariate Bounds}%
\label{sec:multivariate}

In this section we are moving to multivariate bounds.  We show that checking a multivariate bound
can be reduced to the univariate problem.
 We focus on the following decision problem (first defined in Section~\ref{sec:outline}).

\smallskip\noindent
{\bf Problem~\ref{pbm:dec}.}
The \emph{multivariate bound decision problem} is:
Given a core-language program \pgt{P} and a monomial $m(x_1,\dots,x_n) =
x_1^{d_1}\dots x_n^{d_n}$, report
whether $m$ constitutes an attainable bound on the final value of $\X_n$; namely whether constants ${\vec x}_0$, $c>0$ exist such that
\[ \forall {\vec x}\ge {\vec x_0} \ \exists {\vec x}' \ .\  {\vec x} \sempar{P} {\vec x}' \ \land\ x'_n \ge cm(\vec x) \,. \]

\smallskip
We later discuss how to solve the related search problem, asking to find a complete set of attainable bounds.
If there is no polynomial upper bound on $x'_n$, then all monomials are attainable. This case can be detected in polynomial time using the algorithm
of~\cite{BJK08}, so we assume henceforth that $x'_n$ is polynomially bounded. In fact, as already discussed, we may assume that all super-polynomially
growing variables have been excluded.

Given \pgt{P}, we consider the set of attainable monomials (forming positive instances of the problem).
We represent a monomial by
a column vector of degrees: ${\vec d} = \tuple{d_1,\dots,d_n}^T$.
The main idea in this section is to make use of the \emph{geometry} of this set of vectors by viewing the problem as a \emph{linear programming} problem.
Before proceeding, we recall some background knowledge.

\subsection{Polyhedra}

We recall some useful definitions and properties, all
can be found in~\cite{Schrijver86}.

Point ${\vec x}\in \rats^n$ is \emph{a convex combination} of points ${\vec x}_1,\dots,{\vec x}_m$ if
$\vec x = \sum_{i=1}^m a_i\cdot \vec x_i$, where all $a_i \ge 0$ and $\sum_{i} a_i = 1$.
The \emph{convex hull} of a set of points is the set of all their convex combinations.

A \emph{rational convex polyhedron} $\poly{P} \subseteq \rats^n$
(\emph{polyhedron} for short) can be defined in two equivalent ways:
\begin{enumerate}
\item As the set of solutions of a set of
inequalities $A\vec{x} \le \vec{b}$, namely $\poly{P}=\{
\vec{x}\in\rats^n \mid A\vec x \le \vec b \}$, where $A \in \rats^{m
  \times n}$ is a rational matrix of $n$ columns and $m$ rows, $\vec x\in\rats^n$ and $\vec b \in \rats^m$ are column vectors of $n$ and
$m$ rational values respectively.  Each linear inequality (specified by a row of $A$ and the corresponding element of $\vec b$) is known as a
\emph{constraint}.
\item As the convex hull of a finite set of points $\vec x_i$ and rays $\vec y_j$:
\begin{equation} \label{eq:generators}
\poly{P} = \convhull\{\vec x_1,\dots,\vec x_m\} + \cone\{\vec y_1,\dots,\vec y_t\},
\end{equation}
or more explicitly:  $\vec x\in \poly{P}$ if and only if $\vec x =
\sum_{i=1}^m a_i\cdot \vec x_i + \sum_{j=1}^t b_j\cdot \vec y_j$ for some
rationals $a_i,b_j\ge 0$, where $\sum_{i=1}^m a_i = 1$. The vectors $\vec y_1,\dots,\vec y_t$ are  recession directions of $\poly{P}$, i.e., directions in which the polyhedron is unbounded;
in terms of the constraint representation, they satisfy
$A\vec{y}_i \le \vec{0}$.
The points $\vec x_i$ of a minimal set of generators are the \emph{vertices} of $\poly{P}$.
% The set of all recession direction,
% $\{ \vec y \mid A\vec{y} \le \vec{0}\}$, is known as the \emph{recession cone} of $\poly{P}$.
\end{enumerate}

%\paragraph{Integer Polyhedra}
%
\noindent
For any set $\poly{S} \subseteq \rats^n$ we let
$\intof{\poly{S}}$ be $\poly{S} \cap \ints^n$, i.e., the set of
integer points of $\poly{S}$. The \emph{integer hull} of $\poly{S}$,
commonly denoted by $\inthull{\poly{S}}$, is defined as the convex
hull of $\intof{\poly{S}}$.
A polyhedron $\poly{P}$ is \emph{integral} if
$\poly{P} = \inthull{\poly{P}}$.   This is equivalent to stating that it is generated (Eq.~\ref{eq:generators}) by a set of integer points and rays.
In particular, its vertices are integral.

We will be using some results from~\cite{Schrijver86}, regarding
the \emph{computational complexity} of algorithms on polyhedra.  Thus we adhere to definitions used by the author,
which we now recite for completeness.
Schrijver denotes the bit-size of
an integer $x$ y $\bitsize{x} = 1 + \lceil \log (|x|+1)\rceil$; the bit-size of
an $n$-dimensional vector $\vec a$ is defined as $\bitsize{\vec a} = n+\sum_{i=1}^n \bitsize{a_i}$;
and the bit-size of  an inequality
$\vec{a}\cdot \vec{x} \le c$ as $1+\bitsize{c}+\bitsize{\vec{a}}$.
For a polyhedron $\poly{P} \subseteq \rats^n$ defined by $A\vec x \le \vec b$,
the \emph{facet size}, denoted by $\fctsize{\poly{P}}$, is the
smallest number $\phi \ge n$ %
 such that $\poly{P}$ may be described by
some $A\vec x \le \vec b$ where
each inequality in $A\vec x \le \vec b$ fits in $\phi$
bits.
The \emph{vertex size}, denoted by $\vtxsize{\poly{P}}$, is the smallest
number $\psi \ge n$ such that $\poly{P}$ has a generator representation in
which each of $\vec x_i$ and $\vec y_j$ fits in $\psi$ bits (the size of a vector is calculated
as above).
The following theorem~\cite[part of Theorem~10.2]{Schrijver86} relates the two measures:

\begin{thm}%
\label{thm:size}
Let $\poly{P}$ be a rational polyhedron in $\rats^n$; then
% $\vtxsize{\poly{P}} \le 4n^2\fctsize{\poly{P}}$ and
$\fctsize{\poly{P}} \le 4n^2\vtxsize{\poly{P}}$.
\end{thm}

\subsection{Monomial bounds and linear programming}

From this point on, we fix $\setS$ to mean the set of attainable monomials of a given program (as vectors in $\nats^n$).

Let ${\vec u} = \tuple{u_1,\dots,u_n}\in \nats^n$ be a (row) vector representing a univariate input, as in Section~\ref{sec:pspace-uv}.
Let $m(x_1,\dots,x_n) = x_1^{d_1}\dots x_n^{d_n}$.
 Instead of $x_1^{d_1}\dots x_n^{d_n}$ we write, concisely, ${\vec x}^{\vec d}$. Then note that
\[
 m(x^{u_1},\dots, x^{u_n}) = x^{d_1 u_1 + \cdots + d_n u_n} = x^{{\vec u}\cdot {\vec d}} \,.
 \]

 We now state
 \begin{lem}\label{lem:max-on-S}
 Finding a tight upper bound on the final value of $\X_n$ given an initial univariate state $\vec u$ is equivalent to the following optimization problem:
\begin{equation} \label{eq:max-on-S}
 \text{Given $\vec u$, maximize } {\vec u}\cdot {\vec d} \quad \text{subject to } {\vec d}\in \setS \,.
 \end{equation}
 \end{lem}
 \noindent
Note that the multivariate problem is simply to decide $\setS$.
 The statement is by no means trivial. Luckily it comes easily out of the results of \firstpaper.

 \begin{proof}
 According to \firstpaper, given that the values computed by the program are polynomially bounded,
the set $\setS$ of all the attainable \emph{multivariate} monomials provides tight upper bounds as well (more precisely, taking the maximum of these monomials gives
a well-defined function---since the set is finite---and this function is an asymptotic upper bound).
 If we plug $\vec u$ as initial state, the set of bounds becomes $x^{{\vec u}\cdot {\vec d}}$ where ${\vec d}$ ranges over $\setS$.
 These are univariate monomials, so they are fully ordered. The highest bound, namely the maximum value of ${\vec u}\cdot{\vec d}$, is the tight
 worst-case upper bound for the program.
\end{proof}

So far, we have only considered bounds which are polynomials, where the exponents are integer. However, a function
$m(x_1,\dots,x_n) = x_1^{d_1}\dots x_n^{d_n}$ where the $d_i$ are non-negative \emph{rational numbers} is  a perfectly valid candidate for comparison
with the results of a computation. Let $\setT$ be the set of rational-valued vectors $\vec d$ such that the corresponding function $m$ is attainable.
Clearly $\setS \subseteq \setT$; in fact, $\setS = \intof{\setT}$.  The inclusion is clearly strict.  However, when we consider the optimization problem~\eqref{eq:max-on-S}, we have

\begin{lem}\label{lem:TeqvS}
For any vector ${\vec u}\in \nats^n$,
\[
 \max_{{\vec d}\in \setT} {\vec u}\cdot {\vec d} =
 \max_{{\vec d}\in \setS} {\vec u}\cdot {\vec d}.
 \]
 \end{lem}
%  \noindent
% The lemma follows immediately from Lemma~\ref{lem:max-on-S}, since inequality (necessarily a ``greater than") would mean that even the largest
%  attainable bound in $\setS$ is not a valid upper bound.
\begin{proof}
Since $\setS \subseteq \setT$, clearly $\max_{{\vec d}\in \setT} {\vec u}\cdot {\vec d} \ge \max_{{\vec d}\in \setS} {\vec u}\cdot {\vec d}$.
Suppose that the inequality is strict; then there is an attainable function $m = {\vec x}^{\vec d}$ (with ${\vec d}\in \setT$)
 such that $m(\vec u)$ has a greater exponent than $m'({\vec u})$,  for all $m' = {\vec x}^{{\vec d}'}$ where ${\vec d}' \in \setS$. This contradicts Corollary~\ref{cor:monomialbounds}.
 \end{proof}

 We now make a couple of observations on the shape of $\setT$.

 \begin{lem}
 For any ${\vec d}\in \setT$, all integer points of the box $B_{{\vec 0},{\vec d}} = \{ {\vec x} \mid {\vec 0}\le {\vec x}\le {\vec d}\}$ are also in $\setT$.
 \end{lem}

 \begin{proof}
 Immediate since if $\tuple{a_1,\dots,a_n} \le \tuple{b_1,\dots,b_n}$ then $x^{a_1}\cdots x^{a_n} \le x^{b_1}\cdots x^{b_n}$.
 \end{proof}

 We conclude that $\setT$ is a union of boxes. We can say  more.

 \begin{lem}
 If functions $f,g:\nats^n \to \reals$ are attainable, so is $\max(f,g)$.
 \end{lem}

 This can be easily checked against the definition (Definition~\ref{def:attainable}).

 \begin{lem}
 $\setT$ is convex.
 \end{lem}

 \begin{proof}
 Consider a monomial $m = {\vec x}^{\vec d}$ where
${\vec d}$ is the convex
 combination $\sum c_j\cdot {\vec d}_{j}$ of some ${\vec d}_{j} \in \setT$, where $\sum_{j=1}^n c_j = 1$.
We refer to the $i$th component of ${\vec d}_j$ as $d_{ji}$.  Thus
 \[
 m({\vec x}) = {\vec x}^{\sum_j c_j {\vec d}_j} = \prod_{i} \prod_{j} x_i^{c_j d_{ji}} = \prod_j \prod_i x_i^{c_j d_{ji}} = \prod_j {({\vec x}^{{\vec d}_j})}^{c_j} .
 \]
 By the rational-weight form of the classic inequality of means~\cite[Eq.~(2.7)]{Steele2004book}, a weighted geometric mean is
 bounded by the corresponding weighted arithmetic mean:
 \[
  \prod_j {({\vec x}^{{\vec d}_j})}^{c_j} \le \sum_j c_j ({\vec x}^{{\vec d}_j}) \le \max_j ({\vec x}^{{\vec d}_j});
 \]
The latter is an attainable function, therefore so is $m$.
Hence ${\vec d}\in \setT$.
 \end{proof}

 \begin{figure}[t]\label{fig:polytope}
\noindent%
{%\vtop{\vskip 0pt
\begin{minipage}[t]{0.4\textwidth}
\vskip 0pt
\begin{tikzpicture}[scale=0.4]
    % Draw polygon
    \fill[fill=black!10] (11,0) -- (11,1) coordinate (a_1) -- (10,3) coordinate (a_2) -- (7,5) coordinate (a_3) -- (3,6) coordinate (a_4) -- (0,6) -- (0,0) -- cycle;
% Draw axes
    \draw [<->,thick] (0,8) node (yaxis)  {} %[above] {$y$}
        |- (14,0) node (xaxis) {}; % [right] {$x$};
    % Draw vertices
    \fill[black] (a_1) circle (3pt);
    \fill[blue] (a_2) circle (3pt);
    \fill[red] (a_3) circle (3pt);
    \fill[green] (a_4) circle (3pt);
    % Draw rectangles
    \draw (yaxis |- a_1) %node[left] {$y'$}
        -| (xaxis -| a_1);  % node[below] {$x'$};
    \draw[blue] (yaxis |- a_2) %node[left] {$y'$}
        -| (xaxis -| a_2);  % node[below] {$x'$};
    \draw[red] (yaxis |- a_3) %node[left] {$y'$}
        -| (xaxis -| a_3);  % node[below] {$x'$};
    \draw[green] (yaxis |- a_4) %node[left] {$y'$}
        -| (xaxis -| a_4);  % node[below] {$x'$};
%     \draw[dashed] (a_4) -- (0,6);  %dashed black line on top of the green line!
\end{tikzpicture}
\end{minipage}}
\hspace{0.1\textwidth}
{% \vtop{\vskip 0pt
\begin{minipage}[t]{0.4\textwidth}
\vskip 0pt
\begin{tikzpicture}[scale=0.4]
    % Draw polygon
     \fill[fill=black!10] (11,-4) -- (11,1) coordinate (a_1) -- (10,3) coordinate (a_2) -- (7,5) coordinate (a_3) -- (3,6) coordinate (a_4) -- (-4,6) -- (-4,-4) -- cycle;
%     \draw[dotted] (11,-4) -- (11,1) coordinate (a_1) -- (10,3) coordinate (a_2) -- (7,5) coordinate (a_3) -- (3,6) coordinate (a_4) -- (-4,6) -- (-4,-4) -- cycle;
    % Draw axes
    \draw [<->,thick] (0,8) node (yaxis)  {} %[above] {$y$}
        |- (14,0) node (xaxis) {}; % [right] {$x$};
    % Draw vertices
    \fill[black] (a_1) circle (3pt);
    \fill[blue] (a_2) circle (3pt);
    \fill[red] (a_3) circle (3pt);
    \fill[green] (a_4) circle (3pt);
    % Draw rectangles
    \draw (yaxis |- a_1) %node[left] {$y'$}
        -| (xaxis -| a_1);  % node[below] {$x'$};
    \draw[blue] (yaxis |- a_2) %node[left] {$y'$}
        -| (xaxis -| a_2);  % node[below] {$x'$};
    \draw[red] (yaxis |- a_3) %node[left] {$y'$}
        -| (xaxis -| a_3);  % node[below] {$x'$};
    \draw[green] (yaxis |- a_4) %node[left] {$y'$}
        -| (xaxis -| a_4);
%     \draw[dashed] (a_4) -- (0,6);  %dashed black line on top of the green line!
\end{tikzpicture}
\end{minipage}}
\caption{Assuming that the black, blue, red and green dots represent extremal elements of $\mathcal S$, the set $\mathcal T$ is represented by
the convex hull of the union of four boxes (shaded area in the left-hand drawing).
The right-hand drawing illustrates ${\mathcal T}'$.}
\end{figure}

So $\setT$ is convex, includes $\setS$, and the maximum of  ${\vec u}\cdot {\vec d}$ for any ${\vec u} > {\vec 0}$ is obtained at a point of $\setS$.
We conclude that $\setT$ is an integral polyhedron~\cite[\S 16.3]{Schrijver86}. It equals the convex hull of the boxes whose upper right
corners are points of $\setS$ (Figure~\ref{fig:polytope}, left).  It will be convenient in the next proof to extend $\setT$ by allowing negative numbers
as well. This gives a polyhedron $\setT'$ which is unbounded in the negative direction (Figure~\ref{fig:polytope}, right); technically,
$\setT' = \setT \cup \cone(\tuple{-1,0,\dots,0},\tuple{0,-1,\dots,0},\dots,\tuple{0,\dots,0,-1})$.

 Now we can use known results on the complexity of linear programming to obtain our result.

 \begin{lem}\label{lem:size}
 The facet size of $\setT'$ is bounded by $4n^2 (n+ \bitsize{d_{max}})$, where $d_{max}$ is the maximal attainable degree.
 \end{lem}

 This follows directly from Theorem~\ref{thm:size}, using the fact that our polyhedron is integral, i.e., its vertices are members of $\setS$.

 \begin{thm}\label{thm:mv-pspace}
 The multivariate-bound decision problem can be solved in space polynomial  in the size of the given program and the maximal attainable degree $d_{max}$.
 % we have to rely on a bound on d_max, from sec:degree, because the bound of lem:size involves d_max, irrespective of the degree in the problem instance.
 \end{thm}

 \begin{proof}
 As explained above, our approach is to show that we can decide whether ${\vec d}\in \setT$.
 As PSPACE is closed under complement, we can consider the converse question: whether ${\vec d}\notin \setT'$.
 This is equivalent to asking: is there a constraint ${\vec a}\cdot{\vec x} \le b$, satisfied by $\setT'$,
 such that ${\vec a}\cdot{\vec d} > b$.   Our decision procedure is non-deterministic and \emph{guesses} $\vec a$.
 Then we need to find $b$, which amounts to maximizing ${\vec a}\cdot {\vec x}$ over ${\vec x}\in \setT'$.
Note that $\vec a$ will contain no negative numbers, since $\setT'$ does not satisfy
any constraint with negative coefficients.  In case ${\vec a}$ has non-integer rational numbers, we can scale by their common
denominator. Thus, w.l.o.g.~we may assume $\vec a$ to be integer-valued, and
by Lemmas~\ref{lem:max-on-S} and~\ref{lem:TeqvS}, finding whether $b$ exists is equivalent to solving the univariate decision
problem (Problem~\ref{pbm:uv-decision} on p.~\pageref{pbm:uv-decision}) with initial state $\vec a$, querying the degree ${\vec a}\cdot {\vec d}$.

So, we have reduced the multivariate-bound decision problem to the univariate problem. What is the complexity of the resulting algorithm?
Assuming that we know $d_{max}$,
we can impose the bound given by Lemma~\ref{lem:size}.
Since $\vec a$ consists of polynomially-big numbers (in terms of their bit size), our univariate algorithm (Section~\ref{sec:pspace-uv})
 solves the problem in polynomial space as stated.

 What if we do not know $d_{max}$?  In this case we can search for it; we give the procedure below.
\end{proof}

In the case that $d_{max}$ is not known, our complexity result is less satisfying when the monomial presented to the decision problem is of a degree much smaller
than $d_{max}$.  However, if what we really seek is not the decision problem for its own sake, but the discovery of tight upper bounds, then we do want to
reach $d_{max}$, and we can then describe our algorithm's complexity as polynomial space in the \emph{output-sensitive sense} (that is, depending on a
quantity which is not the size of the input but of the output).  First, let us state the complexity of the search procedure.

\begin{thm}
 Given a core-language program $P$, the index of a chosen variable $\X_j$, we can maximize $d_{max} = \sum_i d_i$ over all monomials
 ${\vec x}^{\vec d}$ that constitute an attainable lower bound on the final value of $\X_n$. The algorithm's space complexity is polynomial in the
 size of $P$ plus $\bitsize{d_{max}}$.
 \end{thm}

 \begin{proof}
 Recall that a polynomial-space non-deterministic algorithm can be determinized in polynomial
 space (Savitch's Theorem).
 We search for $d_{max}$ using a deterministic implementation of our non-deterministic algorithm (Section~\ref{sec:pspace-uv}).
 This deterministic algorithm can be used to give a definite answer to the query: is a given degree $d$ attainable for $\X_n$.  We start with
 $d=1$ and repeatedly increase $d$ (reusing the memory) until we hit a negative answer (as usual we assume that the possibility of super-polynomial growth has been
 eliminated first using~\cite{BJK08}). We can actually be quicker and only loop through powers of two, since this would give an approximation to $d_{max}$ which
 is close enough to be used in Theorem~\ref{thm:mv-pspace}.
 \end{proof}

\section{Generating Multivariate Bounds Directly}%
\label{sec:ev-multivariate}

Our analysis algorithm from Section~\ref{sec:pspace-uv} can actually be modified to yield multivariate bounds directly, as we show next.
Given this statement, the reader may wonder why we bothered with the version for univariate bounds, and the reduction of multi-to-uni-variate.
Let us reveal right away, then, that despite the simplicity of the following algorithm, we do not have a direct \emph{proof} for its correctness and we
approach it based on the previous algorithms.

The new algorithm is obtained thanks to a rather simple observation about the analyzer $\evm$, namely, that the only operator applied to our (univariate) monomials
is product.
 Since the product of multivariate monomials is again a (multivariate) monomial, we can replace the data type of our symbolic state $\vec s$
from an $n$-tuple of \emph{univariate} monomials to an $n$-tuple of \emph{multivariate} ones.
Such an $n$-tuple is a special case of AMP, and we also refer to it as \emph{a monomial state} because it is used in the evaluator to abstract a state, just as the
univariate state in the previous algorithm.  Other than this change of datatype,
no change to the code is necessary.
In terms of implementation, a multivariate monomial
is implemented as either an $n$-tuple of degrees, or a special constant representing 0.  The space complexity of this representation remains polynomial
in the number of variables and the maximum degree $d_{max}$.   We call the evaluator, thus modified, $\evmm$.
To illustrate the idea, consider a 3-variable program. Then its initial state is represented by $\idppol = \tuple{x_1,x_2,x_3}$.  Here are two examples
of simple straight-line computations:
\begin{alignat*}{2}
 &\evmm\sempar{{\amul{3}{2}{1}}} {\idppol} &\quad \Rightarrow \quad & (\tuple{x_1,\, x_2,\, x_1 x_2},  E(3) ) \\
 &\evmm\sempar{{\amul{3}{2}{1}}\semi\ \acopy{2}{3}} {\idppol} &\quad \Rightarrow^* \quad & (\tuple{x_1,\, x_1 x_2,\, x_1 x_2},  E(3) A(3,2) )
\end{alignat*}
where the data-flow matrices $A(3,2)$ and $E(3)$ are as in Section~\ref{sec:matrices}.

Proving the correctness of the algorithm falls, as usual, into soundness and completeness.

\subsection{Soundness}

The soundness for straight-line code is straight-forward to prove, so we content ourself with stating it:

\begin{lem}\label{lem:evmm-sl}
Let {\C} be a loop-free command.
Given any monomial state $\vec s$, if
$\evmm \sempar{C} {\vec s} \Rightarrow^* ({\vec g}, M)$ (for any $M$)
 then $\exists {\vec p}\in \sempar{C}^S$ such that ${\vec g} \lepoly {\vec p}\acirc {\vec s}$,
 and $M = \ldfm({\vec p})$.
\end{lem}

This means that if we set the initial state to $\idppol$, the monomials we generate are attainable in the sense of Corollary~\ref{cor:monomialbounds}.
We now wish to show the same property for loops.  As in Section~\ref{sec:proof}, the extension to $\evmm^*$ is immediate:

\begin{lem}\label{lem:evmmstar}
Let {\C} be a loop-free command.
Given any monomial state $\vec s$, if
$\evmm^* \sempar{C} {\vec s} \Rightarrow^* ({\vec g}, M)$
 then $\exists {\vec p}\in \closure{\sempar{C}^S}$ such that ${\vec g} \lepoly {\vec p}\acirc {\vec s}$,
 and $M = \ldfm({\vec p})$.
\end{lem}

As in Section~\ref{sec:proof}, we proceed to cover any command
by structural induction, where all loop-free commands are covered by
the above lemmas, and similarly, given correctness for commands $\C$ and $\D$, correctness for their sequential composition, or non-deterministic choice,
follow straight from definitions.  So
the main task is to prove the correctness for loops, more precisely the
correctness of $\evmm^\oast$; to this end we  compare the possible results of
$GEN(Z, \ell, \, \evmm^*\sempar{C} ({\vec s}[Z\gets 0] ) )$  with our reference result $Gen(\closure{\sempar{C}^S})[x_\ell/\tau]$.

The soundness part of the proof repeats the arguments used with $\evm$ almost verbatim.

\begin{lem}[Soundness of GEN]
For any monomial state $\vec s$, and $Z\subset [n]$,
if $({\vec g},M) \in GEN(Z, \ell, \, \evmm^*\sempar{C} ({\vec s}[Z\gets 0]) )$,  then there is
${\vec r}\in Gen(\closure{\sempar{C}^S})$ such that
${\vec g}  \lepoly ({\vec r}[x_\ell/\tau])\acirc{\vec s}$, and  $\ldfm(\vec r)[x_\ell/\tau] = M$.
\end{lem}
\begin{proof}
Let $({\vec s}', M') =  \evmm^*\sempar{C} ({\vec s}[Z\gets 0])$.
When generalization is not applied, i.e., ${\vec g} = {\vec s}'$ and $M = M'$,
we have, by inductive hypothesis,
${\vec p} \in \closure{\sempar{C}^S}$ such that ${\vec s}'  = {\vec p}\acirc ({\vec s}[Z\gets 0]) \lepoly {\vec p}\acirc {\vec s}$ and $M' = \ldfm({\vec p})$,
which is the same as $\ldfm(\vec r)[x_\ell/\tau]$ (since there is no $\tau$).

Next,
suppose that matrix $M'$ satisfies the condition for generalization ($\forall i\notin Z : M'_{ii}=1$), and
let $\vec p$ be an element of $\closure{\sempar{C}^S}$ such that ${\vec s}' \lepoly {\vec p}\acirc {\vec s}[Z\gets 0]$ and $M' = \ldfm({\vec p})$.
Consider an index $k$ to which generalization applies, i.e., $k\in Z$ and $M'_{kk}=1$.
Thus $x_k$ is self-dependent in $\vec p$.
Let ${\vec p}_Z = {\vec p}\acirc \idppol[Z\gets 0]$. Then ${\vec p}_Z$ is iterative; in fact,
${\vec p}_Z \lepoly \itker{\vec p}$.
Moreover, since $k\in Z$, ${\vec p}_Z[k] \lepoly q$ such that $\itker{\vec p}[k] = x_k + q$.
 Now, in the Closure Algorithm, function $Gen$
 multiplies the polynomial
$q$ by $\tau$ (which is then substituted with $x_\ell$),
 so multiplying ${\vec p}_Z[k]\acirc {\vec s}$ by ${\vec s}'[\ell]$, as GEN does, produces
 \[{\vec s}^{gen}[k] \lepoly  ({\vec p}_Z[k]\acirc {\vec x})\cdot {\vec s}'[\ell] \lepoly
   \left( \itker{\vec p}^\tau[k][x_\ell/\tau] \right) \acirc {\vec x} \,.\]
 We take $\vec r$ to be $\itker{\vec p}^\tau$.  Correctness of $M$ is as in the previous proofs.
\end{proof}

Once we can do the induction step for loops, we easily get

\begin{lem}[Soundness of $\evmm$]\label{lem:evmm-sound}
Let \pgt{P} be a core-language program.
If $\evmm \sempar{P} \idppol \Rightarrow^* ({\vec g}, M)$ (for any $M$)
 then $\exists {\vec p}\in \sempar{P}^S$ such that ${\vec g} \lepoly {\vec p}$.
\end{lem}

\subsection{Completeness}

We formulate our completeness claim
based on the geometric view taken in the previous section.
We maintain the convention that $\setS$ denotes the set of attainable monomials (asymptotic lower bounds on the final value of $\X_n$) for
a program under discussion (arbitrary but fixed through the discussion), and $\setT$, $\setT'$ its extensions to rational-number polyhedra.

% \begin{defi} \label{def:ext-mon}
% A monomial $\vec d \in \setS$ is \emph{extremal} if it
% maximizes the target function in the optimization problem \eqref{eq:max-on-S}, for some $\vec u$.
% \end{defi}

Now, in the previous subsection we have argued for soundness of $\evmm$ in the same way that we did for $\evm$, that is by comparison to the
Closure Algorithm, and established the relation ${\vec g} \lepoly {\vec p}\acirc {\vec s}$ between $\evmm$'s result and some ${\vec p}\in \sempar{C}^S$.
A corresponding completeness claim might be that we can reach any monomial state ${\vec g} \lepoly {\vec p}\acirc {\vec s}$.  This is, however, not the case.
Consider the two assignments:
\begin{Verbatim}[codes={\catcode`$=3\catcode`_=8}]
   X$_2$ := X$_1$ + X$_2$;
   X$_2$ := X$_2$ * X$_2$
\end{Verbatim}
In $\evmm$ (starting with $\idppol$), after the first assignment we have two monomial states, $\tuple{x_1,x_1}$ and $\tuple{x_1,x_2}$; and after the
second, we have $\tuple{x_1,x_1^2}$ and $\tuple{x_1,x_2^2}$.  But the Closure Algorithm generates $\tuple{x_1, \ x_1^2 + x_1 x_2 + x_2^2}$.  Note that
there is a monomial, $x_1 x_2$, which is not obtained by $\evmm$.  But this is OK, because the degree vector $(1,1)$ lies on the line between $(2,0)$ and $(0,2)$; i.e.,
it is not a vertex of the attainable-monomial polyhedron.
In order to guarantee that we find the vertices, it suffices to compare the results on univariate states---this is the main insight of Section~\ref{sec:multivariate}.

\begin{lem}\label{lem:evmm-comp}
Let {\C} be any command.
Given any monomial state $\vec s$,
any ${\vec p}\in \sempar{C}^S$,
and any univariate state $\vec x$, there are
% ${\vec q}\in\sempar{C}^S$ and
a monomial state ${\vec g}$ and matrix $M$ such that
\begin{align*}
& \evmm \sempar{C} {\vec s} \Rightarrow^* ({\vec g},M), \\
& {\vec g}\acirc {\vec x} \ge {\vec p}\acirc {\vec s}\acirc {\vec x} \text{\ and} \\
&M = \ldfm({\vec p}) .
\end{align*}
\end{lem}

\begin{proof}
Fix $\vec s$, $\vec p$ and $\vec x$. By Theorem~\ref{thm:alg-correct}
there is  ${\vec y} \ge {\vec p}\acirc {\vec s}\acirc {\vec x}$ such that
 \[\evm \sempar{C} ({\vec s}\acirc {\vec x}) \Rightarrow^* ({\vec y}, \ldfm({\vec p})) \,.\]
Now, since $\evmm$ simulates the execution of $\evm$ step by step, with the sole difference that it operates on the elements of $\vec s$ instead
of those of ${\vec s}\acirc {\vec x}$, and that the only operations applied to monomials are copying and product,
it should be clear that
\[( \evmm \sempar{C} {\vec s} ) \acirc {\vec x} \]
evaluates to the same set of values as
\[ \evm \sempar{C} ({\vec s} \acirc {\vec x}) \]
and this justifies the lemma.
\end{proof}

\begin{cor}[Completeness of $\evmm$]
Let {\C} be a core-language program. Let $m = x_1^{d_1}\cdots x_n^{d_n}$
 be a monomial representing an attainable bound on the final value of $\X_n$ after executing {\C}, and such that its degree vector $\vec d$ is a vertex
of $\setT'$.
Then $\evmm \sempar{C} \idppol \Rightarrow^* ({\vec g}, M)$ (for some $M$)
such that ${\vec g}[n] = m$.
\end{cor}

\begin{proof}
Since $\vec d$ is a vertex, there is a univariate state $\vec x$ for which $\vec d$ is the unique solution to the optimization problem~\eqref{eq:max-on-S},
meaning that
$m\acirc \vec x$ maximizes the $n$th component in  ${\vec p}\acirc {\vec x}$ when $\vec p$ ranges over $\sempar{C}^S$.
By the last lemma, this can be matched by $\evmm$, and since $\evmm$ is also sound (i.e., it only produces attainable monomials), it must produce $m$.
\end{proof}

Since we can find any vertex of the polyhedron, we have a complete solution in the sense that the \textbf{max} of all these monomials
is a tight multivariate upper bound on the queried variable, as in Corollary~\ref{cor:monomialbounds}.

\subsection{Discussion}

We have given a polynomial-space solution to the problem of \emph{generating} a bound.
Since the algorithm can generate all the vertices of the attainable-monomial polyhedron, it
could be used to implement an alternative \emph{decision} procedure as well, replacing the proof of Theorem~\ref{thm:mv-pspace}.
Note that, ultimately, both proofs rely on the insight expressed by Lemma~\ref{lem:max-on-S}, and even if this is a redundancy,
we thought that presenting both approaches makes the picture more complete.

Looking back at $\evmm$, as we have already noticed that the initial state does not really influence any decision by the algorithm, it is easy to see
that we could rewrite the algorithm to remove this parameter.  The algorithm then abstracts a command to a set in $\absppol\times M_n(\mathbb{B})$,
rather than a function that maps initial states to such sets. Next, we give this version in equational form, in the style of the Closure Algorithm.
The result set may be of exponential size, so the non-deterministic interpreter form of the algorithm is useful
to prove the PSPACE complexity class, but we feel that the following presentation is of a certain elegance.  The abstraction of command {\C} is denoted
$\sempar{C}^M$ (M for Monomials).

\noindent We define $\sempar{C}^M$, first for non-looping commands:
\[ \begin{array}{lcl}
\sempar{{\askip}}^M &=& \{(\idppol, I)\} \\
\sempar{{\acopy{i}{j}}}^M &=& \{({\ampcopy}_{ij}, A(j,i)) \} \\
\sempar{{\asum{i}{j}{k}}}^M &=& \{({\ampcopy}_{ij}, A(j,i)+A(k,i)),\ ({\ampcopy}_{ik}, A(j,i)+A(k,i)) \} \\
\sempar{{\amul{i}{j}{k}}}^M &=& \{ ({\ampmul}_{ijk}, E(i)) \} \\
\lsem\verb+choose C+_1\verb+ or C+_2\rsem^S &=&
  \sempar{C$_1$}^S\cup\sempar{C$_2$}^S \\[1ex]
\lsem{\tt C}_1 {\tt ;C}_2\rsem^S &=&
   \sempar{C$_2$}^S\odot\sempar{C$_1$}^S  \,, \\
\end{array}
\]
where $({\vec p},A)\odot ({\vec q},B) \eqdef ({\vec p}\acirc {\vec q},\ A\cdot B)$.

\noindent
Finally,
\[   \lsem\verb+loop X+_\ell \verb+ {C}+\rsem^M  =  \closure{Gen(\ell,\closure{\sempar{C}^M})}  \]
where $\closure{}$ denotes closure under $\odot$, and
\[  Gen(\ell,\setS) = \bigcup_{\begin{array}{c}\scriptstyle ({\vec p},M)\in \setS  \\ \scriptstyle Z \subset [n] \end{array}}
                  GEN(Z, \ell, \, ({\vec p}\acirc (\idppol [Z\gets 0]),M)) \,,
\]
with function $GEN$ defined just as in $\evmm$ (namely just as in $\evm$, but calculating with multivariate monomials),
as illustrated in the following example.
% \footnote{In fact, GEN could be optimized here, because $Z$ is selected when $M$ is already known, so one could construct just the relevant sets $Z$.}

\begin{exa}
Consider the following loop:
\begin{Verbatim}[codes={\catcode`$=3\catcode`_=8}]
loop X$_5$ {
  choose
    { X$_3$ := X$_1$; X$_4$ := X$_2$ }
   or
      X$_1$ := X$_3$ + X$_4$
}
\end{Verbatim}
The abstraction of the \emph{loop body} clearly yields the following pairs:
\begin{alignat*}{2}
&( \fivetuple{ x_1 }{ x_2 }{ x_1 }{ x_2 }{ x_5 } &, & \begin{bmatrix}
   1 & 0 & 1 & 0 & 0 \\
   0  & 1 & 0 & 1 & 0 \\
   0  & 0 & 0 & 0 & 0 \\
   0  & 0 & 0 & 0 & 0 \\
   0  & 0 & 0 & 0 & 1
\end{bmatrix} ), \\
&( \fivetuple{ x_3 }{ x_2 }{ x_3 }{ x_4 }{ x_5 } &, & \begin{bmatrix}
   0 & 0 & 0 & 0 & 0 \\
   0  & 1 & 0 & 0 & 0 \\
   1  & 0 & 1 & 0 & 0 \\
   1  & 0 & 0 & 1 & 0 \\
   0  & 0 & 0 & 0 & 1
\end{bmatrix} ), \\
&( \fivetuple{ x_4 }{ x_2 }{ x_3 }{ x_4 }{ x_5 } &, & \begin{bmatrix}
   0 & 0 & 0 & 0 & 0 \\
   0  & 1 & 0 & 0 & 0 \\
   1  & 0 & 1 & 0 & 0 \\
   1  & 0 & 0 & 1 & 0 \\
   0  & 0 & 0 & 0 & 1
\end{bmatrix} )
\end{alignat*}

In the closure of the loop body's abstraction we will find compositions of these results; e.g., composing the first with the second (respectively the third)
corresponds to simulating the instructions:
\begin{Verbatim}[codes={\catcode`$=3\catcode`_=8}]
     X$_3$ := X$_1$; X$_4$ := X$_2$;
     X$_1$ := X$_3$ + X$_4$;
 \end{Verbatim}
 and yields the results:
\begin{alignat*}{2}
&( \fivetuple{ x_1 }{ x_2 }{ x_1 }{ x_2 }{ x_5 } &, & \begin{bmatrix}
   1   & 0 & 1 & 0 & 0 \\
   1    & 1 & 0 & 1 & 0 \\
   0    & 0 & 0 & 0 & 0 \\
   0    & 0 & 0 & 0 & 0 \\
   0    & 0 & 0 & 0 & 1
\end{bmatrix} ), \\
&( \fivetuple{ x_2 }{ x_2 }{ x_1 }{ x_2 }{ x_5 } &, & \begin{bmatrix}
   1   & 0 & 1 & 0 & 0 \\
   1    & 1 & 0 & 1 & 0 \\
   0    & 0 & 0 & 0 & 0 \\
   0    & 0 & 0 & 0 & 0 \\
   0    & 0 & 0 & 0 & 1
\end{bmatrix}  ).
\end{alignat*}
To illustrate the effect of $GEN$,
let $Z= \{1, 3, 4\}$. Inspecting the above matrix, we note that $3,4$ must be in the set to permit generalization,
 since they denote non-self-dependent entries; the presence of 1 in $Z$ implies that we treat $\X_1$ as an accumulator.
The above two matrices produce, respectively, the following results
\begin{alignat*}{2}
&( \fivetuple{ 0 }{ x_2 }{ 0 }{ x_2 }{ x_5 } &, & \begin{bmatrix}
   1  & 0 & 1 & 0 & o \\
   0 & 1 & 0 & 1 & 0 \\
   0  & 0 & 0 & 0 & 0 \\
   0  & 0 & 0 & 0 & 0 \\
   0  & 0 & 0 & 0 & 1
\end{bmatrix} ), \\
&( \fivetuple{ x_2 x_5 }{ x_2 }{ 0 }{ x_2 }{ x_5 } &, & \begin{bmatrix}
   1  & 0 & 1 & 0 & 0 \\
   0  & 1 & 0 & 1 & 0 \\
   0  & 0 & 0 & 0 & 0 \\
   0  & 0 & 0 & 0 & 0 \\
   0  & 0 & 0 & 0 & 1
\end{bmatrix}  ).
\end{alignat*}
\end{exa}

We state the correctness of $\sempar{$\cdot$}^M$, which is based on the results established for $\evmm$, by the simple observation that
\[
 {\vec p}\in \sempar{C}^M \iff (\evmm\sempar{C}\idppol \Rightarrow^* ( {\vec p} , \ldfm({\vec p}) )).
\]

\begin{thm}
Let \pgt{P} be a core-language program.  The set $\sempar{P}^M$ satisfies the following properties:
\begin{enumerate}
\item (soundness)
If ${\vec m}\in \sempar{P}^M$
 then $\exists {\vec p}\in \sempar{P}^S$ such that ${\vec m} \lepoly {\vec p}$.
\item (completeness)
Let $m = x_1^{d_1}\cdots x_n^{d_n}$
 be a monomial representing an attainable bound on the final value of variable $\X_n$ after executing \pgt{P}, and such that its degree vector $\vec d$ is a vertex
of $\setT'$.
Then there is ${\vec m}\in \sempar{P}^M$
such that ${\vec m}[n] = m$.
\end{enumerate}
\end{thm}

\noindent
Note that the soundness clause indicates that $\vec m$ is an attainable AMP for \pgt{P};
and that in the completeness clause, the choice of $\X_n$ is arbitrary. The same property holds for any chosen variable (and the corresponding set of attainable
monomial bounds).

\section{A Few Simple Extensions}%
\label{sec:resets}

A natural follow-up to the above results is to extend them to richer programming languages.  In this section we briefly discuss four such extensions.
First, two trivial ones:
\begin{itemize}
\item We can allow any polynomial (with positive coefficients) as the right-hand side of an assignment---this is just syntactic sugar.
\item We can include the weak (non-deterministic) assignment form which says ``let \pgt{X} be a natural number \emph{bounded} by the following expression.''
This construct may be useful if we want to use our language to approximately simulate the behavior of programs that have arithmetic expressions
 which we do not support,
as long as a preprocessor can statically determine polynomial bounds on such expressions.
No change to the algorithm is required.
\end{itemize}

\noindent
Next, two extensions which are not  trivial, but very easy.
\paragraph{\textbf{Resets}}
\newcommand\dzero[0]{{\underline 0}}\cite{B2010:DICE} shows that it is possible to enrich the core language by a reset instruction, \verb/X := 0/,
and still solve the problem of distinguishing polynomially-bounded variables from potentially super-polynomial ones.
Conceptually this is perhaps not a big change, but technically it caused a jump in the complexity of the solution, from PTIME to
PSPACE-complete.  Our tight-bound problem is already PSPACE-complete, and we can extend our solution to handle resets without a further increase
in complexity.  In fact, with the abstract-interpreter algorithm, adding the resets is very smooth.
We add a special constant $\dzero$, which means \emph{definitely zero} --- the symbolic value of a variable that has been reset.
This value is treated using the natural laws, i.e., $\dzero + x = x$,\ \  $\dzero \ast x = \dzero$.
Resetting a variable cuts data-flow, so the abstraction of \verb/X := 0/ reflects this in the LDFM:\@
\begin{alignat*}{2}
&\evm\sempar{\areset{i}} {\vec s} &\quad \Rightarrow \quad & ({\vec s}[i\gets \dzero] ,  E(i) )
\end{alignat*}

Additional changes are necessary in the analysis of loops.
If the loop bound is definitely zero, we know for certain that the loop body is not executed
(unlike the general case, where skipping the loop is just a \emph{possibility}).
Thus we rewrite the corresponding
section of our interpreter as follows:
\[
\evm \lsem\verb/loop X/_\ell \verb/{C}/\rsem {\vec s} =
         \begin{cases}
            ({\vec s}, I) & \text{if ${\vec s}[\ell] = \dzero$}, \\
            \evm^\oast\sempar{C}(\ell, {\vec s}) & \text{otherwise}.
         \end{cases}
\]
and similarly for $\evmm$.
Inside $\evm^\oast$ (respectively $\evmm^\oast$), we have to deal with the case that a definitely-zero variable is selected for the set $Z$ of variables
to be set to zero.
  Let us recall that this set serves two purposes:
\begin{enumerate}
\item To mask out non-self-dependent variables.  If the variable is initially definitely zero, there is of course no need to mask it out, and moreover,
replacing $\dzero$ with an ordinary zero may cause imprecision, in case that an internal loop depends on that variable.
\item To isolate the increment $q$ in computations that set $x_i' = x_i + q$, by putting 0 in $x_i$.  This has to work, even in the case that $x_i$ is initially
$\dzero$; while, as above, replacing $\dzero$ with an ordinary zero would be wrong.
\end{enumerate}
From these considerations, the correct code for $\evm^\oast$ becomes the following.
\[\begin{array}{lcl}
\evm^\oast\sempar{C}(\ell, {\vec s}) & \Rightarrow & ({\vec s}, I) \\[1ex]
\evm^\oast\sempar{C}(\ell, {\vec s}) & \Rightarrow &  \pgt{let } Z \subseteq \{i \mid {\vec s}[i]\ne \dzero\}\pgt{;} \\
&&  \pgt{let }({\vec s}_1, M_1) =  GEN(Z, \ell, \, \evm^*\sempar{C} ({\vec s} [Z\gets 0])) \pgt{;} \\
&&\pgt{let }({\vec s}_2, M_2) = \evm^\oast\sempar{C}(\ell,{\vec s}_1) \\
&&\pgt{in}\quad ({\vec s}_2, M_1\mtimes M_2) \\
\end{array}\]
while in function $GEN$:

\begin{align*}
GEN(Z, \ell, ({\vec s}, M)) = &\text{\ if\ } ( \forall i\notin Z : ({\vec s}[i] = \dzero \lor M_{ii}=1) )  \\
 & \text{\ then\ } ({\vec s}^{gen},M^{gen}) \text{\ else\ } ({\vec s},M) \\
\intertext{where}
 {\vec s}^{gen} &=  {\vec s} \prod_{i  : M_{ii}=1 \land (i\in Z \lor {\vec s}[i] = \dzero )\kern -1ex} {\ampmul}_{i i \ell} \\
 {M^{gen}} &=   (I\land M) M (I\land \overline M) \lor (I\land M)
\end{align*}

\paragraph{\textbf{Unbounded ``value''}}
In~\cite{BAPineles:2016} we proposed to extend our language with an instruction \verb/X := */, sometimes called ``havoc \pgt{X}.''
The intention is to indicate that $\X$ is set to a value which is not expressible in the core language. This may be useful if we want
to use our core-language programs as an abstraction of real programs, that possibly perform computations that we cannot express or
bound by a polynomial.
It may also be used to model a \textbf{while}-loop (as done, for example, in~\cite{JK08}): we do not attempt to analyze the loop for termination
or even for bounds (this is outside the scope of our algorithm; such loops depend on concrete conditionals which we do not model at all).
So all we suggest is to analyze what happens in the loop under the assumption that we cannot bound the number of iterations. This can be simulated
by setting the loop bound to \pgt{*}.

The implementation is very similar to that of reset, and moreover, they can be combined. The rules for calculating with  \pgt{*}
are: $\pgt{*}\cdot 0 = 0$, and $\pgt{*}\cdot m = \pgt{*}$ for any other monomial $m$.
And, of course, if an output variable comes out as \pgt{*} it is reported to have no polynomial upper bound.

\section{Related Work}%
\label{sec:rw}

Bound analysis, in the sense of finding symbolic bounds for data values, iteration bounds and related quantities,
is a classic field of program analysis~\cite{Wegbreit:75,Rosendahl89,ACE}. It is also an area of active research, with tools being
currently (or recently) developed including \tool{COSTA}~\cite{Albert-et-al:TCS:2011}, \tool{AProVE}~\cite{APROVE-JAR2017},
\tool{CiaoPP}~\cite{CiaoPP-TPLP2018} %see arxiv  1803.04451
, $C^4B$~\cite{CHS:pldi2015},
 \tool{Loopus}~\cite{SZV:jar2017} and \tool{CoFloCo}~\cite{FMH14}---and this is just for imperative programs. There is also work on functional and logic programs,
term rewriting systems, recurrence relations, etc.~which we cannot attempt to survey here.

Our programming language, which is based on bounded loops, is similar to Meyer and Ritchie's language of \emph{Loop Programs}~\cite{MR:67},
which differed by being deterministic and including constants; it is capable of computing all the primitive recursive functions and too strong to obtain
decidability results. Similar languages have been used as objects for analysis of growth-rate (in particular, identifying polynomial complexity) in~\cite{KasaiAdachi:80,KN04,NW06,JK08}. These works, too, considered deterministic languages for which precise analysis is impossible, but on reading them,
one can clearly see that there are some clear limits to the aspects of the language that they analyze (such as using a loop counter as a bound, not relying
on the assumption that the iteration count is always completed). Such considerations led to the definition of the weak, non-deterministic language in~\cite{BJK08}.
 Recent work in static analysis of fully-capable programs~\cite{Giesl:toplas2016} combines a subsystem for
loop-bound analysis (via ranking functions) with a subsystem for
growth-rate analysis, which establishes symbolic bounds on data that grow along a loop, based on loop bounds provided by the loop analysis.
This may be seen, loosely speaking, as a process that reduces general programs to bounded-loop programs which are then analyzed.
Our previous paper~\firstpaper was, however, the first that gave a solution for computing tight polynomial bounds which is complete for the class of programs
we consider.

The problem of deciding termination of a weak programming language has been considered in several works;
one well-studied example considers single-path
while loops with integer variables, affine updates, and affine guard conditions~\cite{TIWARI04,Braverman06,Hosseini19}.
Its decidability was first proved in~\cite{Hosseini19}, but it is not shown to what complexity class this
problem belongs. It would be interesting to see whether this can also be shown to be PSPACE-complete.

Dealing with a somewhat different problem,~\cite{Seidl-polynomial-invariants,Ouaknine-polynomial-invariants}
 both check, or find, \emph{invariants} in the form of polynomial equations, so they can identify cases where a final result is
\emph{precisely} polynomial in the input values. We find it remarkable that they give
complete solutions for  weak languages, where the weakness lies in the non-deterministic control-flow, as in our language.
However, their language has no  mechanism to represent bounded loops, whereas in our work, it is all about these loops.

\section{Conclusion}

We have complemented previous  results on the computability of polynomial bounds in the weak imperative language defined in~\cite{BJK08}.
While~\cite{BJK08} established that the existence of polynomial bounds is decidable, \firstpaper showed that tight bounds are computable.
Here we have shown that they can be computed in polynomial space, and that this is optimal in the sense of PSPACE-hardness.
We have thus settled the question of the complexity class of this program-analysis problem.  Interestingly, this improvement required some new
ideas on top of \firstpaper, including the connection of multivariate bounds to univariate bounds based on the geometry of the set of
monomial bounds, and the computation of bounds using an abstraction which is space-economical, including only degree vectors and data-flow
matrices.

Some challenging open problems regarding bound computation remain:
\begin{enumerate}
\item Whether the problem is solvable at all for various extensions to the language (see~\cite{BK11,BH:LMCS:2019} for further discussion).
\item How to compute more precise polynomial bounds (with explicit constant factors); can we make them tight?
\item Computation of tight super-polynomial bounds (currently,
if a variable cannot be bounded by a polynomial, we give no upper bound).
\end{enumerate}

\noindent
Another problem we are interested in is: if one of our programs computes
a polynomially-bounded result, how high can the degree of the bounding polynomial be, in terms of the size of the program? We conjecture that
the degree is at most exponential in the program size.  However, a proof which we attempted (in an earlier technical report) turned out to be wrong, so the
problem stands open.

\subsection*{Acknowledgement}

We thank the anonymous referees, whose diligent reviews have been invaluable to our work on this article.
On the occasion of his $80^{th}$ birthday, we would also like to acknowledge the enormous contributions made by Neil D. Jones in this area,
and his influence on our work.

%appendix: proof of the new closure algorithm (based on the old algorithm)

\appendix

\section{Justification of The Closure Algorithm}%
\label{sec:proof-appendix}

The purpose of this section is to clarify how the Closure Algorithm (Section~\ref{sec:closurealg}) relates to the algorithm we published in \firstpaper,
and justify the correctness of the new version.  The justification amounts to a proof that the algorithm computes a set of lower bounds which is equivalent
to the set computed by \firstpaper; we are thus relying on the correctness proof of \firstpaper.

\subsection{Background}

We begin with a stock of definitions and auxiliary results from \firstpaper. This subsection contains no novel material and is included for completeness.

\begin{defi}[Subsumption order]
For $\tau$-multivariate polynomials $p,q$ we define $p \subsumed q$ to hold if for every monomial $\mathfrak m$ of $p$ some multiple
of $\mathfrak m$ appears in $q$.  We then say that $p$ is subsumed by $q$.
\end{defi}

For example,  we have $x_1x_2 \subsumed \tau x_1 x_2 + x_3$.  Moreover,  $x_1 +\tau x_1 \subsumed \tau x_1$.
It should be clear that this order agrees with the asymptotic growth-rate order commonly denoted by $p = O(q)$.
% If we have an upper bound $\max\{p,q,\dots\}$ on some value (equivalently, we assert that the value is bounded by \emph{some} element of a set of polynomials),
%  the \emph{subsumption rule} eliminates
% subsumed polynomials, in order to reduce the size of the set, while preserving the soundness of the upper bound (up to constant factors).
The definition is extended to MPs  component-wise.
Note that $\subsumed$ is a coarser relation than the fragment relation $\lepoly$, i.e., ${\vec p} \lepoly {\vec q} \Rightarrow {\vec p} \subsumed {\vec q}$.
For example,  $x_1x_2$ is not a fragment of $\tau x_1 x_2$, but is subsumed by it.

\begin{defi}
$\vec p \in \tabsppol$ is called \emph{idempotent} if $\vec p \acirc \vec p = \vec p$.
\end{defi}
\noindent
Note that this is composition in the abstract domain. So, for instance, $\tuple{x_1, x_2}$ is idempotent, and so is
$\tuple{x_1 + x_2, x_2}$, while $\tuple{x_1x_2, x_2}$ and $\tuple{x_1 + x_2, x_1}$ are not.

% \begin{defi} \label{def:MPnotation}
% We define a notational convention for $\tau$-MPs, specifically for self-dependent entries of the MP.  Assuming that $x_i$ appears in ${\vec p}[i]$, we write:
% \[ {\vec p}[i] = x_i + \tau {\vec p}[i]' + {\vec p}[i]''+ {\vec p}[i]''' \, ,\]
% where ${\vec p}[i]'''$ includes all the non-iterative monomials of ${\vec p}[i]$, while the iterative monomials
% (other than $x_i$) are grouped into two sums:
% $\tau{\vec p}[i]'$, including all monomials with a positive degree of $\tau$, and ${\vec p}[i]''$ which includes all the $\tau$-free monomials.
% \end{defi}

% \begin{exa} \label{ex:MPnotation}
% Let $\vec p = \tuple{ x_1 + \tau x_2 + \tau x_3 +  x_3 x_4,\   x_3,\ x_3 ,\ x_4 }$. Then $\sd{\vec p} = \{1,3,4\}$.
%  The self-dependent variables are all but $x_2$.
% Since $x_1$ is self-dependent, we will apply the above definition to ${\vec p}[1]$, so that
%  ${\vec p}[1]' = x_3$, ${\vec p}[1]'' = x_3 x_4$ and ${\vec p}[1]''' = \tau x_2$.
%  Note that a factor of $\tau$ is stripped in ${\vec p}[1]'$.  Had the monomial been $\tau^2 x_3$, we would have ${\vec p}[1]' = \tau x_3$.
%  \end{exa}

Let ${\vec p} \in \tppol$. Recall (Section~\ref{sec:proof}) that its \emph{dependence graph} $G({\vec p})$ is
a directed graph with node set $[n]$, which includes an arc $i\to j$ if and only if
${\vec p}[j]$ depends on $x_i$.

The following lemma (similar to Lemma~\ref{lem:qacyclic}) is from \firstpaper.

\begin{lem}\label{lem:nearly-dag}
Suppose that the SDL $\setS$ is polynomially bounded,
${\vec p}\in \tppol$ is attainable over $\setS$, and $\alpha({\vec p})$ is
idempotent. Then
 $G({\vec p})$ does not have any simple cycle longer than one arc. In other words, $G({\vec p})$ is quasi-acyclic.
\end{lem}

We assume that the loop under analysis is polynomially bounded; therefore we can rely on the above property of
$G({\vec p})$, being quasi-acyclic. We assume, w.l.o.g., that the variables
are indexed in an order consistent with $G({\vec p})$, so that if $x_i$ depends on $x_j$ then $j\le i$.
We refer to an (abstract) MP satisfying this as \emph{neat}.

\subsection{The Old Closure Algorithm}

We now recall the algorithm from \firstpaper, to facilitate its comparison with the current version. We are only considering the SDL analysis component---%
the rest is identical.  The following function from \firstpaper is equivalent in role to function $LC$ (Equation~\ref{eq:newLC}):

\smallskip
\noindent
\procSDL{}$(\setS)$
\begin{enumerate}
\item Set $T = \setS$.
\item Repeat the following steps until $T$ remains fixed:
\begin{enumerate}
\item Closure: Set $T$ to $\closure{T}$.
\item Generalization: For all ${\vec p}\in T$ such that ${\vec p}\acirc {\vec p} = {\vec p}$, add ${\vec p}^\tau$ to $T$.
\end{enumerate}
\end{enumerate}

\noindent
The reader may have noticed that in this algorithm, generalization is applied to idempotent elements, instead of iterative elements. As an idempotent element
is not necessarily iterative, the definition of generalization we used (Definition~\ref{def:generalize}) does not suffice.

%old definition
\begin{defi}[extended generalization]\label{def:oldgen}
Let ${\vec p}\in \tabsppol$; define
${\vec p}^\tau$ by: \[
%the following is a good definition only once we show that \itker{\vec p} is defined
%  {\vec p}^\tau = {\vec p}^\tau + \itker{\vec p}^\tau
 {\vec p}^\tau[i] = \begin{cases}
   {\vec p}[i] + \tau {\vec p}[i]''  &  \text{if $i\in \sd{\vec p}$} \\
   {\vec p}[i] & \text{otherwise},
\end{cases}
\]
where ${\vec p}[i]''$ is defined as the sum of all iterative monomials of ${\vec p}[i]$, except for $x_i$ (a monomial in $\vec p$ is iterative if it consists of variables self-dependent in $\vec p$.
We employ the unmotivated notation ${\vec p}[i]''$  just for consistency with \firstpaper).
 \end{defi}
 \noindent
 Note that the addition in this definition is abstract (see example below).
% We remark that this definition is not a verbatim copy of the definition from \firstpaper, but it is
% easy to verify their equivalence.
 It is easy to verify that Definition~\ref{def:generalize} agrees with this definition (up to subsumption) if $\vec p$ is iterative (and, in particular,
 $\tau$-free;  monomials that include $\tau$ are not considered iterative).

\begin{exa}\label{ex:oldgen}  We illustrate Definition~\ref{def:oldgen} with an example.
Let $\vec p = \fourtuple{ x_1}{ x_2}{ x_2 }{ x_4 + \tau x_2 + \tau x_1 +  x_1 x_2}$. Then $\sd{\vec p} = \{1,2,4\}$.
  This $\tau$-AMP is not iterative, because of the $\tau$-monomials in ${\vec p}[4]$. However,
 it is idempotent, and applying Definition~\ref{def:oldgen},
${\vec p}^\tau = \fourtuple{ x_1}{ x_2}{ x_2 }{ x_4 + \tau x_2 + \tau x_1 + \tau x_1 x_2}$.
Note also that $\vec p$ is neat: $G({\vec p})$ is shown in Figure~\ref{fig:depgraph}.
 \begin{figure}[htb]
\[
\xymatrix@u@R=30pt{
x_1 \ar@(dl,ul)[]\ar@<2pt>@/^15pt/[ddd]  & \\ % chktex 25
x_2  \ar@(dr,dl)[]\ar@/^4pt/[d]\ar@/_15pt/[dd] &  \\ % chktex 25
x_3  \ar@/^4pt/[d] & \\
x_4  \ar@(dr,dl)[] &
}
\]
\bigskip
\caption{Dependence graph for the example MP\@. Nodes are labeled $x_j$ rather than just $j$ for readability. }\label{fig:depgraph}
\end{figure}
 \end{exa}

\begin{exa} Next, we give an example for a complete computation of \procSDL\@.  We consider the loop:
\begin{center}
$\verb/loop X/_3\verb/ { X/_1\verb/:= X/_1\verb/ + X/_2\verb/; X/_2\verb/:= X/_2\verb/ + X/_3\verb/; X/_4\verb/:= X/_3\verb/ }/$
\end{center}
\noindent
The body of the loop is evaluated symbolically and yields the abstract multi-polynomial:
\[
 \vec p = \fourtuple{ x_1+x_2}{ x_2+x_3}{ x_3}{ x_3 }
 \]
 This singleton is the initial set $T$. To obtain $\closure{T}$, we include the identity MP, and also compute the compositions
 \begin{align*}
{({\vec p})}^{\acirc (2)} &=  {\vec p} \acirc {\vec p} = \fourtuple{ x_1+  x_2+x_3}{ x_2+ x_3}{ x_3}{ x_3 } ; \\
 {({\vec p})}^{\acirc (3)} &=  {({\vec p})}^{\acirc (2)}  \,.
 \end{align*}
 Here the closure computation stops. Next, in the generalization phase,
since $ {\vec p}^{\acirc (2)} $ is idempotent, we compute:
\[
{\vec q} = {({({\vec p}) }^{\acirc (2)})}^\tau  = \fourtuple{ x_1+\tau x_2 + \tau x_3 }{ x_2+\tau x_3 }{ x_3}{ x_3 }
\]
So, the following set results:
\begin{align*}
&\fourtuple{ x_1}{ x_2}{ x_3}{ x_4 },
&\fourtuple{ x_1+x_2}{ x_2+x_3}{ x_3}{ x_3 } \\
&\fourtuple{ x_1+  x_2+x_3}{ x_2+  x_3}{ x_3}{ x_3 } ,
&\fourtuple{ x_1+\tau x_2 + \tau x_3 }{ x_2+\tau x_3 }{ x_3}{ x_3 }
\end{align*}
applying closure again, we obtain some additional results:
\[\begin{array}{ll}
{\vec q}  \acirc {{\vec p} } &= \fourtuple{ x_1+ x_2 + x_3 + \tau x_2 + \tau x_3 }{ x_2+ x_3 +\tau x_3 }{ x_3}{ x_3 } \\
{({\vec q})}^{\acirc (2)} &= \fourtuple{ x_1+ \tau x_2 + \tau x_3  + \tau^2 x_3 }{ x_2 +\tau x_3 }{ x_3}{ x_3 } \\
{({\vec q})}^{\acirc (2)}  \acirc {{\vec p} }
&= \fourtuple{ x_1+  x_2 + x_3 + \tau x_2  + \tau x_3 + \tau^2 x_3}{ x_2 +  x_3 + \tau x_3}{ x_3}{ x_3 }
\end{array}
\]
The last element is idempotent but applying generalization does not generate anything new. Thus the final set $T$ has been reached.
 The reader may reconsider the source code to verify that we have indeed obtained tight bounds for the loop.
 Note that $\vec p$ itself is not idempotent, but it is iterative. Our improved version of the Closure Algorithm (Section~\ref{sec:closurealg})
 would generalize $\vec p$, adding ${\vec p}^\tau = \fourtuple{ x_1+ \tau x_2}{ x_2+ \tau x_3}{ x_3}{ x_3 }$.  This result is subsumed by $\vec q$
 above, showing that the final sets of bounds resulting from the two algorithms are equivalent.
\end{exa}

Our new algorithm is simpler than the old one as it performs only one round of generalization---it does not ``iterate until fixed-point'' like \procSDL{}.
This is a simplification, since we do not have to consider the issue of correctly generalizing MPs that already include $\tau$.  It can be proved that
every idempotent AMP has an iterative kernel; but many AMPs which are not idempotent also have an
iterative kernel. One could say that the new algorithm is more eager to generalize.
The motivation for this change was to elimnate the idempotence condition:
 in order to judge that an AMP is idempotent we need full information about it. Our PSPACE algorithm operates with very partial information
and we derive its soundness by establishing that what it generalizes is an iterative fragement of \emph{some} AMP in the closure, which we cannot
guarantee to be idempotent.  Thus, we made this change in order to justify
the PSPACE algorithm. Nonetheless, we believe that our current variant may be more efficient in practice than the old one, mostly because of the elimination
of the outer ``iterate until fixed point'' loop.
It would be our choice,
should we prefer to implement a closure-based algorithm (perhaps
because determinization of the PSPACE algorithm, which is non-deterministic, would result in a deterministic algorithm worse in practice).

\subsection{Equivalence of the algorithms}

We now prove that our revised algorithm gives results equivalent to the former version.  Basically, we prove subsumption of the two sets of results
in each other.

\begin{lem}\label{lem:oldLEnew}
Every ${\vec p}\in  \procSDL{}(\setS)$ is subsumed by a member of $LC(\setS)$.
\end{lem}

\begin{proof}
We prove inductively, based on the structure of the algorithm, that the set $T$, computed by \procSDL{}, is subsumed by $LC(\setS)$.

\noindent
\emph{Base case}: $T = \setS$ is, of course, contained in $\setS$.

\noindent
\emph{Induction Step~1}: Setting $T$ to $\closure{T}$ preserves the relation because $LC(\setS)$ is closed under composition, as one can clearly see from its definition.

\noindent
\emph{Induction Step~2}: This step adds ${\vec p}^\tau$ to $T$ for idempotent AMPs $\vec p$.  We shall now prove that ${\vec p}^\tau \subsumed LC(\setS)$.
We consider two sub-cases.  If $\vec p$ is $\tau$-free, we infer that ${\vec p}\in\closure{\setS}$. Then it is clearly also in $LC(\setS)$.
If $\vec p$ has $\tau$'s, we observe that ${\vec p}[1/\tau]$ is what one gets by performing the same \emph{composition} operations
that yielded $\vec p$, but skipping the generalizations.  So ${\vec p}[1/\tau] \in \closure{\setS} \subseteq LC(\setS)$.
It is easy to see that ${\vec p}[1/\tau]$ is idempotent as well; thus $\itker{{\vec p}[1/\tau]}^\tau$ is formed by function $Gen$.
 Also, by IH, there is
${\vec q}\in LC(\setS)$ such that ${\vec p} \subsumed {\vec q}$.   Using
a few lemmas proved below, we conclude that  ${\vec q}\acirc \itker{{\vec p}[1/\tau]}^\tau \acirc {\vec p}[1/\tau] \subsumes {\vec p}^\tau$;
specifically,
\begin{align*}
{\vec q}\acirc \itker{{\vec p}[1/\tau]}^\tau \acirc {\vec p}[1/\tau] &\subsumes
{\vec p}\acirc \itker{{\vec p}[1/\tau]}^\tau \acirc {\vec p}[1/\tau] && \text{(Lemma~\ref{lem:subsume-comp})} \\
&\subsumes {\vec p} \acirc {({\vec p}[1/\tau])}^\tau && \text{(Lemma~\ref{lem:itker})} \\
&\subsumes {\vec p}^\tau && \text{(Lemma~\ref{lem:retractTau})}
\end{align*}
\noindent
The above product is in $LC(\setS)$, as we have established that each of its factors is present in the set $Gen(\closure{\setS})$;
we conclude ${\vec p}^\tau \subsumed LC(\setS)$.
\end{proof}

\begin{lem}\label{lem:subsume-comp}
For $p, q\in \tabspol$ and ${\vec r} \in \tabsppol$,
if ${\vec q} \subsumes {\vec p}$ then ${\vec q}\acirc {\vec r} \subsumes {\vec p}\acirc {\vec r}$.
\end{lem}

\begin{proof}
Let $\mathfrak m$ be a monomial of $p\acirc {\vec r}$. It is obtained by substituting components ${\vec r}[i]$
into variables $x_i$ of a monomial ${\mathfrak p}$ of $p$.  Now, $q$ is known to contain some multiple
${\mathfrak q}$ of ${\mathfrak p}$, and clearly substituting $\vec r$ in $\mathfrak q$ yields a multiple of $\mathfrak m$.
\end{proof}

\begin{lem}\label{lem:retractTau}
Let ${\vec p} \in \tabsppol$ be idempotent and neat.  Then ${\vec p}\acirc {({\vec p}[1/\tau])}^\tau \subsumes {\vec p}^\tau$.
\end{lem}

\begin{proof}
We prove, by induction on $i$, that the inequality holds for the first $i$ components of the respective AMPs;
we write this using ``array slice notation'' as $({\vec p}\acirc {({\vec p}[1/\tau])}^\tau)[1..i] \subsumes {\vec p}^\tau[1..i]$.

\emph{Base case}:
Since $\vec p$ is neat we know that ${\vec p}[1] = x_1$.
It is also clear that ${({\vec p}[1/\tau])}^\tau [1] = x_1$.  Consequently,
$({\vec p}\acirc {({\vec p}[1/\tau])}^\tau)[1] = x_1 = {\vec p}^\tau [1]$.

\emph{Induction step}: let  $i > 1$.
If $i \notin \sd{\vec p}$, then ${\vec p}[i]$ only depends on variables $x_j$ with $j<i$. This allows us to use the induction hypothesis:
\begin{align*}
 {\vec p}\acirc {({\vec p}[1/\tau])}^\tau[i] &=
 {\vec p}\acirc {\vec p}\acirc {({\vec p}[1/\tau])}^\tau[i]  &&\text{by idempotence} \\ &=
 {\vec p}[i] \acirc {\vec p}\acirc {({\vec p}[1/\tau])}^\tau \\ &\subsumes
 {\vec p}[i] \acirc {\vec p}^\tau  &&\text{by IH} \\ &\subsumes
 {\vec p}[i] \acirc {\vec p} = {\vec p}[i] = {\vec p}^\tau [i]  &&\text{by Def.~\ref{def:oldgen}} \,.
\end{align*}

If $i\in \sd{\vec p}$, then  ${\vec p}[i] = x_i + p_i(\vec x, \tau)$ where $p_i$ only depends on variables $x_j$ with $j<i$.
For conciseness, define $\vec q$ to be ${\vec p}\acirc {({\vec p}[1/\tau])}^\tau$.
Now by idempotence of $\vec p$:
\begin{align*}
({\vec p}\acirc {({\vec p}[1/\tau])}^\tau)[i] &=
 ({\vec p}\acirc {\vec p}\acirc {({\vec p}[1/\tau])}^\tau)[i] \\ &=
 {\vec p}[i] \acirc {\vec q} \\ &=
 {\vec q}[i] + p_i \acirc {\vec q} \\
&\subsumes
 {({\vec p}[1/\tau])}^\tau [i] + p_i \acirc {\vec q}  && \hspace{-1cm}{\text{since ${\vec p}[i] \subsumes x_i$}} \\
 &\subsumes
 x_i + \tau ({\vec p}[1/\tau])[i]''+ p_i \acirc {\vec q} && \hspace{-1cm}{\text{by Def.~\ref{def:oldgen}}} \\
&\subsumes
 x_i + \tau {\vec p}[i]''+ p_i \acirc {\vec q}  && \hspace{-1cm}{\text{since ${\vec p}[i]''$ is $\tau$-free}} \\
&=
 x_i + p_i \acirc {\vec q} + \tau {\vec p}[i]'' \\
&\subsumes
 x_i + p_i \acirc {\vec p}^\tau + \tau {\vec p}[i]''  && \hspace{-1cm}{\text{by IH}}  \\
&\subsumes
 x_i + p_i \acirc {\vec p} + \tau {\vec p}[i]'' \\ &=
({\vec p}\acirc {\vec p})[i]  + \tau {\vec p}[i]''  \\ &\subsumes
{\vec p}[i]  + \tau {\vec p}[i]''  \quad = \quad
{\vec p}^\tau [i] \,.
\tag*{\qedhere}
 \end{align*}
\end{proof}

\begin{lem}
Assume that ${\vec p}\in \absppol$ is taken from the closure set of a polynomially-bounded SDL\@.
If ${\vec p}$ is idempotent, then $\itker{\vec p}\acirc {\vec p}  = {\vec p}$.
\end{lem}

\begin{proof}
$G({\vec p})$ is quasi-acyclic by Lemma~\ref{lem:nearly-dag} above.
Thus, w.l.o.g., it is neat (that is, the variables are indexed to match the DAG order).
Clearly, $\itker{\vec p} \lepoly {\vec p}$ and consequently $\itker{\vec p}\acirc {\vec p}  \lepoly {\vec p} \acirc {\vec p} = {\vec p}$.
Now, assume that equality does not hold. Let $\mathfrak m$ be a lexicographically biggest monomial%
\footnote{Monomials are ordered lexicographically by viewing $x_1^{d_1}\dots x_n^{d_n}$ as $\tuple{d_n,\dots,d_1}$. The reversal of the list
has the effect that dropping the highest-numbered variable makes a monomial smaller, even if other degrees go up.}
 that appears in $\vec p$ and is missing in $\itker{\vec p} \acirc {\vec p}$.  Consider the calculation of ${\vec p}\acirc {\vec p}$.
 It generates $\mathfrak m$ by substituting monomials from the right-hand instance of $\vec p$ for the variables of a certain monomial $\mathfrak n$
taken from the left-hand instance.  Because $\vec p$ is neat, substitution replaces a variable $x_i$ either by $x_i$ itself or by a lexicographically-smaller monomial.
 If all the variables of $\mathfrak n$ could be replaced by themselves, then $\mathfrak n = \mathfrak m$ would be a iterative monomial and
 therefore present in $\itker{\vec p}$ and in $\itker{\vec p} \acirc {\vec p}$.  We conclude that some variables in $\mathfrak n$ are not self-dependent.
 It is further easy to see that $\mathfrak m$ is lexicographically smaller than $\mathfrak n$. By the choice of $\mathfrak m$, it follows that
 $\mathfrak n$ appears in $\itker{\vec p} \acirc {\vec p}$.  But then $\mathfrak m$ is present in
 $(\itker{\vec p} \acirc {\vec p}) \acirc {\vec p}$ which equals $\itker{\vec p} \acirc {\vec p}$ (by associativity of composition and idempotence of $\vec p$).
 A contradiction has been shown.
 \end{proof}

\begin{lem}\label{lem:itker}
Assume that ${\vec p}\in \absppol$ is taken from the closure set of a polynomially-bounded SDL\@.
If ${\vec p}$ is idempotent, then $\itker{\vec p}^\tau \acirc {\vec p}  \subsumes {\vec p}^\tau$.
\end{lem}

\begin{proof}
We already know that every monomial $\mathfrak m$ of $\vec p$ appears in $\itker{\vec p} \acirc {\vec p}$ and therefore either $\mathfrak m$
or $\tau \mathfrak m$ appears in $\itker{\vec p}^\tau \acirc {\vec p}$.  So we have covered all the monomials of ${\vec p}^\tau$ except
those affected by generalization, namely those included in the polynomials ${\vec p}[i]''$. By definition of ${\vec p}[i]''$,
they are self-dependent and also appear in $\itker{\vec p}$. In fact,
$\itker{\vec p}[i] = x_i + {\vec p}[i]''$ and
$\itker{\vec p}^\tau [i] = x_i + \tau{\vec p}[i]''$.  Substituting, we find that
$(\itker{\vec p}^\tau \acirc {\vec p})[i] \gepoly x_i + \tau{\vec p}[i]''$.
The statement of the lemma follows.
\end{proof}

The proof of Lemma~\ref{lem:oldLEnew} is now complete, and we move on to the converse subsumption. We will begin this time with an auxiliary result.

\begin{lem}\label{lem:idempotentPower}
Let $\setS$ be a polynomially-bounded SDL, and let ${\vec p}\in\closure{\setS}$.
Then there is a $k>0$ such that ${\vec p}^{(k)}$, the AMP representation of a $k$-fold application of $\vec p$, is idempotent.
\end{lem}

\begin{proof}
This follows from the fact that $\closure{\setS}$ is finite (an easy consequence of polynomial boundedness).
Finiteness implies that the sequence  ${\vec p}^{(i)}$ must repeat elements.  If  ${\vec p}^{(i)} =  {\vec p}^{(i+k)}$, $k>0$, then for any $j\ge i$,
 ${\vec p}^{(j)} =  {\vec p}^{(j+k)}$.  Assuming (w.l.o.g.) $k\ge i$, we have
  ${\vec p}^{(k)} =  {\vec p}^{(2k)}$, so ${\vec p}^{(k)}$ is idempotent.
\end{proof}

\begin{lem}\label{lem:itPowers}
Let $\setS$ be a polynomially-bounded SDL, and let ${\vec q}$ be an \emph{iterative} fragment of any ${\vec p}\in\closure{\setS}$.
Then \begin{enumerate}
 \item ${\vec q}^{(i)}$ is also iterative;
\item ${\vec q}^{(i)} \lepoly {\vec p}^{(i)}$;
\item ${\vec q}^{(i)} \lepoly {\vec q}^{(i+1)}$.
\end{enumerate}
\end{lem}

\begin{proof}
\begin{enumerate}
\item
It easy to see that if $i\in \sd{\vec q}$, then $i\in \sd{{\vec q}^{(i)}}$ for any $i$. On the other hand, as each component of ${\vec q}$ only depends
on the self-dependent variables, the same is true for ${\vec q}^{(i)}$; thus ${\vec q}^{(i)}$ is iterative. In fact, it has the same set of self-dependent variables.
\item
This is true because ${\vec q}\lepoly {\vec p}$ implies ${\vec q}\acirc {\vec r}\lepoly {\vec p}\acirc {\vec r}$ for any $\vec r$.
\item
Let $\mathfrak m$ be a monomial of ${\vec q}^{(i)}$. When we compose ${\vec q}^{(i)}$ over $\vec q$ (to obtain ${\vec q}^{(i+1)}$), we
 substitute ${\vec q}[j]$ for each variable $x_j$ in $\mathfrak m$.  Since $j\in \sd{\vec q}$, we have $x_j \lepoly {\vec q}[j]$.  Thus, among the monomials
 resulting from this substitution, there is a monomial identical to $\mathfrak m$.
 \qedhere
\end{enumerate}
\end{proof}

\begin{lem}\label{lem:newLEold}
Every ${\vec p}\in LC(\setS)$ is subsumed by a member of $\procSDL{}(\setS)$.
\end{lem}

\begin{proof}
We proceed by three steps, following the definition of $LC$.  First, if ${\vec p}\in \closure{\setS}$ then it is clearly also in $\procSDL{}(\setS)$.
Next, consider ${\itker{\vec p}}^\tau$ for  ${\vec p}\in \closure{\setS}$.
By Lemma~\ref{lem:idempotentPower},
choose $k$ such that ${\vec p}^{(k)}$ is idempotent. Hence ${({\vec p}^{(k)})}^\tau \in \procSDL{}(\setS)$.
Applying Lemma~\ref{lem:itPowers} to ${\itker{\vec p}}$, we obtain:
\[
 \itker{\vec p} \lepoly {\itker{\vec p}}^{(k)} \lepoly {\vec p}^{(k)}
 \]
 Moreover, it is not hard to extend these inequalities to
 \[
 {\itker{\vec p}}^\tau \lepoly {({\itker{\vec p}}^{(k)})}^{\tau} \lepoly {({\vec p}^{(k)})}^\tau
 \]
 Thus ${\itker{\vec p}}^\tau$  is subsumed by a member of  $\procSDL{}(\setS)$.

 Finally, generalized AMPs are subjected to composition-closure again, but its results will certainly be subsumed by corresponding compositions in
  $\procSDL{}(\setS)$, as the latter is closed under composition, and composition respects the subsumption order.
  \end{proof}

\bibliographystyle{alpha}
\bibliography{icc,sct,integer-loops}

\end{document}